\documentclass[useAMS,usenatbib]{mn2e}

\usepackage{url,times,graphicx,amsmath,amsfonts,amssymb,aas_macros,color,epsfig,epstopdf}

\usepackage[dvipsnames,svgnames,hyperref]{xcolor}
\usepackage{footnote}

\usepackage[
    pdftex,% get better results and can break links but can't use psfrag
    a4paper=true,
    breaklinks=true,
    bookmarks=true,
    bookmarksopen=false,
    bookmarksopenlevel=2
    bookmarksnumbered=true,
    bookmarkstype=toc,
    colorlinks=true,
    citecolor=NavyBlue,
    linkcolor=black,
    menucolor=green,
    urlcolor=blue,
]{hyperref}
\usepackage{comment}

\usepackage{multirow} %Tabellen Zeilenumbruch
\usepackage{multicol}
\usepackage{subcaption}
\usepackage{rotating}

\usepackage{xspace}

\newcommand{\galacticus}{\textsc{Galacticus}\xspace}

\newcommand{\sag}{\textsc{Sag}\xspace}
\newcommand{\sage}{\textsc{Sage}\xspace}

\newcommand{\consistenttree}{\textsc{Consistent Trees}}

\newcommand{\rockstar}{\textsc{Rockstar}}

\newcommand{\Mbnd}{{\ifmmode{M_{\rm bnd}}\else{$M_{\rm bnd}$}\fi}}
\newcommand{\Mfof}{{\ifmmode{M_{\rm fof}}\else{$M_{\rm fof}$}\fi}}
\newcommand{\Mcrit}{{\ifmmode{M_{\rm 200c}}\else{$M_{\rm 200c}$}\fi}}
\newcommand{\Mmean}{{\ifmmode{M_{\rm 200m}}\else{$M_{\rm 200m}$}\fi}}
\newcommand{\MBN}{{\ifmmode{M_{\rm BN98}}\else{$M_{\rm BN98}$}\fi}}

\newcommand{\Rcrit}{{\ifmmode{R_{\rm 200c}}\else{$R_{\rm 200c}$}\fi}}

\newcommand{\hGpc}{{\ifmmode{h^{-1}{\rm Gpc}}\else{$h^{-1}$Gpc}\fi}}
\newcommand{\hMpc}{{\ifmmode{h^{-1}{\rm Mpc}}\else{$h^{-1}$Mpc}\fi}}
\newcommand{\hkpc}{{\ifmmode{h^{-1}{\rm kpc}}\else{$h^{-1}$kpc}\fi}}
\newcommand{\hMsun}{{\ifmmode{h^{-1}{\rm {M_{\odot}}}}\else{$h^{-1}{\rm{M_{\odot}}}$}\fi}}
\newcommand{\Mstar}{{\ifmmode{M_{*}}\else{$M_{*}$}\fi}}
\newcommand{\SPS}{\textsc{\textit{SPS}}\xspace}
\newcommand{\SMF}{\textsc{\textit{SMF}}\xspace}
\newcommand{\SMFs}{\textsc{\textit{SMFs}}\xspace}
\newcommand{\LF}{\textsc{\textit{LF}}\xspace}
\newcommand{\SFR}{\textsc{\textit{SFR}}\xspace}
\newcommand{\SFRs}{\textsc{\textit{SFRs}}\xspace}
\newcommand{\SFRF}{\textsc{\textit{SFRF}}\xspace}

\newcommand{\sSFR}{\textsc{\textit{sSFR}}\xspace}

\newcommand{\cSFRD}{\textsc{\textit{cSFRD}}\xspace}
\newcommand{\SHMF}{\textsc{\textit{SHMF}}\xspace}

\newcommand{\BHBM}{\textsc{\textit{BHBM}}\xspace}
\newcommand{\CGF}{\textsc{\textit{CGF}}\xspace}

\newcommand{\Mcold}{$M_{\rm cold}$\xspace}

\newcommand{\tao}{\textsc{Theoretical Astrophysical Observatory}\xspace}
\newcommand{\TAO}{\textsc{TAO}\xspace}
\newcommand{\MD}{\textsc{MultiDark}\xspace}
\newcommand{\MDG}{\textsc{MultiDark-Galaxies}\xspace}
\newcommand{\Planck}{\textsc{Planck}\xspace}
\newcommand{\MDPL}{\textsc{MDPL2}\xspace}
\newcommand{\Mhalo}{{\ifmmode{M_{\rm Halo}}\else{$M_{\rm Halo}$}\fi}}
\newcommand{\Ngal}{{\ifmmode{N_{\rm gal}}\else{$N_{\rm gal}$}\fi}}
\newcommand{\Norph}{{\ifmmode{N_{\rm orphan}}\else{$N_{\rm orphan}$}\fi}}
\newcommand{\Nxorph}{{\ifmmode{N_{\rm non-orphan}}\else{$N_{\rm non-orphan}$}\fi}}
\newcommand{\Zsolar}{{\ifmmode{Z_{\odot}}\else{$Z_{\odot}$}\fi}}
\newcommand{\Msun}{{\ifmmode{{\rm {M_{\odot}}}}\else{${\rm{M_{\odot}}}$}\fi}}
\newcommand{\ltsima}{$\; \buildrel < \over \sim \;$}
\newcommand{\gtsima}{$\; \buildrel > \over \sim \;$}
\newcommand{\lsim}{\lower.5ex\hbox{\ltsima}}
\newcommand{\gsim}{\lower.5ex\hbox{\gtsima}}
\def\LCDM{$\Lambda$CDM}

\def\lesssim{\mathrel{\hbox{\rlap{\hbox{\lower4pt\hbox{$\sim$}}}\hbox{$<$}}}}
\def\gtrsim{\mathrel{\hbox{\rlap{\hbox{\lower4pt\hbox{$\sim$}}}\hbox{$>$}}}}

\newcommand{\Tab}[1]{Table~\ref{#1}}
\newcommand{\Sec}[1]{Section~\ref{#1}}
\newcommand{\App}[1]{Appendix~\ref{#1}}
\newcommand{\Eq}[1]{Eq.~(\ref{#1})}
\newcommand{\Fig}[1]{Fig.~\ref{#1}}
\newcommand{\beq}{\begin{equation}}
\newcommand{\eeq}{\end{equation}}
\def\beqa{\begin{eqnarray}}
\def\eeqa{\end{eqnarray}}
\def\Mpc{${\rm Mpc}$}

\def\hMpc{$h^{-1}\,{\rm Mpc}$}
\def\hkpc{$h^{-1}\,{\rm kpc}$}
\def\LCDM{\ensuremath{\Lambda}CDM}

\title[\MDG]
{\MDG: data release and first results}
\author[Knebe et. al]
       {Alexander Knebe,$^{1,2,3}$ %\thanks{E-mail: alexander.knebe@uam.es}
        Doris Stoppacher,$^{4,1,\dagger}$
        Francisco Prada,$^{5}$
        Christoph Behrens,$^{6}$ 			  		% Galacticus
\newauthor
        Andrew Benson,$^{7}$    					% Galacticus
        Sofia A. Cora,$^{8,9}$					% SAG
        Darren J. Croton,$^{10}$			                 % SAGE
        Nelson D. Padilla,$^{11,12}$ 				% SAG
\newauthor
        Andr\'es N. Ruiz,$^{13,14}$                                 % SAG
        Manodeep Sinha,$^{10}$			                 % SAGE
        Adam R. H. Stevens,$^{15,10}$		                 % SAGE
\newauthor
        Cristian A. Vega-Mart\'inez,$^{8}$ 			 % SAG
        Peter Behroozi,$^{16,\ddagger}$					% halo catalogues and merger trees
	Violeta Gonzalez-Perez,$^{17}$			% very useful contributions
\newauthor
        Stefan Gottl\"ober,$^{18}$	              			% simulation time and database
        Anatoly A. Klypin,$^{19}$                                     % MultiDark member
        Gustavo Yepes,$^{1,2,3}$					 % simulations
	Harry Enke,$^{18}$                                             % database manager
\newauthor
	Noam I. Libeskind,$^{18}$                                  % MultiDark database administration
        Kristin Riebe,$^{18}$						% database upload
        Matthias Steinmetz$^{18}$                                 % database access
\\
\\
$^{1}$Departamento de F\'isica Te\'{o}rica, M\'{o}dulo 15, Facultad de Ciencias, Universidad Aut\'{o}noma de Madrid, 28049 Madrid, Spain\\
$^{2}$Centro de Investigaci\'{o}n Avanzada en F\'isica Fundamental (CIAFF), Facultad de Ciencias, Universidad Aut\'{o}noma de Madrid, 28049 Madrid, Spain\\
$^{3}$Astro-UAM, UAM, Unidad Asociada CSIC\\
$^{4}$ Instituto de F\'{\i}sica Te\'orica, (UAM/CSIC), Universidad Aut\'onoma de Madrid, Cantoblanco, E-28049 Madrid, Spain\\
$^{5}$ Instituto de Astrof\'{\i}sica de Andaluc\'{\i}a (CSIC), Glorieta de la Astronom\'{\i}a, E-18080 Granada, Spain\\
$^{6}$Institut f\"{u}r Astrophysik, Georg-August Universit\"at G\"{o}ttingen, Friedrich-Hund-Platz 1, 37077, G\"{o}ttingen, Germany\\
$^{7}$Carnegie Observatories, 813 Santa Barbara Street, Pasadena, CA 91101, USA\\
$^{8}$Instituto de Astrof\'isica de La Plata (CCT La Plata, CONICET, UNLP), Paseo del Bosque s/n, B1900FWA, La Plata, Argentina\\
$^{9}$Facultad de Ciencias Astron\'omicas y Geof\'{\i}sicas, Universidad Nacional de La Plata, Paseo del Bosque s/n, B1900FWA, La Plata, Argentina\\
$^{10}$Centre for Astrophysics and Supercomputing, Swinburne University of Technology, Hawthorn, Victoria 3122, Australia\\
$^{11}$Instituto de Astrof\'isica, Universidad Cat\'olica de Chile, Santiago, Chile\\
$^{12}$Centro de Astro-Ingenier\'ia, Universidad Cat\'olica de Chile, Santiago, Chile\\
$^{13}$Instituto de Astronom\'ia Te\'orica y Experimental (CONICET-UNC), Laprida 854, X5000BGR, C\'ordoba, Argentina\\
$^{14}$Observatorio Astron\'omico de C\'ordoba, Universidad Nacional de C\'ordoba, Laprida 854, X5000BGR, C\'ordoba, Argentina\\
$^{15}$International Centre for Radio Astronomy Research, University of Western Australia, 35 Stirling Highway, Crawley, Western Australia 6009, Australia\\
$^{16}$Department of Astronomy, University of California at Berkeley, Berkeley, CA 94720, USA\\
$^{17}$Institute of Cosmology and Gravitation, University of Portsmouth, Portsmouth PO1 3FX, UK\\
$^{18}$Leibniz-Institut f\"ur Astrophysik Potsdam (AIP), An der Sternwarte 16, D-14482 Potsdam, Germany\\
$^{19}$Astronomy Department, New Mexico State University, Dept.4500, Las Cruces, NM 88003-0001, USA\\
$^{\dagger}$Severo Ochoa IFT-CSIC Scholar\\
$^{\ddagger}$Hubble Fellow\\
}

\begin{document}

\date{Accepted XXXX . Received XXXX; in original form XXXX}

\pagerange{\pageref{firstpage}--\pageref{lastpage}} \pubyear{2010}

\maketitle

\label{firstpage}

% this will move the abstract to the second page
%\clearpage

%%%%%%%%%%%%%%%%%%%%%%%%%%%%%%%%%%%%%%%%%%%%%%%%%%%
\begin{abstract}

We present the public release of the \MDG: three distinct galaxy catalogues derived from one of the Planck cosmology \MD simulations (i.e. \MDPL, with a volume of (1 \hGpc)$^3$ and mass resolution of $1.5\times 10^{9}\hMsun$) by applying the semi-analytic models \galacticus, \sag, and \sage\ to it. We compare the three models and their conformity with observational data for a selection of fundamental properties of galaxies like stellar mass function, star formation rate, cold gas fractions, and metallicities -- noting that they sometimes perform differently reflecting model designs and calibrations. We have further selected galaxy subsamples of the catalogues by number densities in stellar mass, cold gas mass, and star formation rate in order to study the clustering statistics of galaxies. We show that despite different treatment of orphan galaxies, i.e. galaxies that lost their dark-matter host halo due to the finite mass resolution of the $N$-body simulation or tidal stripping, the clustering signal is comparable, and reproduces the observations in all three models -- in particular when selecting samples based upon stellar mass. Our catalogues provide a powerful tool to study galaxy formation within a volume comparable to those probed by on-going and future photometric and redshift surveys. All model data consisting of a range of galaxy properties -- including broad-band SDSS magnitudes -- are publicly available.

\end{abstract}
%%%%%%%%%%%%%%%%%%%%%%%%%%%%%%%%%%%%%%%%%%%%%%%%%%%
\noindent

%\clearpage
%%%%%%%%%%%%%%%%%%%%%%%%%%%%%%%%%%%%%%%%%%%%%%%%%%%
\section{Introduction} \label{sec:introduction}
%%%%%%%%%%%%%%%%%%%%%%%%%%%%%%%%%%%%%%%%%%%%%%%%%%%
Galaxy formation is one of the most complex phenomena in astrophysics, as it involves scales from the large-scale structure of the Universe down to the sizes of black holes \citep[e.g.][]{Silk12,Silk13}. And during the last few decades we have witnessed great steps in the field of galaxy formation within a cosmological context. On the one hand, through directly accounting for the baryonic component (gas, stars, supermassive black holes, etc.) in cosmological simulations that include hydrodynamics and gravity, and on the other hand through `semi-analytic galaxy formation' modelling (SAM). The former approach has left us to date with excellent cosmological simulations such as Illustris \citep{Vogelsberger14}, Horizon-AGN \citep{Dubois14}, EAGLE \citep{Schaye15}, Magneticum \citep{Dolag15}, and MassiveBlack~II \citep{Khandai15}  -- just to name the full box simulations, i.e. simulations with a unique mass resolution across the whole volume modelled. However, these volumes are still much smaller than those covered by on-going and upcoming large surveys (see below). There are also groups that focus the computational time on individual objects, still within a cosmological volume, but increasing the mass resolution to a level suitable to model galaxy formation only within a much smaller sub-volume \citep[e.g.][]{Governato10,Guedes11,Hopkins14,Wang15,Grand17} -- of which some are even constraining their initial conditions in a way to model the actual observed Local Universe \citep{Gottloeber10,Yepes14,Sawala16}.

Besides of advances in hydrodynamical simulation, the last few decades have also seen great improvements in  aforementioned semi-analytic galaxy formation modelling in which the distribution of dark-matter haloes and their merger history -- mostly extracted from $N$-body cosmological simulations these days -- is combined with simplified yet physically motivated prescriptions to estimate the distribution and physical properties of galaxies. Those models date back to the work of \citet{White78} who used a synthesis of the theory of \citet{Press74} to describe the hierarchy of gravitationally bound structures, and gas cooling arguments to motivate the first ideas of galaxy formation. White \& Rees proposed a two-stage process for galaxy formation: dark-matter haloes form first via gravitational collapse and then provide the potential wells for gas to cool and subsequently form galaxies. This idea was picked up later by \citet{white91} where it was developed into a  semi-analytic method for studying the formation of galaxies by gas condensation within dark matter haloes. Their model included gas cooling, star formation, evolution of stellar populations, stellar feedback, and chemical enrichment. This has been refined and improved over the following years leading to highly successful semi-analytic models \citep[for a review see][]{baugh_review_2006,Benson10,Somerville14}.

The strong point of SAMs over direct hydrodynamical simulations is that they are computationally far less expensive. This allows the construction of a multitude of galaxy catalogues exploring parameter space \citep[e.g.][]{Henriques09,ruiz_2015,Rodrigues17}. A SAM further facilitates the addition of new physics without the need of re-running the cosmological simulation as would be the case for a hydrodynamical simulation. But any model of galaxy formation depends on prescriptions for all the physical processes we believe are relevant for galaxy formation. These recipes are not precisely known but are each regulated by several parameters that are chosen to satisfy one or more observational constraints. While in the past this has been primarily accomplished by means of one-point functions (like the stellar mass function, the black hole-bulge mass relation, the star formation rate density, etc.), more recent studies have extended their recipes for galaxy formation to two-point functions \citep[e.g. the two-point correlation function of galaxies; see][]{Kauffmann99a,Kauffmann99b,Benson00,vanDaalen16}. 

SAMs can be considered the most versatile tool when it comes to studying the multitude of galaxy properties such as sizes, masses, metallicities, luminosities, etc. as well as their individual components like disc, bulge, halo, black hole, etc. However, when interpreting and using the resulting galaxy catalogues from SAMs one needs to bear in mind that these models are primarily tools: our understanding of galaxy formation is still not advanced enough to `predict' every possible galaxy property. For that reason one needs to distinguish between actual model `predictions' and `descriptions', i.e. model parameters have to be tuned to reproduce selected observational data. But this calibration is a highly degenerate process and may also depend on the scientific question to be addressed. \citet{Knebe15} have  shown that there exist significant model-to-model variations when applying different SAMs to the same cosmological dark-matter-only simulation (especially when not re-calibrating the parameters, Knebe et al., in prep.). And if the SAM parameters have been tuned to a certain observation this particular galaxy property is then `described' rather than `predicted'. But this process also allows to adjust the model to the actual needs and objectives of any galaxy study. If the aim is to investigate, for instance, galaxy clustering, one might refrain from using the observed two-point correlation function during the calibration of the model parameters so that it becomes a clear prediction. Further, models might also put a different emphasis on certain galaxy properties aiming at predicting (or describing) them better than other properties. We will return to this point later (in \hyperref[sec:differences]{\Sec{sec:differences+similarities}}) when we highlight the similarities and differences between the three models used in this study. But we like to already stress here that our galaxy catalogues are diverse enough to provide the community with predictions/descriptions that fit the needs of users with assorted interests in galaxies as we chose to not only apply one but three well-tested SAMs to one of the \MD\ dark-matter-only cosmological simulations in a flat \LCDM\ Planck cosmology.

While the field of galaxy formation is very much driven by observations where cosmological simulations provide the gravitational scaffolding for it, semi-analytic modelling of galaxy formation now combines both providing the framework for theoretically interpreting, understanding, and even predicting new results verifiable observationally. Access to such models attracts an ever growing interest and relevance with galaxy surveys nowadays routinely mapping millions of galaxies. Extracting information from on-going and upcoming surveys (such as eBOSS, DES, J-PAS, DESI, LSST, Euclid, WFIRST) requires theoretical models and galaxy catalogues comparable in volume to the sizes of these surveys, which still is a highly demanding task and not feasible by means of hydrodynamical simulations yet. The MultiDark simulations have been especially helpful in designing current cosmological surveys, such as SDSS-IV/eBOSS \citep{Favole16,Rodriguez-Torres17,Comparat17,Favole17}. But so far all these works have been using empirical models together with the MultiDark simulation. The new catalogues are providing the opportunity to have physically motivated models to populate the simulation and thus, they can be useful for exploring the physical properties of cosmological tracers of current and future surveys. Moreover, given that \galacticus\ and \sage\ are publicly available codes, it also provides the opportunity to re-run these models on this simulation, but varying their parameters to explore particular aspects of the galaxies clustering and the dependence on their physical properties.

Within this work we present a study of the properties of the three distinct galaxy catalogues which is divided into three main parts: \hyperref[sec:models]{\Sec{sec:models}} primarily introduces the SAMs highlighting their differences and similarities. In that Section we also present the \MD \Planck2 simulation \citep[\MDPL][]{Klypin16} with a cubical volume of (1000~\hMpc)$^3$. The mass and temporal resolution of the simulation is sufficiently high to allow for post-processing with semi-analytic galaxy formation models \citep[see][]{Guo11,Benson12b}. In \hyperref[sec:comparison]{\Sec{sec:comparison}} we present the \MDG\ by calculating distributions and correlations of the most fundamental properties (see \hyperref[tab:plots]{\Tab{tab:plots}} for an overview), and compare them to observational and computational data. In \hyperref[sec:clustering]{\Sec{sec:clustering}}, we then select subsamples by number density cuts using stellar mass, cold gas mass, and star formation rate to study the two-point correlation function (2PCF). We further present a comparison to the observed projected two-point correlation function (p2PCF). A summary and discussion can be found in \hyperref[sec:discussion]{\Sec{sec:discussion}}.

All further and more detailed studies of the \MD semi-analytic catalogues would be beyond the scope of this paper which is mainly written to present our models and provide some first results which verify the validity and show possible limitations of the catalogues. The simulation itself and its associated dark matter haloes, merger trees, and the catalogues of the \MDG\ are publicly available.

%%%%%%%%%%%%%%%%%%%%%%%%%%%%%%%%%%%%%%%%%%%%%%%%%%%
\section{The Simulation and Galaxy Formation Models} \label{sec:models}
%%%%%%%%%%%%%%%%%%%%%%%%%%%%%%%%%%%%%%%%%%%%%%%%%%%
In this Section we present -- in addition to the underlying cosmological simulation in \Sec{sec:simulation} -- the three semi-analytic models (\galacticus, \sag, and \sage) used to generate the three distinct galaxy catalogues \MDPL-\galacticus, \MDPL-\sag, and \MDPL-\sage. We briefly describe the implementation of physical processes for each model individually (\Sec{sec:galacticus}-\Sec{sec:sage}) before highlighting any differences and/or similarities in \Sec{sec:differences+similarities}.

\subsection{Simulation Data} \label{sec:simulation}
%%%%%%%%%%%%%%%%%%%%%%%%%%%%%%%%%%%%%%%%%%%%%%%%%%%
The simulation used in this work forms part of the aforementioed \textsc{CosmoSim} database. The original \MD\ (and Bolshoi) simulations as well as the structure of the database have been described in \citet{Riebe13}. Here we use a simulation from the \MD\ suite which follows the evolution of 3840$^3$ particles in a cubical volume of side-length 1475.6~\Mpc\ (1000~\hMpc) described in \citet{Klypin16}. The adopted cosmology consists of a  flat $\Lambda$CDM model with the \Planck cosmological parameters: $\Omega_m = 0.307, \Omega_B = 0.048, \Omega_\Lambda = 0.693, \sigma_8 = 0.823, n_s = 0.96$ and a dimensionless Hubble parameter $h = 0.678$ \citep{Planck2015}. This leaves us with a mass resolution of $m_p=1.51\times 10^{9}\hMsun$ per dark matter particle and a force resolution of 13 \hkpc\ (high $z$) to 5 \hkpc\ (low $z$). The catalogues are split into 126 snapshots between redshifts $z=17$ and $z=0$.

Haloes and subhaloes have been identified with \rockstar\ \citep{Behroozi12} and merger trees constructed with \consistenttree\ \citep{Behroozi12b}. All models follow and trace substructures explicitly from the $N$-body simulation. It has been demonstrated that both of these choices guarantee highly reliable halo catalogues and merger trees \citep{Knebe13b,Avila14,Behroozi15,Wang16}. Like most SAMs, the models operate on merger trees of dark matter haloes. A galaxy is potentially formed within each branch of each merger tree, and is defined by a set of properties. Some of these properties are determined by direct measurements from the $N$-body simulation (such as halo position, velocity, and spin). Most of the remaining properties are typically evolved using a set of differential equations. This differential evolution is sometimes interrupted by stochastic events (such as galaxy mergers). Finally, some properties (such as galaxy sizes) are determined under assumptions of equilibrium.

Below we now describe each of the SAM models as applied to the \MDPL\ simulation.

\subsection{\galacticus} \label{sec:galacticus}
%%%%%%%%%%%%%%%%%%%%%%%%%%%%%%%%%%%%%%%%%%%%%%%%%%%
As \galacticus\ is primarily described in \citet{Benson12}, we only summarise its salient features here.

{\bf Cooling:}
Cooling rates from the hot halo are computed using the traditional cooling radius approach \citep{white91}, with a time available for cooling equal to the halo dynamical time, and assuming a $\beta$-model profile with isothermal temperature profile (at the virial temperature) $\rho_{\mathrm h}(r) = \rho_{\rm h,0} \left[ r^2 + r_\beta^2 \right]^{3\beta/2}$, where $\beta=2/3$, $r_\beta=f_\beta r_{\rm v}$, $f_\beta=0.3$, and $\rho_{\rm h,0}$ is determined by normalizing to the total mass, $M_{\rm h}$, within radius $r_{\rm h}$. Metallicity dependent cooling curves are computed using \textsc{CLOUDY} \citep[v13.01,][]{Ferland13} assuming collisional ionisation equilibrium; we note that the differences with respect to \citet{sutherland93} for low-metallicities are very low whereas they can reach factors of up to 3 for metallicities of 0.1 solar and above.

{\bf Star formation:}
Star formation in discs is modelled using the prescription of \citet[][i.e. their Eq.~(1) for the star formation rate surface density, and Eq.~(2) for the molecular fraction]{Krumholz09}, assuming that the cold gas of each galaxy is distributed with an exponential radial distribution. The scale length of this distribution is computed from the disc's angular momentum by solving for the equilibrium radius within the gravitational potential of the disc+bulge+dark matter halo system \citep[accounting for adiabatic contraction using the algorithm of][]{Gnedin04}.

%The star formation rate in disks is computed using,
%\begin{equation}
%\phi_{\rm d} = \int_0^\infty {\rm d}R 2 \pi R \dot{\Sigma}_{\rm sf, d}(R),
%\end{equation}
%where $\dot{\Sigma}_{\rm sf, d}(R)$ is the star formation rate surface density 
%and is computed using the expression given by \cite{Krumholz09}:
%\begin{equation}
%\dot{\Sigma}_\star(R) = \nu_{\mathrm SF} f_{\mathrm H_2}(R)\Sigma_{\mathrm 
%HI, disk}(R) \left\{ \begin{array}{ll} (\Sigma_{\mathrm HI}/\Sigma_0)^{-1/3}, 
%&  \hbox{ if } \Sigma_{\mathrm HI}/\Sigma_0 \le 1 \\ (\Sigma_{\mathrm HI}/
%\Sigma_0)^{1/3}, & \hbox{ if } \Sigma_{\mathrm HI}/\Sigma_0 > 1 \end{array} 
%\right. ,
%\end{equation}
%where $\nu_{\mathrm SF}=0.385~\hbox{Gyr}^{-1}$ is a frequency and $\Sigma_0=85 
%M_\odot \hbox{pc}^{-2}$. The molecular fraction is given by
%\begin{equation}
%f_{\mathrm H_2} = 1 - \left( 1 + \left[ { 3 s \over 4 (1+\delta)} 
%\right]^{-5} \right)^{-1/5},
%\end{equation}
%where
%\begin{equation}
%\delta = 0.0712 \left[ 0.1 s^{-1} + 0.675 \right]^{-2.8},
%\end{equation}
%and
%\begin{equation}
%s = {\ln(1+0.6\chi+0.01\chi^2) \over 0.04 \Sigma_{\mathrm comp,0} Z^\prime},
%\end{equation}
%with
%\begin{equation}
%\chi = 0.77 \left[ 1 + 3.1 Z^{\prime 0.365} \right],
%\end{equation}
%and $\Sigma_{\mathrm comp,0}=c \Sigma_{\mathrm HI}/M_\odot \hbox{pc}^{-2}$ where $c=5$ is a density enhancement factor relating the surface density of molecular complexes to the gas density on larger scales.

{\bf Metal treatment:}
Metal enrichment is followed using the instantaneous recycling approximation, with a recycled fraction of 0.46 and yield of 0.035. Metals are assumed to be fully mixed in all phases, and so trace all mass flows between phases.

{\bf Supernova feedback and winds:}
The wind mass loading factor, $\beta$, is computed as $\beta = (V_{\rm disc}/250 {\rm km/s})^{-3.5}$ where $V_{\rm disc}$ is the circular velocity at the disc scale radius. Winds move cold gas from the disc back into the hot halo. For satellite galaxies, the ouflowing gas is added to the hot halo of the satellite's host.

{\bf Gas ejection \& reincorporation:}
Gas removed from galaxies by winds is retained in an outflowed reservoir. This reservoir gradually leaks mass back into the hot halo on a timescale of $t_{\rm dyn}/5$ where $t_{\rm dyn}$ is the dynamical time of the halo at the virial radius. As with all parameter values, the $1/5$ was chosen to give a reasonable match to a variety of datasets (see below). While the value is small (so reincorporation is fast), the results are not highly sensitive to this (e.g. if the value was $0$ instead of $1/5$ the results would not be dramatically different).

{\bf Disc instability:}
%The \citet{Efstathiou82} criterion is used to judge when discs are unstable, with a stability threshold that depends on the gas fraction in the disc (0.7 for pure gas discs, 1.1 for pure stellar, linearly interpolated in between). When a disc is unstable, it begins to transfer stars and gas from the disc to the bulge on a timescale that equals the disc dynamical time for a maximally unstable disc, and increases to infinite timescale as the disc approaches the stability threshold.

Material is transferred from the disc to the spheroid on an instability timescale $\tau_{\rm ins}$ which is given by
\begin{equation}
\tau_{\rm ins} = \left\{ \begin{array}{ll} (\Delta\epsilon_{\rm iso} / \Delta
\epsilon) \tau_{\rm d} & \hbox{ if } \epsilon < \epsilon_{\rm stab} \\ \infty 
& \hbox{ otherwise,} \end{array} \right.
\end{equation}
where $\tau_{\rm d}=R_{\rm d}/V_{\rm d}$ is the dynamical timescale of the disc, $\Delta \epsilon = \epsilon_{\rm stab} - \epsilon$, $\Delta\epsilon_{\rm iso}=\epsilon_{\rm stab}-\epsilon_{\rm iso}$, $\epsilon_{\rm stab}=(\epsilon_{\rm stab, g} M_{\rm g, d} +\epsilon_{\rm stab, \star} M_{\rm \star,d})/( M_{\rm g, d} + M_{\rm \star,d})$, $\epsilon_{\rm stab, g}=0.7$, $\epsilon_{\rm stab, \star}=1.1$,  $\epsilon= V_{\rm d, max} / [{\rm G} (M_{\rm \star, d}+M_{\rm g, d})/r]^{1/2}$ is the stability parameter defined by \cite{Efstathiou82}, $V_{\rm d, max} = \chi_{\rm d} V_{\rm d}$ is the maximum of the disc rotation curve, $\chi_{\rm d}\approx 1.18$ converts velocity at the scale-radius to the maximum velocity (assuming an exponential disc which is the only source of gravitational potential), and $\epsilon_{\rm iso} \approx 0.622$ is the stability parameter attained for an exponential disc which is the only source of gravitational potential). In this way, discs are unstable if $\epsilon < \epsilon_{\rm stab}$, and the timescale for instability decreases from infinity at the stability threshold to the dynamical timescale for a maximally unstable disc.

{\bf Starburst:}
There is no special `starburst' mode in \galacticus. Instead, gas in the spheroid forms stars at a rate $\dot{M}_\star = 0.04 M_{\rm gas}/t_{\rm dyn} (V/200 {\rm km/s})^{-2}$, where $t_{\rm dyn}$ is the dynamical time of the spheroid at its half mass radius, and $V$ its circular velocity at the same radius.

{\bf AGN feedback:}
%The mass and spin of black holes are followed in detail, assuming black holes accrete from both the hot gas halo and the ISM of the spheroid component. We compute the power of the jet produced by the black hole using the method of \citet{Benson09}, which offsets the cooling luminosity in hydrostatic hot haloes, thereby reducing the cooling rate onto the galaxy. Additionally, a radiatively-driven wind is launched from the spheroid by the black hole, assuming that a fraction 0.0024 of its radiative output couples efficiently to the outflow. 

The mass and spin of black holes are followed in detail, assuming black holes accrete from both the hot gas halo and the ISM of the spheroid component at rates
\begin{eqnarray}
\dot{m}_{\rm acc, h} &=& \hbox{min}[ \mathcal{C}_{\rm h} \dot{m}_{\rm Bondi}
(m_{\rm BH},\rho_{\rm h},T_{\rm h}), \dot{m}_{\rm Edd} / \epsilon_{\rm rad} ], 
\\
\dot{m}_{\rm acc, s} &=& \hbox{min}[ \mathcal{C}_{\rm s} \dot{M}_{\rm Bondi}
(m_{\rm BH},\rho_{\rm s},T_{\rm s}), \dot{m}_{\rm Edd} / \epsilon_{\rm rad} ],
\end{eqnarray}
resulting in the black hole gaining mass at rates
\begin{eqnarray}
\dot{m}^\prime_{\rm acc, h} &=& (1-\epsilon_{\rm rad}-\epsilon_{\rm jet}) 
\dot{m}_{\rm acc, h}, \\
\dot{m}^\prime_{\rm acc, s} &=& (1-\epsilon_{\rm rad}-\epsilon_{\rm jet}) 
\dot{m}_{\rm acc, s}.
\end{eqnarray}
In the above $\mathcal{C}_{\rm h}=6$ and $\mathcal{C}_{\rm s}=5$ are numerical factors, $\dot{m}_{\rm Bondi}(M,\rho,T)$ is the Bondi accretion rate for gas of density $\rho$, and temperature $T$ onto a stationary black hole of mass $m_{\rm BH}$, $\dot{m}_{\rm Edd}$ is the Eddington accretion rate for the black hole, $\epsilon_{\rm rad}$ is the radiative efficiency of the accretion disc feeding the black hole, and $\epsilon_{\rm jet}$ is the jet efficiency (defined as the jet power divided by the accretion power, $\sum_i \dot{m}_{{\rm acc}, i} {\rm c}^2$). For the Bondi accretion rate from the spheroid, $\rho_{\rm s}$ is the density of gas in the spheroid at the larger of the Bondi radius and Jeans length, and we assume $T_{\rm s}=100$~K. For accretion from the hot halo, $T_{\rm h}=T_{\rm v}$, and $\rho_{\rm h}$ is computed at the Bondi radius but reduced by a factor $f_{\rm h}$ as we assume accretion can only occur from the fraction of the hot halo mass actually in the hot mode. 

We do not explicitly model whether haloes are undergoing hot or cold mode accretion, and so instead impose a simple transition from cold-mode to hot-mode behaviour at the point where a halo (were it in the hot mode) is able to cool out to the virial radius \citep[see details in][]{benson_cold_2010}.

%We therefore use 
%\begin{equation}
%f_{\rm h}(x) = \left\{ \begin{array}{ll} 1 & \hbox{ if } x < 0, \\ x^2 (2x-3) 
%+1 & \hbox{ if } 0 \le x \le 1, \\ 0 & \hbox{ if } x > 1, \end{array} \right.
%\end{equation}
%where
%\begin{equation}
%x=\frac{r_{\rm cool}/r_{\rm v}-x_{\rm h}}{1-x_{\rm h}},
%\end{equation}
%with $x_{\rm h}=0.9$, to give a smooth, but reasonably sharp transition between hot-mode dominated and cold-mode dominated haloes as the cooling radius approaches the virial radius.

As discussed in detail by \cite{begelman_accreting_2014}, accretion flows with accretion rates close to the Eddington limit will be radiatively inefficient as they struggle to radiate the energy they release, while flows with accretion rates that are much smaller than Eddington ($\dot{M}_{\rm acc} < \alpha^2 \dot{M}_{\rm Edd}$, where $\alpha\sim 0.1$ is the usual parameter controlling the rate of angular momentum transport in a \cite{shakura_black_1973} accretion disc) are also radiatively inefficient as radiative processes are too inefficient at the associated low densities to radiate energy at the rate it is being liberated. Therefore, accretion disc structure is assumed to be a radiatively-efficient, geometrically thin, \cite{shakura_black_1973} accretion disc if the accretion rate is between $0.01\dot{M}_{\rm Edd}$ and $0.3\dot{M}_{\rm Edd}$, and an advection dominated accretion flow (ADAF) otherwise \citep{begelman_accreting_2014}.

For thin discs and high-accretion rate ADAFs the radiative efficiency is given by $\epsilon_{\rm rad}=1-E_{\rm ISCO}$, where $E_{\rm ISCO}$ is the specific energy of the innermost stable circular orbit (in dimensionless units) for the given black hole spin. For low accretion rate ADAFs the radiative efficiency is matched to that of the thin disc solution at the transition point ($0.01\dot{M}_{\rm Edd}$) and is decreased linearly with accretion rate below that.

For the jet efficiency in thin accretion discs we use the results of \cite{meier_association_2001}, interpolating between their solutions for Schwarzchild black holes and rapidly rotating Kerr black holes. For the case of ADAF accretion flows we use the jet efficiency computed by \cite{Benson09}. Note that the only role of black hole spin is to determine the jet power for a given accretion rate.

{\bf Merger treatment:}
A merger between two galaxies is deemed to be `major' if their (baryonic) mass ratio exceeds 1:4. In major mergers, the stars and gas of the two merging galaxies are rearranged into a spheroidal remnant. In other, minor mergers, the merging galaxy is added to the spheroid of the galaxy that it merges with, while the disc of that galaxy is left unaffected.

{\bf Orphans:}
When a subhalo can no longer be found in the $N$-body merger trees, a `subresolution merging time' is computed for the subhalo \citep[based on its last known orbital properties and the algorithm of][]{Boylan-Kolchin08}. The associated galaxy is then an orphan, which continues to evolve as normal (although we have no detailed knowledge of its position within its host halo) until the subresolution merging time has passed, at which point it is assumed to merge with the central galaxy of its host halo.

{\bf Calibration method:}
The parameters of galaxy formation physics in \galacticus\ have been chosen by manually searching parameter space and seeking models which provide a reasonable match to a variety of observational data, including the $z=0$ stellar mass function of galaxies \citep{li_distribution_2009}, $z=0$ K and b$_{\rm J}$-band luminosity functions \citep{cole_2df_2001,norberg_2df_2002}, the local Tully-Fisher relation \citep{pizagno_tully-fisher_2007}, the colour-magnitude distribution of galaxies in the local Universe \citep{weinmann_properties_2006}, the distribution of disc sizes at $z=0$ \citep{de_jong_local_2000}, the black hole--mass bulge mass relation \citep{haring_black_2004}, and the star formation history of the Universe \citep{hopkins_evolution_2004}. However, we need to remind the reader that the model has \textit{not} been recalibrated to the \MDPL\ simulation used for this project.

\subsection{\sag} \label{sec:sag}
%%%%%%%%%%%%%%%%%%%%%%%%%%%%%%%%%%%%%%%%%%%%%%%%%%%
\sag\ model originates from a version of the Munich code \citep{springel_subfind_2001} and has been further developed and improved as described in \citet{cora_sam_2006}, \citet{lagos_sam_2008}, \citet{tecce_sam_2010}, \citet{orsi_sam_2014}, \citet{munnozarancibia_sam_2015} and \citet{gargiulo_sam_2015}. The latest version of the model is presented by Cora et al. (in prep.). The major changes introduced are related to supernova and AGN feedback, gas ejection and reincorporation, and environmental effects, coupled to a detailed treatment of the orbits of orphan galaxies. 

{\bf Cooling:}
Radiative cooling of the hot gas in the halo is treated as in \citet{white91}, but with the metal-dependent cooling function estimated by considering the radiated power per chemical element obtained from the plasma modelling code \textsc{AtomDB v2.0.2} \citep{foster_xray_2012}. Gas inflows generate gaseous discs with an exponential density profile. Both central and satellite galaxies acquire gas through cooling processes. Galaxies keep their hot gas halo when they become satellites, which are gradually removed by the action of tidal stripping and ram pressure stripping (RPS); the latter is modelled according to \citet{mccarthy2008}. The amount of gas stripped is determined by the stronger effect. When a significant fraction ($90$ per cent) of the hot halo is removed, the cold gas disc can also be affected by RPS following the criterion from \citet{gg72}, as explained in detail in \citet{tecce_sam_2010}.

{\bf Star formation:}
An event of quiescent star formation takes place when the mass of the cold gas disc exceeds a critical limit ($M_{\rm cold,crit}$), as in \citet{Croton06}, according to the star formation law ${\dot{M}_\star} = \alpha {{M_{\rm cold} - M_{\rm cold,crit}}/{t_{\rm dyn}}}$, with $M_{\rm cold,crit} = 3.8\times10^9 \left({V_{\rm vir}}\over{200 {\rm km}\,{\rm s}^{-1}}\right) 
\left({3R_{\rm disc}} \over{\rm~10kpc}\right) \Msun$, where $t_{\rm dyn}=V_{\rm vir}/3R_{\rm disc}$ is the dynamical time of the galaxy, $V_{\rm vir}$ is the circular velocity at the virial radius and $R_{\rm disc}$ the disc scale length calculated as described in \cite{tecce10}. The star formation effciciency is given by the free parameter $\alpha$.

{\bf Metal treatment:}
The chemical model included in SAG follows the detailed implementation described in \citet{cora_sam_2006} in which stars in different mass ranges can contaminate the cold and hot gas because of mass loss during their stellar evolution and metal ejection at the end of their lives. Their stellar yields have been updated as detailed in \citep{gargiulo_sam_2015}. Namely, for low and intermediate-mass stars (mass interval $1-8 \, M_{\odot}$), the code considers yields given by \citet{Karakas2010}, while it adopts results from \citet{Hirschi2005} and \citet{Kobayashi2006} for the mass loss of pre-supernova stars (He and CNO elements) and the explosive nucleosynthesis of core collapse supernovae (SNe CC), respectively; all of these yields correspond to stars with solar metallicities. Rates of supernovae type Ia (SNe Ia) are estimated using the single degenerate model \citep{Lia2002} with ejecta given by \citet{Iwamoto99}. Metals are recycled back to the gas phase taking into account stellar lifetimes \citep{Padovani93}. Thus, the model tracks the production and circulation of eight chemical elements (H, $^{4}$He, $^{12}$C, $^{14}$N, $^{16}$O, $^{24}$Mg, $^{28}$Si, $^{56}$Fe) generated by stars with masses distributed in 27 mass ranges, from $1$ to $100\,M_{\odot}$, relaxing the instantaneous recycling approximation. Initially, the hot gas has primordial abundances (76 per cent of hydrogen and 24 per cent of helium), but becomes chemically enriched as a result of gas reheating by supernovae explosions that transfers contaminated cold gas to the hot phase, which calls for the use of  metal-dependent cooling rates. Gas cooling in turn influences the level of star formation which is ultimately responsible for the chemical pollution.

{\bf Supernova feedback and winds:}
The energy released by SNe CC determines the amount of reheated cold gas that is transferred to the hot gas phase of the galaxy host halo. The reheated mass is inversely proportional to the square of the halo virial velocity. The mass transfer takes place when SNe CC explode, to be consistent with the chemical model implemented \citep{cora_sam_2006}. For satellite galaxies, the hot gas halo is reduced by gas cooling and environmental effects but can also be rebuilt by the transfer of reheated gas, receiving mass and metals proportionally to its mass. This takes place whenever the fraction of hot gas with respect to the total baryonic content of the galaxy is above a certain fraction considered as a free parameter of the model. The estimation of reheated mass is modified by adding a dependence on redshift and an additional modulation with virial velocity, according to a fit to hydrodynamical simulations results presented by \citet{muratov_2015}, so that the current prescription is $\Delta M_{\rm reheated} = \frac{4}{3} \epsilon {\frac{\eta E_\text{SN}}{V_{\rm vir}^2}} \,(1+z)^{\beta}\,\left(\frac{V_{\rm vir}}{60\,{\rm km}\,{\rm s}^{-1}}\right)^{\alpha}\Delta M_{\star}$, where the exponent $\alpha$ takes the values $-3.2$ and $-1.0$ for virial velocities smaller and larger than $60\,{\rm km}\,{\rm s}^{-1}$, respectively. The efficiency $\epsilon$ and the exponent $\beta$ are  free parameters of the model; the latter takes a value of $2$ during its calibration, a bit higher than the one corresponding to the fit provided by \citet{muratov_2015}.

{\bf Gas ejection \& reincorporation:}
Some of the hot gas is ejected out of the halo as a result of the energy input by massive stars according to the energy conservation argument presented by \citet{Guo11}, that is $\Delta M_{\rm ejected}= ({\Delta E_\text{SN} - 0.5\,\Delta M_{\rm reheated}\,V_{\rm vir}^2})/({0.5\,V_{\rm vir}^2})$, where $\Delta E_\text{SN}$ is the energy injected by massive stars which includes the same explicit redshift dependence and the additional modulation with virial velocity as the reheated mass (see above), with its corresponding efficiency $\epsilon_\text{ejec}$. It also involves the factor $0.5\,V_\text{SN}^2$ which is the mean kinetic energy of SN ejecta per unit mass of stars formed, being $V_\text{SN}=1.9\,V_\text{vir}^{1.1}$ \citep{muratov_2015}. The ejected gas mass is re-incorporated back onto the corresponding (sub)halo within a timescale that depends on the inverse of (sub)halo mass following \citet{henriques_mcmc_2013}.

{\bf Disc instability:}
Galactic discs with high surface densities become unstable against small perturbations according to the criterion of \citet{Efstathiou82}. \sag\ model considers that the presence of a neighbouring galaxy perturbs the unstable disc triggering the instability; this condition involves the mean separation between galaxies in a main host halo. When the instability is triggered, stars are transferred to the bulge component along with the cold gas that is consumed in a starburst.

{\bf Starburst:}
Starbursts take place in both mergers and triggered disc instabilities; these mechanisms are channels of bulge formation. The cold gas available for starbursts is kept in a separate reservoir and is gradually consumed as described in \citet{gargiulo_sam_2015}. This gas reservoir is affected by recycling and reheated processes in the same way as the cold gas disc.
 
{\bf AGN feedback:}
%AGNs are produced from the growth of central black holes that take place through two channels: i) infall of gas towards the galactic centre, induced by merger events or disc instabilities; ii) accretion of gas during the cooling process, which produces radio mode feedback that injects energy into the hot atmosphere reducing the amount of gas that can cool. The former process is implemented as described by \citet{lagos_sam_2008}, following \citet{Croton06}, while the latter is replaced by the formulation proposed by \citet{henriques_mcmc_2015}.
AGNs are produced from the growth of central black holes that take place through two channels: i) infall of gas towards the galactic centre, induced by merger events or disc instabilities; ii) accretion of gas during the cooling process, which produces radio mode feedback that injects energy into the hot atmosphere reducing the amount of gas that can cool as $\dot{M}_{\rm cool}^{'}= \dot{M}_{\rm cool}-{L_{\rm BH}}/{(V_{\rm vir}^{2}/2)}$, where $L_{\rm BH}=\eta \, \dot{M}_{\rm BH} c^{2}$ is the black hole luminosity (the mechanical heating generated by the BH accretion), being ${M}_{\rm BH}$ the black hole mass, $c$ the speed of light, and $\eta$ the standard efficiency of energy production that occurs in the vicinity of the event horizon, which takes a value of $0.1$. The former process is implemented as described by \citet{lagos_sam_2008} (following \citet{Croton06}), that is, $\Delta M_{\rm BH} = f_{\rm BH} {({M_{\rm sat}}/{M_{\rm cen}})} {({M_{\rm cold,sat} + M_{\rm cold,cen})}/{(1 + 280 {\rm km s^{-1}} / V_{\rm vir})^2}}$, where $M_{\rm cen}$ and $M_{\rm sat}$ are the masses of the merging central and satellite galaxies, and $M_{\rm cold, cen}$ and $M_{\rm cold,sat}$ are their corresponding cold gas masses.  In the case of disc instabilities, only the host galaxy is involved. The parameter $f_{\rm BH}$ is the fraction of cold gas accreted onto the central BH. The latter is replaced by the formulation proposed by Henriques et al. (2015), so that $\dot{M}_{\rm BH} = \kappa_{\rm AGN}{({M_{\rm hot}}/{10^{11} \Msun})} {({M_{\rm BH}}/{10^8 \Msun})}$ where $M_{\rm hot}$ is the mass of the hot gas atmosphere and $\kappa_{\rm AGN}$ is the efficiency of cold gas accretion onto the BH during gas cooling. Both $f_{\rm BH}$ and $\kappa_{\rm AGN}$ are free parameters of the model.

{\bf Merger treatment:}
Orphan satellites inhabiting a subhalo are assumed to merge with the corresponding central galaxy when the pericentric distance of their orbits becomes less than $10$ percent the virial radius of the host subhalo. If the (baryonic) mass ratio between satellite and central galaxies is larger than $0.3$, then the merger is considered a major one. In this case, the stars and cold gas in the disc of the remnant galaxy are transferred to the bulge, where the gas is consumed in a starburst. In minor mergers, the stars of the merging satellite are transferred to the bulge component of the central galaxy. A starburst is triggered depending on the fraction of cold gas in the disc of the central, consuming all the cold gas from both merging galaxies, as implemented in \citet{lagos_sam_2008}; even if there is enough cold gas available, the starburst is prevented if the mass ratio between satellite and central is less than $5$ per cent.

{\bf Orphans:}
Orphan galaxies emerge when their subhaloes are no longer identified. Their positions and velocities are obtained from a detailed treatment of their orbital evolution, taking into account mass loss by tidal stripping and dynamical friction effects. This treatment allows to apply the position based merger criterion and to obtain an adequate radial distribution of satellite galaxies (Vega-Mart\'inez et al., in prep.).

{\bf Calibration method:}
Calibrations of \sag\ are performed using the Particle Swarm Optimisation (PSO) technique presented in \citet{ruiz_2015}. The PSO consists in a set of particles which explore the parameter space comparing the model's results with a given set of observables and sharing information between them, thus determining new exploratory positions from both their individual and collective knowledge. The result of this exploration is a set of best-fitting values for the free parameters that allows the model to achieve the best possible agreement with the imposed observational constraints. For the current calibration we consider nine parameters as free in the model related to the star formation and supernova feedback efficiencies, the power of the redshift dependent factor involved in the estimation of the reheated mass, the ejection of hot gas and its reincorporation, the growth of black hole masses and efficiency of radio-mode AGN feedback, the disc instability events and the circulation of the reheated cold gas. The observational constraints used are the stellar mass function at $z=0$ and $z=2$ (data compilations of \citealt{henriques_mcmc_2015}), the star formation rate function at $z=0.14$ \citep{Gruppioni15}, the fraction of mass in cold gas as a function of stellar mass at $z=0$ \citep{Boselli14}, and the black hole mass to bulge mass relation at $z=0$ (combination of the datasets from \citealt{McConnell13}, and \citealt{Kormendy13}). This set of observational data is called `CARNage set' and is presented in more detail in Knebe et al. (in prep.).

\subsection{\sage} \label{sec:sage}
%%%%%%%%%%%%%%%%%%%%%%%%%%%%%%%%%%%%%%%%%%%%%%%%%%%
The Semi-Analytic Galaxy Evolution (\sage) model is a major update to that described in \citet{Croton06}. \sage\ was rebuilt from that version to be modular and customisable; it is described in full in \citet{Croton16}. It runs on any $N$-body simulation whose trees are organized in a supported format and has basic set of halo properties. Key changes with respect to 2006 cover the treatment of gas cooling and AGN heating, quasar mode feedback, ejected gas reincorporation, satellite galaxies, mergers, and intra-cluster stars.

{\bf Cooling:}: 
Cooling is handled as in the original \citet{Croton06} model and assumes a singular isothermal density profile. It is basically the same as the \citet{white91} algorithm, but has undergone some evolution (e.g. definition of cooling time) since then. The cooling rate estimated from a simple continuity equation \citep{Bertschinger89}, where it was shown that -- under this assumption -- the rate at which gas is deposited at the centre is proportional (and close to 1) to the rate at which it crosses the cooling radius. 

{\bf Star formation:} 
\sage\ calculates the mass of cold gas in the disc that is above a critical surface density for star formation. New stars then form from this gas using a Kennicutt-Schmidt type relation \citep{Kennicutt89,Kauffmann96,Kennicutt98}.   

{\bf Metal treatment:}
\sage\ uses the simple metal treatment introduced in \citet{DeLucia04b}. A yield of metals is produced from each star formation event and is recycled instantly back to the cold gas from short-lived stars.

{\bf Supernova feedback and winds:} 
Feedback from supernova in \sage\ is a two step process. Firstly, a parametrised mass loading factor blows cold gas out of the disc and into the hot halo following the simple prescription $\dot{m}_{\rm reheated}=\epsilon_{\rm disk} \dot{m}_{*}$ (where $\epsilon_{\rm disk}=3$ here). Secondly, if the thermal energy from supernova added to the hot halo by this gas exceeds the binding energy of the hot halo, some of the hot gas becomes unbound and is removed to an ejected reservoir \citep[see Sec. 8 of][]{Croton16}.

{\bf Gas ejection \& reincorporation:} 
Gas can be ejected from the halo potential through supernova or quasar winds. Ejected gas is reincorporated back into the hot halo at a rate proportional to the dynamical time of the dark matter halo, i.e. the reincorporation mass scales at $\dot{m}_{\rm reinc}=(V_{\rm vir}/V_{\rm crit}-1) m_{\rm ejected}/t_{\rm dyn}$ where $V_{\rm vir}$ ($V_{\rm esc}$) is the virial (escape) velocity of the halo \citep[see][]{Croton16}. Here we used $V_{\rm crit}/V_{\rm esc}=0.15$.

{\bf Disc instability:}
The \sage\ model applies the idealised \citet{Mo98} model to determine when a disc has become unstable. If $V_{\rm circ}/\sqrt{G m_{\rm disk}/r_{\rm disk}}<1$, existing stars are transferred to the bulge to make the disc stable again, along with any new stars as a result of an instability-triggered starburst.

{\bf Starburst:}  
\sage\ uses the collisional starburst model of \citet{Somerville01}, in which bursts of star formation are triggered by galaxy-galaxy mergers, to determine the mass of cold gas that becomes new stars as a result of a merger.

{\bf AGN feedback:}  
As described in detail in \citet{Croton16}, \sage\ uses a modified version of the radio-mode AGN heating model introduced by \citet{Croton06}, which invokes an additional heating radius based on previous AGN activity, where hot gas internal to this has its cooling ceased. \sage\ also includes a new quasar-mode wind model. In the radio mode the central black hole accretes gas at a rate $\dot{m}_{\rm BH}=\kappa_R (15/16) \pi G \bar{\mu} m_p (kT/\Lambda) m_{\rm BH}$ where $\kappa_R=0.08$ is the `radio-mode efficiency factor'. The resulting heating rate from this feedback mode is then $\dot{m}_{\rm heat}=\eta \dot{m}_{\rm BH} c^2/(0.5 V^2_{\rm vir})$ where $\eta=0.1$ is the standard efficiency. The effect of mergers (as well as disk instabilities) on black hole growth -- as modelled by the `quasar mode' -- is modelled phenomenological as $\Delta m_{\rm BH}=f_{\rm BH} (m_{\rm sat}/m_{\rm central}) m_{\rm cold}/(1+(280 {\rm km/s/V_{\rm vir}})^2)$ where $f_{\rm BH}$ controls the accretion efficiency. This change in black hole mass (due to some rapid gas accretion) then leads to an energy input into the surrounding medium, too.

{\bf Merger treatment:}
Mergers are treated using the method described in \citet{Croton16}. Major mergers are defined by a threshold for the (baryonic) mass ratio of 0.3. Satellites are either merged with the central galaxy or added to the halo's intra-cluster stars, depending on the subhalo survival time relative to an average expected based on its infall properties. Briefly, upon becoming a satellite an (analytic) expected time to merge is calculated using the dynamical friction model of \citet{Binney87}. The satellite-subhalo system is then followed until the dark-to-baryonic mass ratio falls below a critical threshold (chosen to be 1.0). At this point, if the system has survived longer than the (analytic) expected merger time we say it is more resistant to disruption and merge the satellite with the central galaxy. Otherwise the satellite is disrupted and its stars added to an intracluster mass component around the central galaxy. This is described in more detail in Sec. 10 of \citet{Croton16}. 

{\bf Orphans:}
\sage\ does not contain orphan galaxies. Before a galaxy can become an orphan a decision is made about its fate based on its actual survival time and the average survival time for subhaloes that have similar properties.

{\bf Calibration method:} 
\sage\ is calibrated by hand primarily using the $z=0$ stellar mass function \citep{Baldry08}, and secondarily using the stellar metallicity--mass relation \citep{Tremonti04}, baryonic Tully-Fisher relation \citep{Stark09}, black hole--bulge mass relation \citep{Scott13}, and cosmic star formation rate density \citep{Somerville01}.

\subsection{SAM Differences \& Similarities} \label{sec:differences+similarities}
%%%%%%%%%%%%%%%%%%%%%%%%%%%%%%%%%%%%%%%%%%%%%%%%%%%
We already know that model-to-model variations in galaxy catalogues exist when different SAMs are applied to the same simulation \citep{Knebe15}, but are currently investigating the influence of re-calibration on this scatter. This has its origin not only in different calibration approaches \citep[Knebe et al., in prep.; see also][where this has been partially addressed, too]{Guo13,gp14,Guo16}, but also in the model design and implementation of the actual physical phenomena \citep{Hirschmann16}. This certainly also applies to the three models presented here. \sag, for instance, is a model with strengths in providing reasonable gas fractions and metallicity relations; \sage\ fits multiple observables simultaneously, first and foremost the stellar mass function and stellar-to-halo mass relation; and \galacticus\ has its strength in the star formation rate function and evolution. Therefore it appears important to not only have a single but multiple galaxy formation models available exploring different approaches to galaxy formation physics. 

\begin{savenotes}
\begin{table*}
  \caption{We list the following acronyms and the intrinsic constraints adopted for the calibration of the parameters of our models: black-hole bulge-mass relation \BHBM, stellar mass function \SMF, luminosity function \LF, star formation rate function \SFRF, cosmic star formation rate density history \cSFRD, (baryonic,local) Tully-Fisher relation {\it (b,l)TF}, mass-metallicity relation {\it MZ}, cold gas fraction {\it CGF}, local colour-magnitude {\it lCMD} at $z=0.1$ and disc size distribution of galaxies {\it DSD}. Unless specified otherwise constraints are used at redshift $z=0$. Note that all our SAM models assume a \citet{Chabrier03} initial mass function (IMF), but use different mass definitions for dark matter haloes. Further, they all provide different information for orphan galaxies. Next to the column stating whether the model parameters have been re-calibrated to the \MDPL\ simulation, we also assign a name to each catalogue that combines the particular simulation and SAM name.}
	\label{tab:models}\vspace{-0.2cm}
	\setlength{\tabcolsep}{1.0pt}
		\begin{center}
			\begin{tabular}{lllcccl}
				\hline
				{Model Name} \ \ \	& {Reference} \ \ \ &  {Intrinsic constraints} & \ \ \ {Mass definition} \ \ \ & {Orphans} &  \ \ \ {Re-calibrated} \ \ \ & {Catalogue name}\\
				\hline
				\hline
				 \galacticus	& \citet{Benson12}			& BHBM, SMF, LFs (K- \& $b_j$-bands), & \MBN 	& yes, but & no & \MDPL-\galacticus  \\
				 & & lCMD ($z=0.1$), DSD, lTF, cSFRD & & without ${\bf x},{\bf v}$  & &	 \\
				 \hline
				 \sag			& Cora et al. (in prep.)		& BHBM, SMF ($z$=0 \& $z$=2), CGF,	& \Mcrit  	& yes 						& yes & \MDPL-\sag \\
				 			& 						& SFRF ($z$=0.14)					& 	 	&  						& 	 & 	 \\
				\hline
				 \sage		& \citet{Croton16} \ \ \ \ \		& BHBM, SMF, bTF, MZ, cSFRD		& \Mcrit 	& no 							& yes & \MDPL-\sage \\
				\hline
				\vspace{-0.6cm}
			\end{tabular}
		\end{center}
	\end{table*}
\end{savenotes}

\paragraph*{Calibration:} While \sag\ and \sage\ modellers have re-tuned their model parameters to the \MDPL simulation, \galacticus\ was run with its standard calibration. While \sage\ relies on a manual tuning of its parameters, \sag\ applies a Particle Swarm Optimisation technique. \galacticus\ uses seven observational data sets during calibration. The model further has a large set of parameters to chose from depending on the desired implementation. \sag has left nine of its parameters free during the calibration to five observations, whereas \sage has five observational constraints and 14 parameters out of which seven have been varied during the calibration. All of the five observations used by \sag\ for calibration (see \hyperref[tab:models]{\Tab{tab:models}}) coincide with galaxy properties used throughout this paper for comparison; while \sage\ also calibrates to some of these properties, this model uses observational data sets different to the ones employed here. In that regards also note that all models have been calibrated to the BHBM relation and the stellar mass function, but again, not necessarily using the same observational data as presented in the respective plots below.

\paragraph*{Initial mass function:}
%%%%%%%%%%%%%%%%%
For the processing of the \MDPL simulation all our SAM models assume a \citet{Chabrier03} initial mass function (IMF). But whenever we compare the models to observations based upon a different IMF we apply the following conversion to that reference data \citep[][]{Lacey16}: 
\begin{equation} \label{eq:IMFcorrection}
\begin{array}{lcll}
\log_{10}(M^{\rm Chabrier}_{*}) & = & \log_{10}(M^{\rm Salpeter}_{*})    & - \ 0.240\\
\log_{10}(M^{\rm Chabrier}_{*}) & = & \log_{10}(M^{\rm Kroupa}_{*})    & - \ 0.039\\
%\log_{10}(M^{\rm Chabrier}_{*}) & = & \log_{10}(M^{\rm diet-Salpeter}_{*})    & - \ 0.090\\
%\log_{10}(M^{\rm Chabrier}_{*}) & = & \log_{10}(M^{\rm Kennicutt}_{*})  & + \ 0.089\\
\end{array}
\end{equation}
These numbers certainly depend on the assumed stellar population synthesis (\SPS) model, age and metallicity as well as the estimation of stellar masses from broad band photometry, but have proven to be sufficiently accurate for average galaxies \citep{Mitchell14}.

\paragraph*{Mass definition:}
%%%%%%%%%%%%%%%
The mass of a dark matter halo is a not well-defined quantity \citep[see, for instance,][]{Diemer13} and various possible definitions exist \citep[see, for instance, discussion in Sec. 2.5 of][]{Knebe13}. The \rockstar\ halo finder -- used for the \MDPL\ simulation -- provides us with a variety of masses:
\begin{equation} \label{eq:virialradius}
  M_{\rm ref}(<R_{\rm ref}) = \Delta_{\rm ref} \rho_{\rm c} \frac{4\pi}{3} R_{\rm ref}^3 \ ,
\end{equation}
where 
\begin{equation} \label{eq:massdefinition}
\begin{tabular}{ll}
$\Delta_{\rm ref}=200$				& for \Mcrit \ , \\
$\Delta_{\rm ref}=\Delta_{\rm BN98}$	& for \MBN \ , \\
\end{tabular} \\
\end{equation}

and $\rho_{\rm c}$ is the critical and background density of the Universe. $\Delta_{\rm BN98}$ is the virial factor as given by Eq.(6) in \citet{Bryan98}, and $R_{\rm ref}$ is the corresponding halo radius for which the interior mean density matches the desired value on the right-hand side of \Eq{eq:virialradius}.

The models presented here apply two different mass definitions to define the dark matter haloes that formed their halo merger tree. \galacticus\ uses \MBN\ whereas \sag, and \sage\ apply \Mcrit. But as can be verified in Appendix~B of \citet{Knebe15}, this will have little impact on the properties of the galaxies.

\paragraph*{AGN feedback:}
%%%%%%%%%%%%%%%
All models include AGN feedback caused by accretion of gas onto a central black hole via various channels. \galacticus\ and \sag\ both model radio-mode feedback caused by the accretion of cooling gas from the hot halo onto the black hole. \sage\ additionally features a new quasar-mode wind \citep[see][]{Croton16}.

\paragraph*{Mergers:}
%%%%%%%%%%%%%%%
All three models treat minor and major mergers a bit differently, also using varying thresholds for this separation: \sag\ and \sage\ use 0.3 as the threshold for the mass ratio to separate major from minor mergers, while \galacticus\ uses a slightly lower value of 0.25. For \sage, the satellite survival time determines whether the galaxy will contribute to the central or the intra-cluster light. In \galacticus\ a major merger calls for a rearrangement of the spheroid whereas a minor merger simply leads to adding the satellite to the existing spheroid leaving the disc unaffected. And in \sag, all mergers contribute to the bulge formation through the transfer of stars and gas from the disc to the spheroid. However, the gas transfer and subsequent starburst depend on the mass ratio of the merging galaxies and their gas content.

\paragraph*{Orphan galaxies:}
%%%%%%%%%%%%%%%
Besides the different implementations of the underlying physics and differences in the choices of parameter calibration, there is one fundamental difference between the three models presented here: the treatment of orphans galaxies. \sage\ does not feature any orphans at all; \galacticus\ creates orphan galaxies and assigns physical properties to them, but does not integrate their orbits, i.e. no phase-space information is provided; \sag\ not only provides galaxy properties but also full position and velocity information for orphans as their orbits are integrated in a pre-processing step previous to the application of \sag (Vega-Mart\'inez et al., in prep.). This will then certainly have implications for studies such as clustering, where positions directly enter.

Another important difference in the orphan treatment for \sage\ is that the stellar mass of disrupted satellites (see \hyperref[sec:sage]{\Sec{sec:sage}}) can be added to an intra-cluster component (ICC): what would otherwise end up as an orphan in other models can either be merged with the central or go to the ICC in \sage, depending on how long its subhalo had survived (compared to the average for a subhalo of its general properties). This is rather distinct from the other models. In the case of \sag, such component is built up by the contribution of tidally stripped material of the stellar components of satellite galaxies that could be smaller than expected if tidal stripping is not efficient enough. And \galacticus\ does not track intra-cluster stars in the version used here. We will see later that this will have an impact on the galaxy stellar mass function.

\subsection{Public Release of \MDG} \label{sec:publicrelease}
%%%%%%%%%%%%%%%%%%%%%%%%%%%%%%%%%%%%%%%%%%%%%%%%%%%
As mentioned before, all galaxy catalogues (as well as halo catalogues and merger trees) are publicly available in the \textsc{CosmoSim} database\footnote{\url{http://www.cosmosim.org}}. Direct downloads of the data products are also available from the `Skies \& Universes' site\footnote{\url{http://www.skiesanduniverses.org}}. The uploaded galaxy properties, their units, and further information are given in \App{app:database}.

%%%%%%%%%%%%%%%%%%%%%%%%%%%%%%%%%%%%%%%%%%%%%%%%%%%
\section{\MDG Properties} \label{sec:comparison}
%%%%%%%%%%%%%%%%%%%%%%%%%%%%%%%%%%%%%%%%%%%%%%%%%%%
In this section we present a comparison of the three \MDG catalogues. We restrict our work to studying some of the more basic properties of galaxies\footnote{Please check \hyperref[app:plotsummary]{\App{app:plotsummary}} for a summary of the all the plots presented in this Section.} and leave further details regarding the SAMs to the accompanying model papers and references listed in \hyperref[tab:models]{\Tab{tab:models}}.

%As all our SAM models adopt a \citet{Chabrier03} IMF we convert all reference data sets (e.g. observations) which assume a different IMF.
If the data we refer to is spread over a certain redshift range we choose the value which lies in the middle, unless redshift evolution is studied. Note that the Hubble parameter $h=0.678$ is included in the numerical value of the data presented in plots or calculations, therefore we will not further refer to $h$. When binning data we always use median values (except in the luminosity function) and estimate a \textit{`median absolute deviation'} as our preferred error estimator which is the median of the absolute deviations of the data points about the median. If there are no error bars given in the plot, the bars are considered negligible.
%This estimator is very robust and insensitive to outliers or sample size \cite[p. 111]{MAD}.
For the contour-plots we use throughout this work the following confidence levels in per cent: [4.55, 10.0, 20.0, 31.74, 50.0, 68.26, 80.0, 90.0, 95.45, 99.9] and -- in case we are presenting SAM results -- apply a stellar mass cut of $\Mstar > 10^8$\Msun\ for better readability. For observational data retrieved from other works we use the following contour levels (in per cent, too): [4.55, 31.74, 50.0, 68.26, 90.0, 99.9].

Before discussing any plots, we present in \hyperref[tab:ngal]{\Tab{tab:ngal}} an overview of the number of galaxies (measured in millions) each model's catalogue contains; we also provide the fraction of orphans in parenthesis (noting that \sage\ does not feature orphans). The number of galaxies in the first column (`all') refers to the total number of galaxies provided. In the second and third column we list the numbers above a certain stellar mass threshold: even though the subsequent plots use all supplied galaxies (if not indicated otherwise), a mass threshold of $\Mstar \gtrsim 10^9$\Msun\ seems appropriate for simulations with a resolution comparable to the Millennium simulation \citep[like the \MDPL\ simulation used here, see][]{Guo11}. In practical terms we can consider $M_{*} \gtrsim 10^{9}$\Msun\ the completeness limit of our galaxy catalogues. We have verified that implementing such a cut does not change the conclusions from any of the plots, but it does greatly facilitate the handling of the data. 

\begin{table}
	\caption{The table presents for the three \MDG catalogues the number of galaxies (as measured in millions)  for various redshifts and stellar mass cuts. The numbers in parenthesis give the fraction of orphans. \textit{`all'} represents the total number of objects in the catalogue and $\Mstar > M_{\rm cut}$ (where \Mstar\ is measured in \Msun) stands for a selection at that particular threshold $M_{\rm cut}$.}
	\label{tab:ngal} \vspace{-0.2cm}
	\setlength{\tabcolsep}{2.5pt}
	\begin{center} 
		\begin{tabular}{lccccc}
		\hline
		\textsc{catalogue} & $z$ & \multicolumn{4}{c}{\textsc{number of galaxies} [$10^6$]} \vspace{0.1cm} \\
				 & & all & $\Mstar > 10^9$ & $\Mstar > 10^{10}$ & $\Mstar > 10^{11}$ \\
		 \hline \hline
		 \MDPL- 		& 0.0 	& $189$ ($0.33$)  & $60$ ($0.30$) & $26$ ($0.15$) & $0.8$ ($0.04$)\\
		 \galacticus	& 0.1 	& $191$ ($0.32$)  & $59$ ($0.29$) & $25$ ($0.16$) & $0.8$ ($0.02$)\\
		 			& 0.14 	& $190$ ($0.32$)  & $59$ ($0.28$) & $25$ ($0.16$) &$0.7$ ($0.02$) \\
		 \hline
		 \MDPL-\sag 	& 0.0 	& $194$ ($0.34$)  & $40$ ($0.12$) & $11$ ($0.05$) & $1.0$ ($0.03$) \\
		 			& 0.1 	& $197$ ($0.34$)  & $38$ ($0.11$) & $10$ ($0.05$) & $0.9$ ($0.02$) \\
		 			& 0.14 	& $196$ ($0.34$)  & $37$ ($0.11$) & $10$ ($0.05$) & $0.9$ ($0.02$) \\
		 \hline
		 \MDPL-\sage 	& 0.0 	& $127$ ($0.00$)  & $58$ ($0.00$) & $19$ ($0.00$) & $1.6$ ($0.00$) \\
		 			& 0.1 	& $130$ ($0.00$) & $59$ ($0.00$) & $19$ ($0.00$) & $1.5$ ($0.00$) \\
		 			& 0.14 	& $130$ ($0.00$)  & $60$ ($0.00$) & $19$ ($0.00$) & $1.5$ ($0.00$) \\		 
		 \hline
		 \vspace{-0.6cm}
		\end{tabular}
	\end{center}
\end{table}

%%%%%%%%%%%%%%%%%%%%%%%%%%%%%%%%%%%%%%%%%%%%%%%%%%%
\subsection{Stellar Mass Function (\SMF)} \label{sec:smf}
%%%%%%%%%%%%%%%%%%%%%%%%%%%%%%%%%%%%%%%%%%%%%%%%%%%
 \begin{figure}
   \includegraphics[width=\columnwidth]{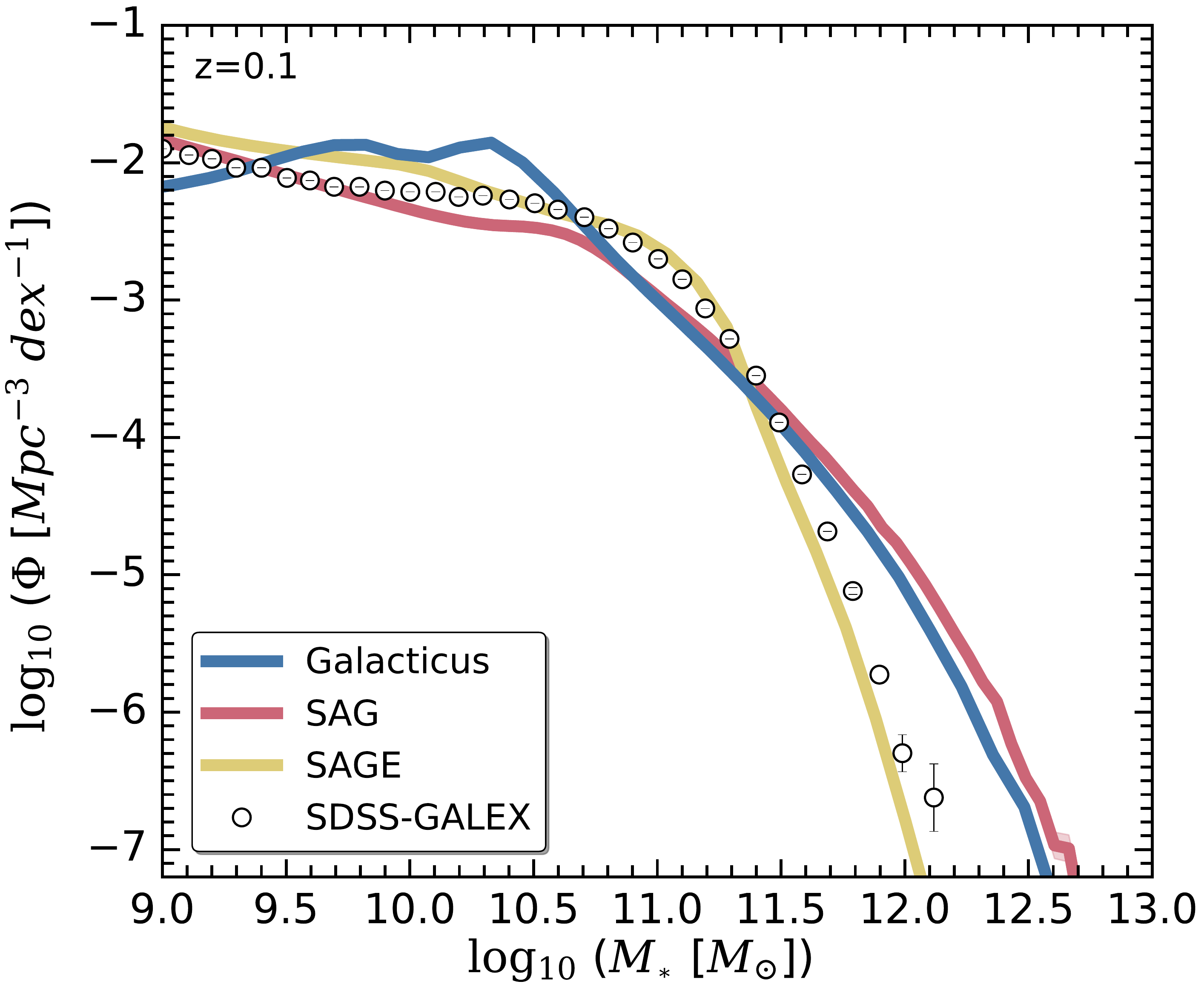}\vspace*{-0.2cm}
   \caption{Stellar mass function of all three models in comparison with the SDSS-GALEX observation at $z=0.1$}
 \label{fig:SMF}
 \vspace{-0.3cm}
 \end{figure}
 
One of the most fundamental properties of galaxies is the stellar mass and its distribution into individual galaxies, as measured by the galaxy stellar mass function (\SMF). Generally, \SMFs\ also play a central role for the calibration of the models, i.e. model parameters are fine-tuned to reproduce a given observationally measured \SMF.

In \hyperref[fig:SMF]{\Fig{fig:SMF}} we compare each of the three SAMs to the stellar mass function of the SDSS-GALEX survey at redshift $z=0.1$ \citep{Moustakas2013_SMF}. We observe a similar yet smaller model-to-model variation as already reported by \citet{Knebe15}: all models presented here provide a valid reproduction of the observed stellar mass function, but all with individual features, e.g. \galacticus\ shows a `bump' at medium masses -- a feature that will affect some of the other results shown below -- and a flattening at smaller masses. \sage provides the closest match to the observational data. This is unsurprising given it is the strongest constraint used for that model, even though they did not use the observational data shown here, but \citet{Baldry08} instead. However, both \galacticus and \sag\ over-predict galaxies at the very high-mass end. \citet{Croton06} and \citet{bower_agn_2006} relate such an excess to a radio-mode AGN feedback not being efficient enough to suppress star formation in these massive galaxies \citep[see also][]{Hirschmann16}. However, this has also been investigated in more detail with the \sag\ model here, but changing some aspects of the AGN feedback to avoid the excess at the high-mass end of the \SMF\ did not lead to an improvement: when calibrating the code, the values of the free parameters change to compensate for those modifications, and the results are eventually the same. But we need to remind the reader that \sag\ simultaneously calibrates to the \SMF\ at redshift $z=0$ \textit{and} $z=2$ (cf. \Tab{tab:models}).\footnote{\sag\ uses the compilation of observed \SMFs\ of the `CARNage set' (Knebe et al., in prep.) for which the agreement is better -- especially at redshift $z=2$ (not explicitly shown here, but see Cora et al., in prep.)} And this is non-trivial for any SAM model \citep[][Knebe et al., in prep., Asquith et al., in prep.]{Henriques15,Hirschmann16,Rodrigues17}. We also like to mention at this point that the excess of massive galaxies for \sag\ and \galacticus\ is not readily explained by a too high star formation rate -- at least not when considering the local star formation rate function (see \Sec{sec:SFRF} below) where we find that all models reproduce the observed \SFRF\ sufficiently well. This is further supported by aforementioned agreement of the \SMF\ at $z=2$ with observational data and abrupt decay of \SFR\ from $z=2$ for galaxies with masses larger than $10^{11}$\hMsun\ (Cora et al., in prep.). 

We conclude that the results seen here in \hyperref[fig:SMF]{\Fig{fig:SMF}} have to be attributed to other aspects such as mergers or the treatment of orphans (see \Sec{sec:differences+similarities}). In particular, one of the features of \sage\ is to tidally disrupt satellite galaxies when they become orphans adding their stars to the intra-cluster light. As \sage\ keeps track of this component, we confirm that adding it back to the mass of the galaxy substantially lifts the \SMF\ for masses $\log_{10}(M_{*})>11.3$, i.e. above the knee (not shown here though), to a level where it is approximately 1.5dex larger than the other two models at $\log_{10}(M_{*})>12.5$. This exercise hints at possible inefficiencies in the mechanism of tidal stripping implemented in \sag. This process removes stellar mass from the disc and bulge of satellite galaxies that is deposited in the intra-cluster component. However, it seems that the stripped mass is being underpredicted, preventing tidal stripping from alleviating the discrepancy between model and observations at the high mass end of the \SMF\ at $z=0$. This will be discussed in more detail in an accompanying paper that focuses on the treatment of orphan galaxies in the \sag\ model (Vega-Martínez et al., in prep.).

%%%%%%%%%%%%%%%%%%%%%%%%%%%%%%%%%%%%%%%%%%%%%%%%%%%
\subsection{Star Formation} \label{sec:starformation}
%%%%%%%%%%%%%%%%%%%%%%%%%%%%%%%%%%%%%%%%%%%%%%%%%%%
While a fraction of the galaxy mass is expected to be ejected by stellar winds and new mass being accreted via mergers, different amounts of stellar mass across semi-analytic models -- as found in the previous sub-section -- also has to relate to different star formation rates (\SFRs) and star formation histories, respectively. We investigate such differences in star formation across our models in this sub-section and compare them to observational data, too.

%%%%%%%%%%%%%%%%%%%%%%%%%%%%%%%%%%%%%%%%%%%%%%%%%%%
\subsubsection{The Star Formation Rate Function (\SFRF)} \label{sec:SFRF}
\begin{figure}
   \includegraphics[width=\columnwidth]{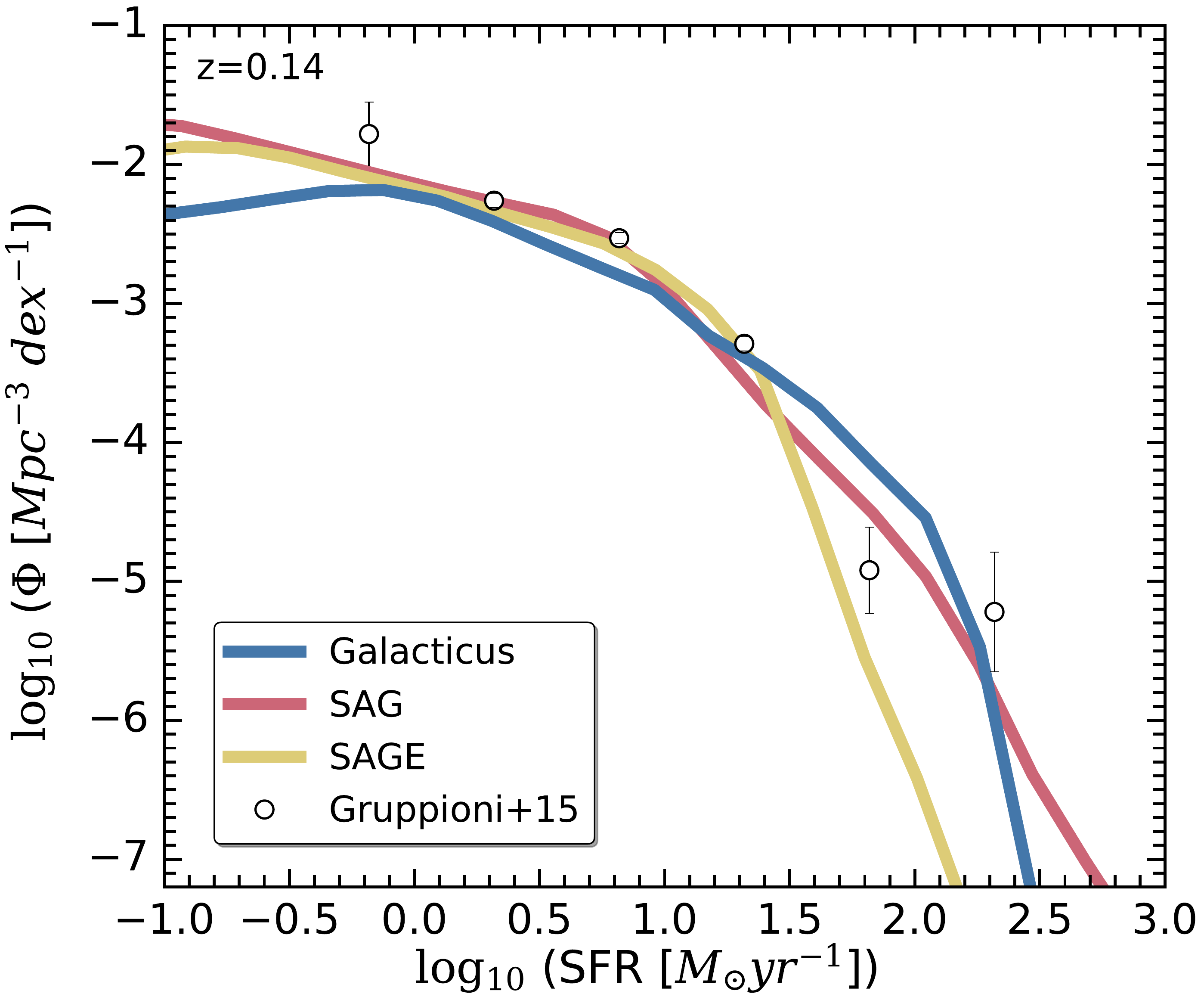}\vspace*{-0.2cm}
   \caption{Star formation rate function for of all three models at $z=0.14$ compared to observations from \citet{Gruppioni15}.}
 \label{fig:SFR}
 \vspace{-0.3cm}
 \end{figure}

%%%%%%%%%%%%%%%%%%%%%%%%%%%%%%%%%%%%%%%%%%%%%%%%%%%
We start with showing in \hyperref[fig:SFR]{\Fig{fig:SFR}} the star formation rate function (\SFRF), i.e. the number of galaxies per unit volume with a given \SFR. The models are contrasted to observations from \citet{Gruppioni15} who determined the \SFR function in the redshift interval $z\in[0.0,0.3]$ whereas the SAM data are shown for redshift $z=0.14$. We find that all models reproduce it rather well although we again observe some scatter from model to model (bearing in mind that \sag\ used the \SFRF\ as a constraint during parameter calibration). We further note that \galacticus\ and \sag\ have more galaxies with higher \SFR as compared to \sage. They nevertheless both match the observational data and hence -- as mentioned before -- the \SFR alone does not explain their excess of high-mass galaxies seen in \hyperref[fig:SMF]{\Fig{fig:SMF}}, i.e. they form stars at the correct (i.e. observed) rate -- at least during the epoch $0.0<z<0.3$ which is the redshift range of the observational data shown in \hyperref[fig:SFR]{\Fig{fig:SFR}}. Their overabundance of high-mass galaxies as previously seen in \hyperref[fig:SMF]{\Fig{fig:SMF}} must be related to other phenomena as already discussed before.

%%%%%%%%%%%%%%%%%%%%%%%%%%%%%%%%%%%%%%%%%%%%%%%%%%%
\subsubsection{The Specific Star Formation Rate to Stellar Mass Relation} \label{sec:ssfr2mstar}
%%%%%%%%%%%%%%%%%%%%%%%%%%%%%%%%%%%%%%%%%%%%%%%%%%%
\begin{figure}
	\begin{subfigure}{\columnwidth}
	  \centering
	  \includegraphics[width=\columnwidth]{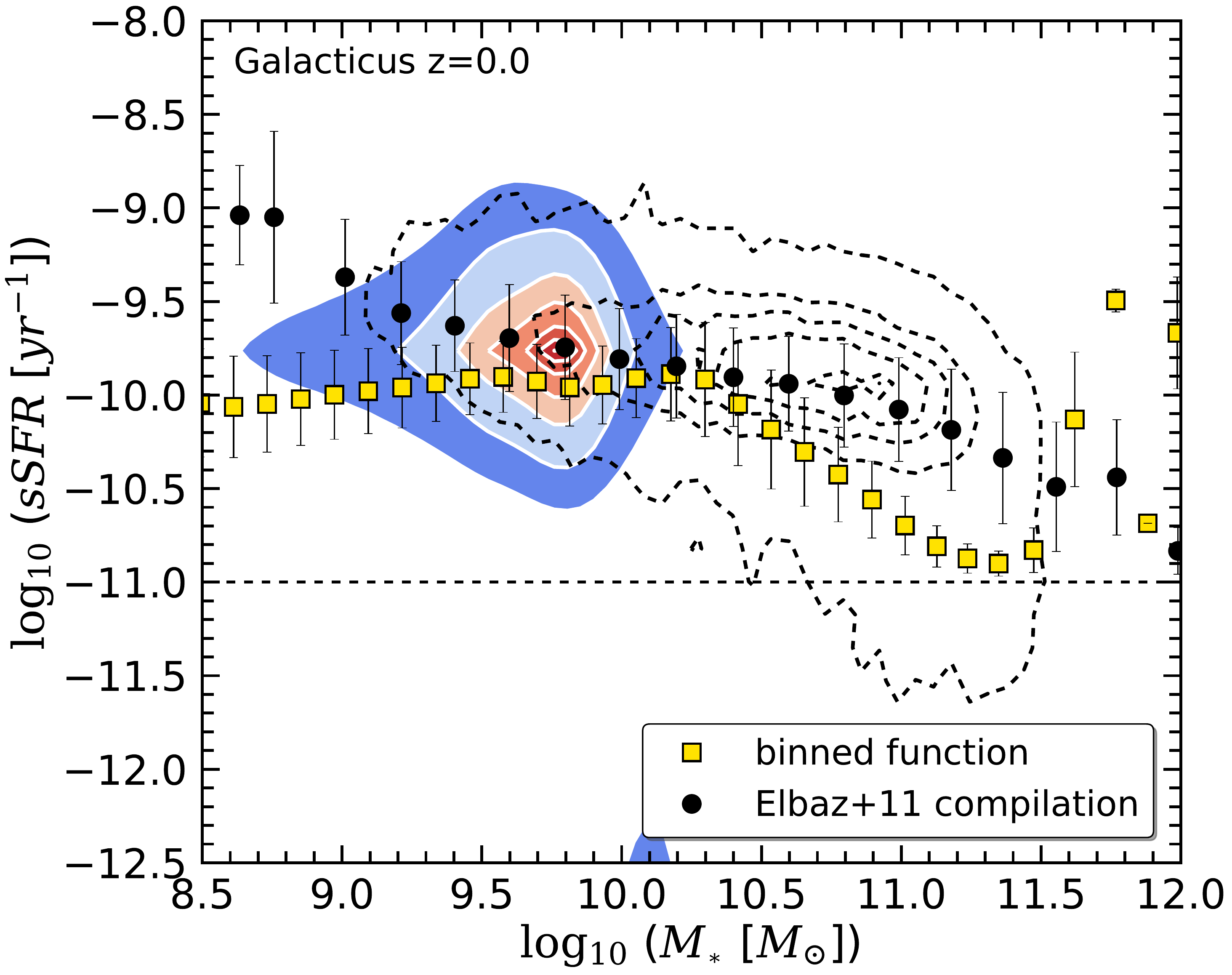}
	  %\caption{\galacticus: \sSFR to stellar mass relation.}
	  %\label{fig:sfig1} \hspace*{2.5em}
	\end{subfigure}
	\begin{subfigure}{\columnwidth}
	  \centering
	  \includegraphics[width=\columnwidth]{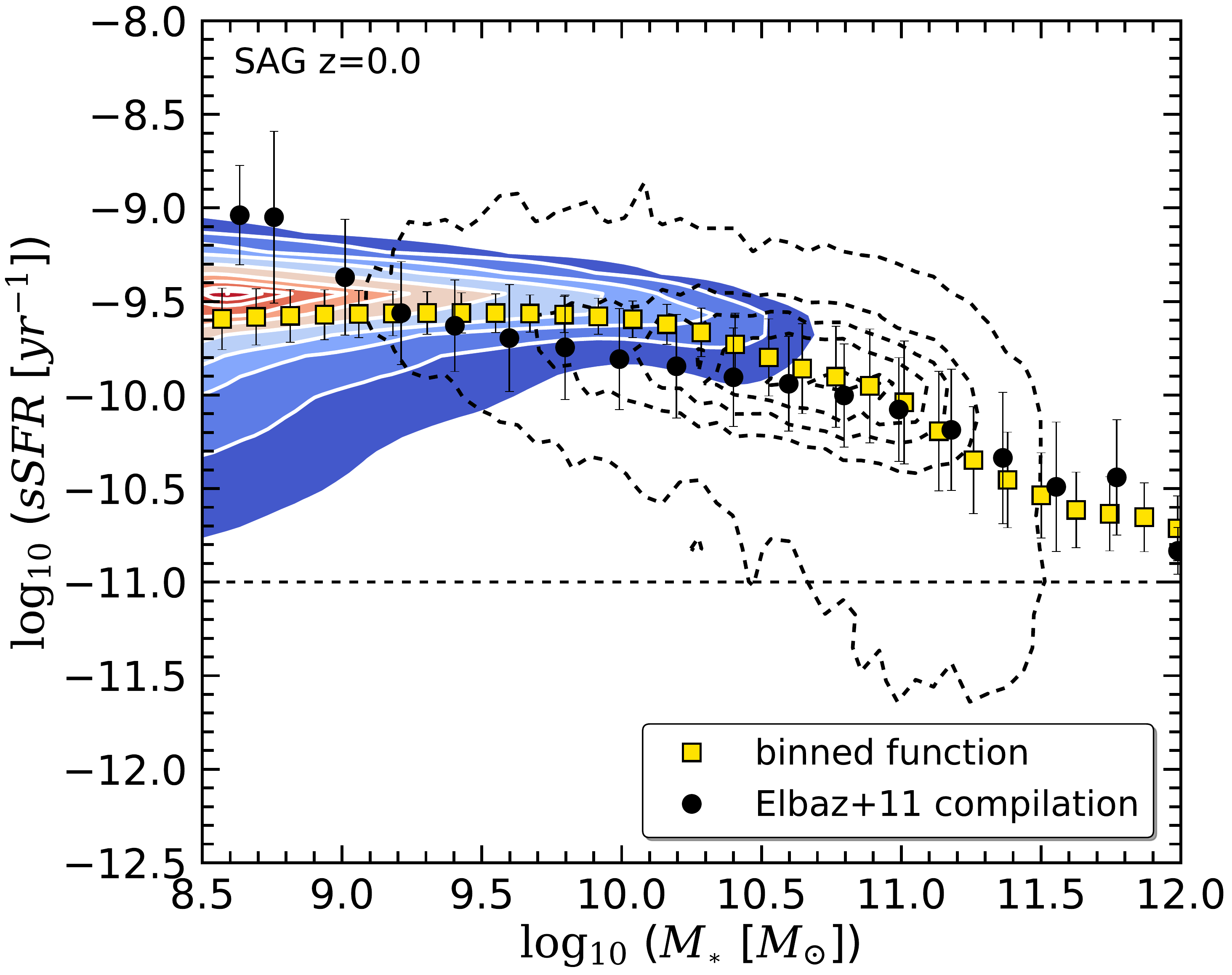}
	  %\caption{\sag: \sSFR to stellar mass relation.}
	  %\label{fig:sfig2} \hspace*{2.5em}
	\end{subfigure}
	\begin{subfigure}{\columnwidth}
	  \centering
	  \includegraphics[width=\columnwidth]{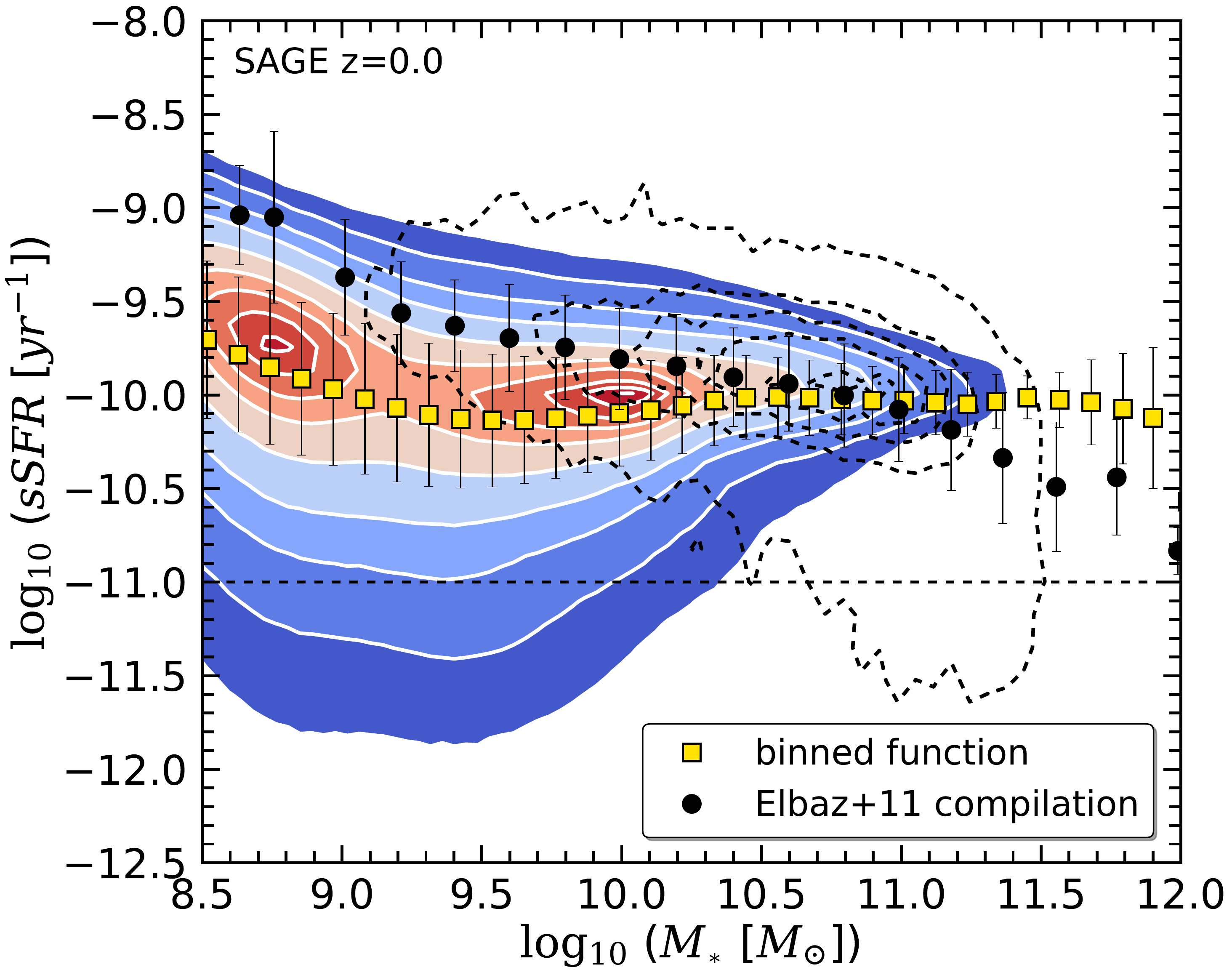}
	  %\caption{\sage: \sSFR to stellar mass relation.}
	  %\label{fig:sfig3}
	\end{subfigure}
 	\caption{The specific star formation rate vs. stellar mass contours (coloured with white lines) and binned function for star-forming galaxies (yellow squares) at $z=0$ for \galacticus (\textit{top panel}), \sag (\textit{middle panel}) and \sage (\textit{bottom panel}). As a reference we include a compilation of observations of star-forming galaxies \citep[left panel of Fig. 16]{Elbaz11} which is presented here as dashed black contours as well as the a binned function (black dots) at $z \sim 0$. The dashed black line represents a commonly used separation of active and passive galaxies \protect{$\log_{10}(sSFR~[yr^{-1}])>\log_{10}(0.3/t_{Hubble(z=0)}[yr^{-1}])~\sim~-11$} \citep[][]{Franx08}.}
	\label{fig:sSFR2Mstar}
	\vspace{-0.3cm}
\end{figure}

Not all galaxies form stars at the same rate and the \SFR certainly depends on the actual (stellar) mass of the galaxy. Thus, it is instructive to have a closer look at the specific \SFR (\sSFR), i.e. the star formation rate per unit stellar mass. Assuming a constant star formation rate, we like to remark that the inverse of the \sSFR can serve as a proxy for galaxy age. We show \sSFR vs. stellar mass as a contour plot (coloured with white lines) for our models at $z=0$ in \hyperref[fig:sSFR2Mstar]{\Fig{fig:sSFR2Mstar}}. The dashed black line represents a commonly used separation of active and passive galaxies \protect{$\log_{10}(sSFR~[yr^{-1}])>\log_{10}(0.3/t_{Hubble(z=0)}[yr^{-1}])~\sim~-11$} \citep[][]{Franx08}. From this sample we calculate the binned function of active galaxies for our models, represented as yellow squares in the figure. We compare the model results to a compilation of star-forming galaxies from \citet{Elbaz11} at $z \sim 0$ presented here both as black dashed contour lines and binned data (black dots). \hyperref[fig:sSFR2Mstar]{\Fig{fig:sSFR2Mstar}} gives us wider insight into the galaxy stellar masses as compared to studying the \SFRF (\hyperref[fig:SFR]{\Fig{fig:SFR}}) only. While the \SFRF agreed impressively well with observations in the range $1~<~\SFR[\Msun yr^{-1}]~<~30$, the \sSFR as a function of stellar mass shows that the distribution of star-formation across galaxies follows a marginally different mass-trend as found in observations (especially for \galacticus\ and \sage). When interpreting the panels we need to bear in mind that observations are likely incomplete at the low-mass end -- a region where all models still provide data. But as the specific star formation can be viewed as a proxy for the (inverse of the) age of a galaxy, all models agree with the observations in the sense that more massive galaxies tend to be older -- at least in terms of stellar ages -- a phenomenon also referred to as \textit{down-sizing} \citep[e.g.][]{Cowie96,Neistein06,Fontanot09}. However, this trend is not as pronounced for \sage\ as for the other models as the highest-mass galaxies in \sage\ are too star-forming. These galaxies also have discs that are relatively too massive and bulges that are relatively too low in mass \citep[see Fig.6 in][]{Stevens16}, something to be remembered when discussing the black hole--bulge mass relation below. 

%%%%%%%%%%%%%%%%%%%%%%%%%%%%%%%%%%%%%%%%%%%%%%%%%%%
\subsubsection{The Cosmic Star Formation Rate Function (\cSFRD)} \label{sec:cSFR}
%%%%%%%%%%%%%%%%%%%%%%%%%%%%%%%%%%%%%%%%%%%%%%%%%%%
 \begin{figure}
   \includegraphics[width=\columnwidth]{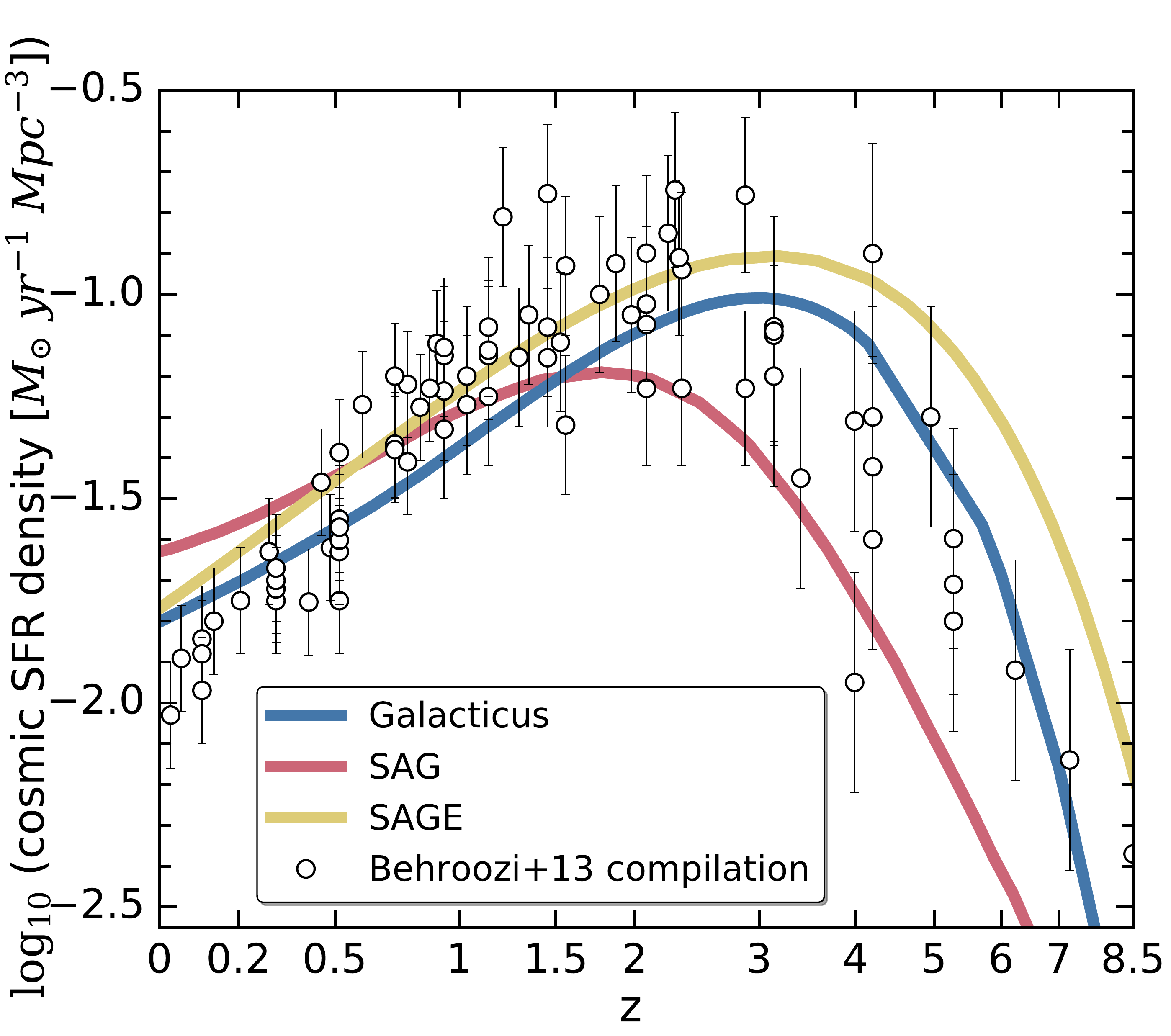}\vspace{-0.2cm}
   \caption{Cosmic star formation rate density for all galaxies as a function of redshift compared to a compilation of observations from \citet[][Table 4]{Behroozi13c_cSFRF}.}
 \label{fig:SFRdensity}
 \vspace{-0.3cm}
 \end{figure}
 
In \hyperref[fig:SFRdensity]{\Fig{fig:SFRdensity}}, we close this sub-section with a presentation of the evolution of the \SFR\ density across cosmic time (cosmic star formation rate density, \cSFRD), i.e. the so-called {`Madau-Lilly plot'} \citep{Lilly96,Madau96,Madau14}. We confirm that all three models show a pronounced peak around redshift $z\sim2-3$ and approximately follow the observational data compiled by \citet{Behroozi13c_cSFRF} (shown here as open circles with error bars) within the error bars. However, their individual curves are rather distinct. \galacticus\ and \sage\ show approximately the same shape but appear shifted in amplitude with respect to each other, whereas \sag\ shows a marginally different shape. Up to redshift $z\sim 1$ (i.e. approximately 40 per cent of the present age of the Universe) the \sag\ model shows a substantially lower \SFR. From that time onwards the model follows the same trend as \galacticus, albeit a marginally larger amplitude now. While \sag\ forms in total very few stars (due to the low \SFR\ in the early Universe) it nevertheless provides roughly the same number of galaxies as \galacticus (see \hyperref[tab:ngal]{\Tab{tab:ngal}}), especially above $\log_{10}(\Mstar [\Msun])>10.5$ (see \hyperref[fig:SMF]{\Fig{fig:SMF}}): that can be explained by the fact that there are lots of galaxies with low stellar mass, i.e. $\Mstar\leq10^{8.5}$\Msun\ (not explicitly shown here, but can be concluded from the numbers in \hyperref[tab:ngal]{\Tab{tab:ngal}}). And despite \sage\ having the highest integrated \SFR\ the total number of galaxies is lowest for this model (see \hyperref[tab:ngal]{\Tab{tab:ngal}}). \sage\ forms -- in total -- the fewest number of galaxies below $\Mstar<10^{8}$\Msun\ (as can be inferred again from \hyperref[tab:ngal]{\Tab{tab:ngal}}, noting also that \sage\ does not feature orphans). Therefore the question remains why \sage\ -- with reasonable matches to both the observed \SFRF\ and \SMF\ -- shows a consistently higher \cSFRD\ for redshifts $z>0.5$. While we leave a more detailed study of high-redshift galaxies to future work, we have seen -- at least at redshift $z=0$ -- that \sage\ features a marginal excess of \sSFR\ as seen for high-mass galaxies in \hyperref[fig:sSFR2Mstar]{\Fig{fig:sSFR2Mstar}}: for stellar masses $\log_{10}(\Mstar)>11$, \sage\ shows the highest \sSFR\ amongst all models.
%When exploring the result using distinct stellar mass bins, we find that for \sage\ the contribution to the integrated \SFR\ at low-redshift primarily comes from galaxies with $\log_{10}(\Mstar)\in [10,11]$ whereas at high-redshift mainly $\log_{10}(\Mstar)\in [9,10]$ contribute.

One can now raise the question about the interplay and simultaneous interpretation, respectively, of the four plots presented in this Section. For instance, the integral over all masses of the \SMF\ at a fixed redshift corresponds to the integral of the \cSFRD\ up to that redshift. Further, the integral over all \SFR\ values in the \SFRF\ gives the point in the \cSFRD\ at the corresponding redshift. However, this relation has to be viewed with care because of the recycle fraction of exploding stars and/or produced by stellar winds which has to be considered during that integration. \hyperref[fig:SFRdensity]{\Fig{fig:SFRdensity}} now tells us that at redshift $z=0.14$  \sag\ has a higher (integrated) \SFR\ than the \galacticus\ model. But when comparing this to \hyperref[fig:SFR]{\Fig{fig:SFR}} one needs to bear in mind that the excess seen there for \sag\ and \galacticus\ at the high-\SFR\ end hardly contributes to such an integral. And the fact that the \sage\  model gives the smallest number of galaxies (cf. \hyperref[tab:ngal]{\Tab{tab:ngal}}) is also not inconsistent with the fact that its \cSFRD\ is highest (at least for redshifts $z>0.5$): it simply means that all those stars generated over the course of the simulation are forming part of the lower mass galaxies (note that for stellar masses $\log_{10}{(M_{*}[\Msun])}<11.3$ \sage\ provides the highest \SMF, cf. \hyperref[fig:SMF]{\Fig{fig:SMF}}).

Therefore, while the set of plots presented in this Section clearly show consistency, they are not sufficient to explain, for instance, an excess of high-mass galaxies in the \SMF\ plot. But it is apparent that for both \sag\ and \galacticus\ those objects with high \SFR\ (as seen in \hyperref[fig:SFR]{\Fig{fig:SFR}}) have to be high in stellar mass, too. Similarly, the deficit of objects with high \SFR\ for \sage\ evidently helps the model to better reproduce the high-mass end of the \SMF, even though those high-mass galaxies have rather high \sSFR's, according to \hyperref[fig:sSFR2Mstar]{\Fig{fig:sSFR2Mstar}}. Further, for the \sag\ model we also confirm that galaxies with stellar mass $M_{*}<10^{11}$\Msun\ are actually responsible for the `excess' seen in the \cSFRD.

%%%%%%%%%%%%%%%%%%%%%%%%%%%%%%%%%%%%%%%%%%%%%%%%%%%
\subsection{The Black Hole to Bulge Mass Relation (\BHBM)} \label{sec:BHBM}
%%%%%%%%%%%%%%%%%%%%%%%%%%%%%%%%%%%%%%%%%%%%%%%%%%%

 \begin{figure}
	\begin{subfigure}{\columnwidth}
	  \centering
	  \includegraphics[width=\columnwidth]{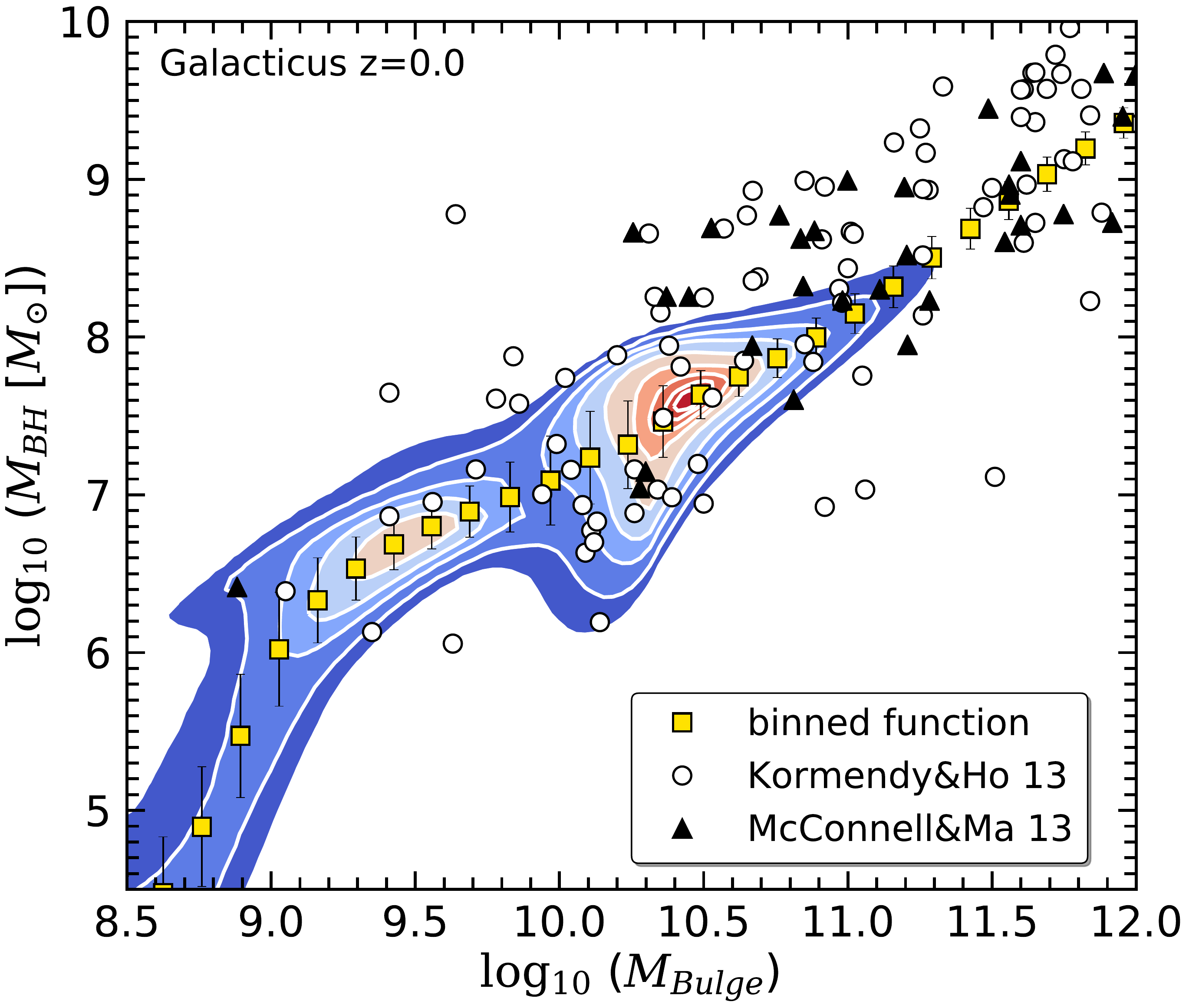}
	\end{subfigure}
	\begin{subfigure}{\columnwidth}
	  \centering
	  \includegraphics[width=\columnwidth]{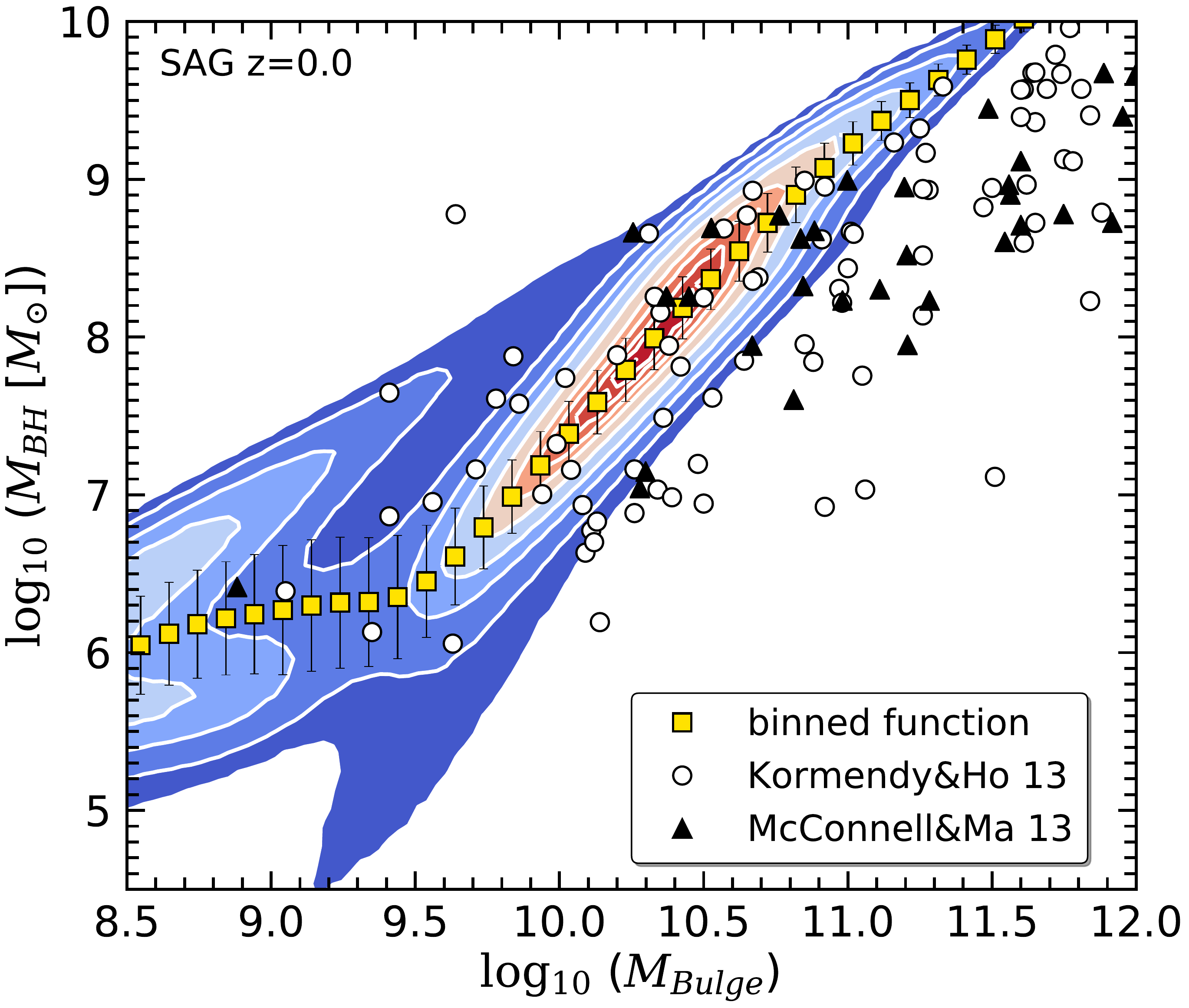}
	\end{subfigure}
	\begin{subfigure}{\columnwidth}
	  \centering
	  \includegraphics[width=\columnwidth]{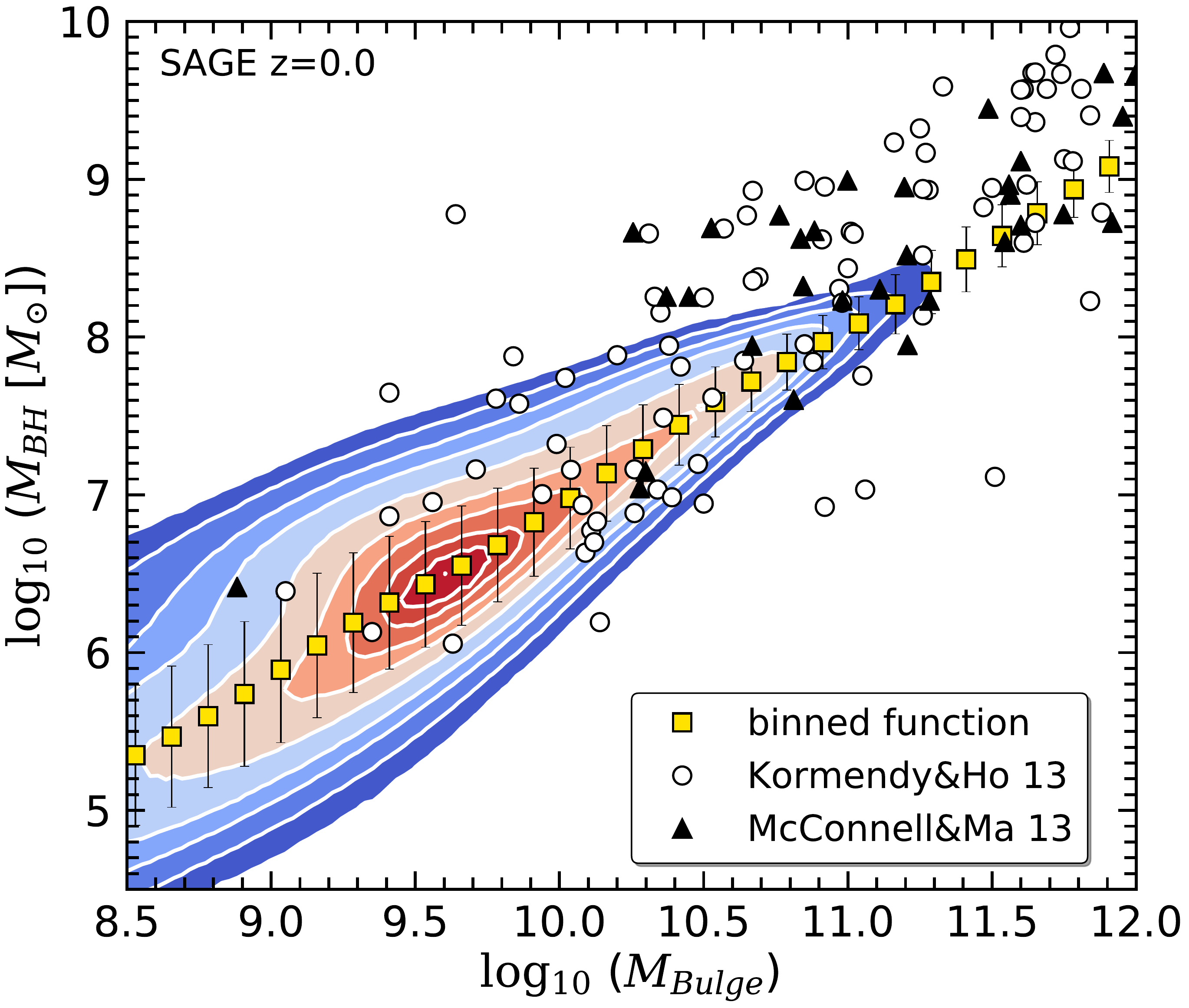}
	\end{subfigure}
	 \caption{The contours show the relation of the black hole mass to stellar bulge mass at redshift $z=0$ compared to observations from \citet{Kormendy13} (open circles) and \citet{McConnell13} (filled triangles) for \galacticus (\textit{top panel}), \sag (\textit{middle panel}) and \sage (\textit{bottom panel}). The yellow squares represent the binned data points of the same relation for a certain model.}
	 \label{fig:mbh2mstarsph}
	 \vspace{-0.3cm}
 \end{figure}
 
It is very challenging to observe black hole (BH) masses in galaxies especially in a lower mass regime. Therefore SAMs provide a helpful and valuable tool to study possible correlations of galaxy properties -- even at scales not yet well probed observationally. And black hole growth and growth in stellar mass are connected via feedback mechanisms (e.g. AGN feedback); therefore, BH growth plays a critical role in galaxy evolution \citep{Croton06,bower_agn_2006,Croton16}. For more than a decade the picture that BHs and bulges co-evolve by regulating each others growth was mainly accepted. However, more recent studies support a more advanced picture claiming that BHs correlate differently with different galaxy components \citep{Kormendy13}. In \hyperref[fig:mbh2mstarsph]{\Fig{fig:mbh2mstarsph}}, we present the \BHBM for \galacticus (\textit{top panel}), \sag (\textit{middle panel}) and \sage (\textit{bottom panel}) at redshift $z=0$ as coloured contours and binned data points (yellow squares). All three models are in excellent agreement with the observations reported by \citet{Kormendy13} and \citet{McConnell13}; they all favour the almost linear relation (in log-space, i.e. a power-law in linear-space) between black hole and stellar mass. We note that all our models are tuned to match the \BHBM\ relation, and hence the agreement reported here is expected.

However, we observe for \sag\ that for large bulge masses, the black holes are more massive than in the other models. This is due to the restriction imposed on the high mass end of the \SMF\ at $z=0$. The \sag\ model tries to avoid the excess in the high-mass end making the AGN feedback as effective as possible by large accretions onto BHs (high values of $f_{BH}$ in the related equations; see formula in \hyperref[sec:sag]{\Sec{sec:sag}}) which leads to their high masses. However, despite this strong effect, model predictions do not satisfy this particular aspect of the observational constraint, indicating that other processes must be revised, like tidal stripping and disruption of satellites galaxies, since the dry mergers at low redshifts with massive satellites seem to produce the excess at the high mass end.

The reason why the correlation for \sage\ does not extend to larger bulge masses -- as, for instance, \sag\ -- relates back to what we have already noted in \hyperref[fig:sSFR2Mstar]{\Fig{fig:sSFR2Mstar}}, i.e. the star-forming massive galaxies in \sage\ have discs that are relatively too massive and bulges that are relatively too low in mass. There are, therefore, fewer galaxies with massive bulges than expected, meaning there are fewer galaxies hosting massive black holes (because the model is constrained for the black hole and bulge masses to meet the observed trend). While a thorough treatment of disc evolution in an extension of \sage\ has been presented in \citet{Stevens16}, we leave the application of that particular model (\textsc{Dark Sage}) to MultiDark for future work.

%%%%%%%%%%%%%%%%%%%%%%%%%%%%%%%%%%%%%%%%%%%%%%%%%%%
\subsection{The Cold Gas Fraction (\CGF)} \label{sec:mcold2mstar}
%%%%%%%%%%%%%%%%%%%%%%%%%%%%%%%%%%%%%%%%%%%%%%%%%%%
  \begin{figure}
 	\begin{subfigure}{\columnwidth}
 	  \centering
 	  \includegraphics[width=\columnwidth]{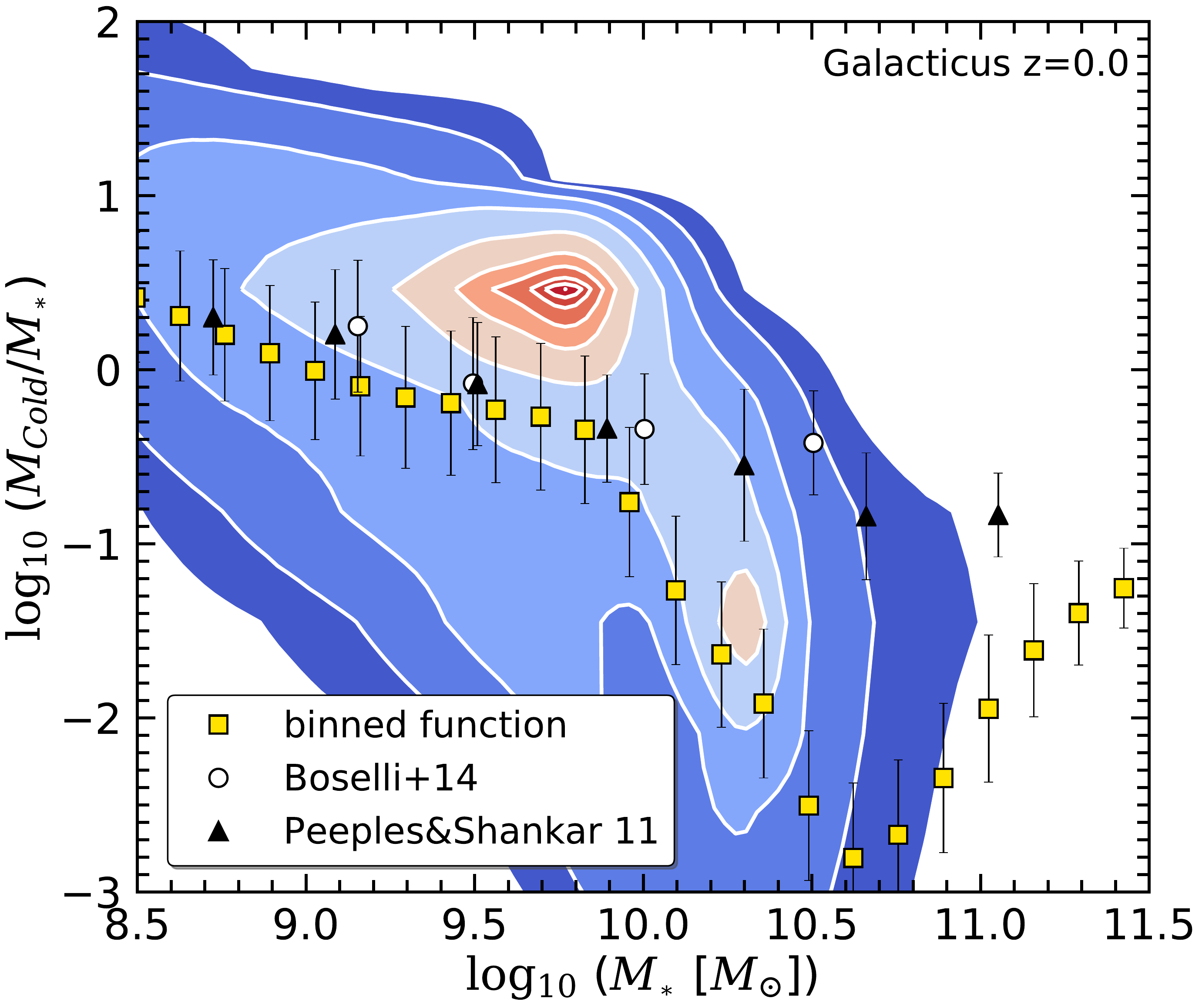}
 	\end{subfigure}
 	\begin{subfigure}{\columnwidth}
 	  \centering
 	  \includegraphics[width=\columnwidth]{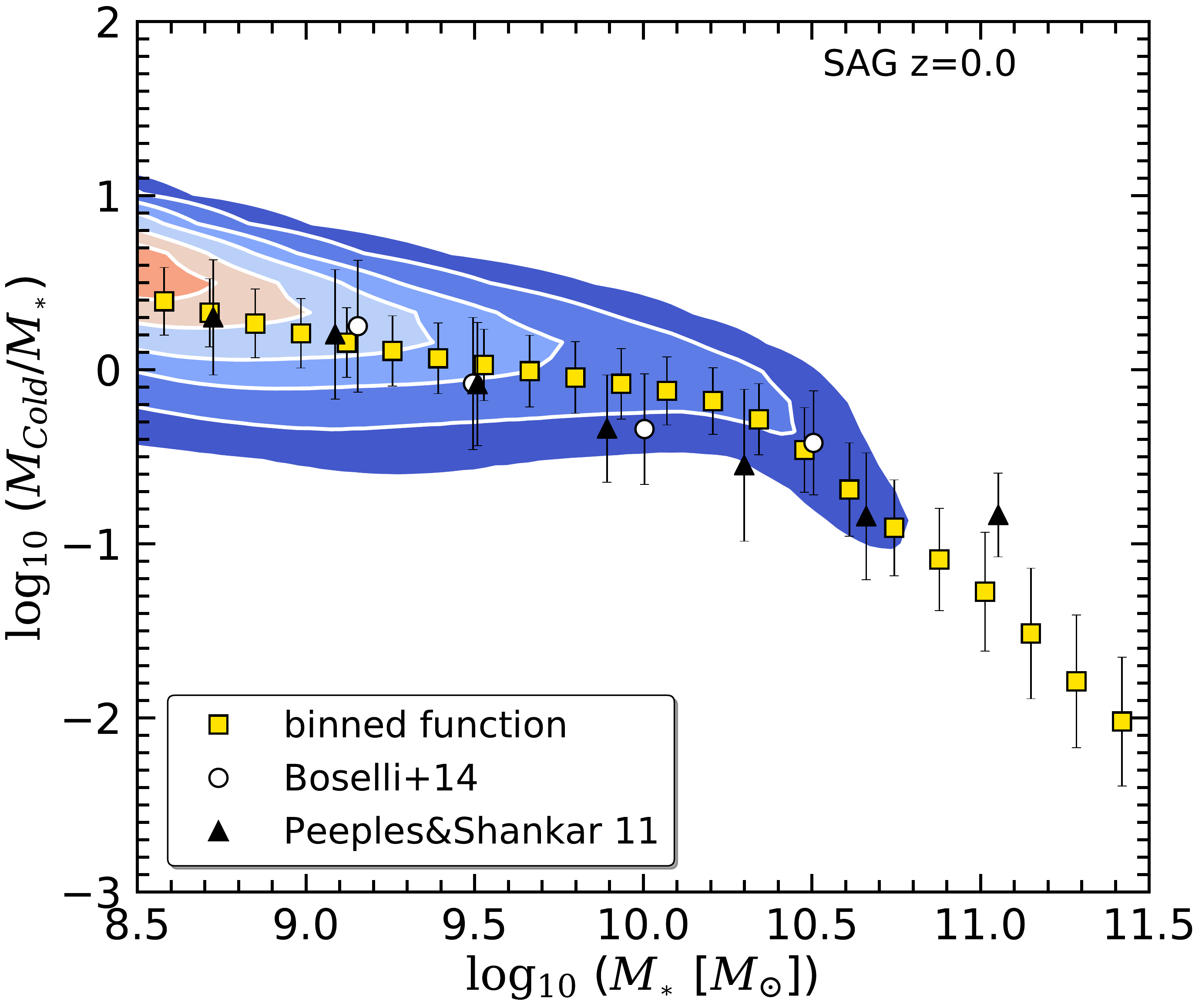}
 	\end{subfigure}
 	\begin{subfigure}{\columnwidth}
 	  \centering
 	  \includegraphics[width=\columnwidth]{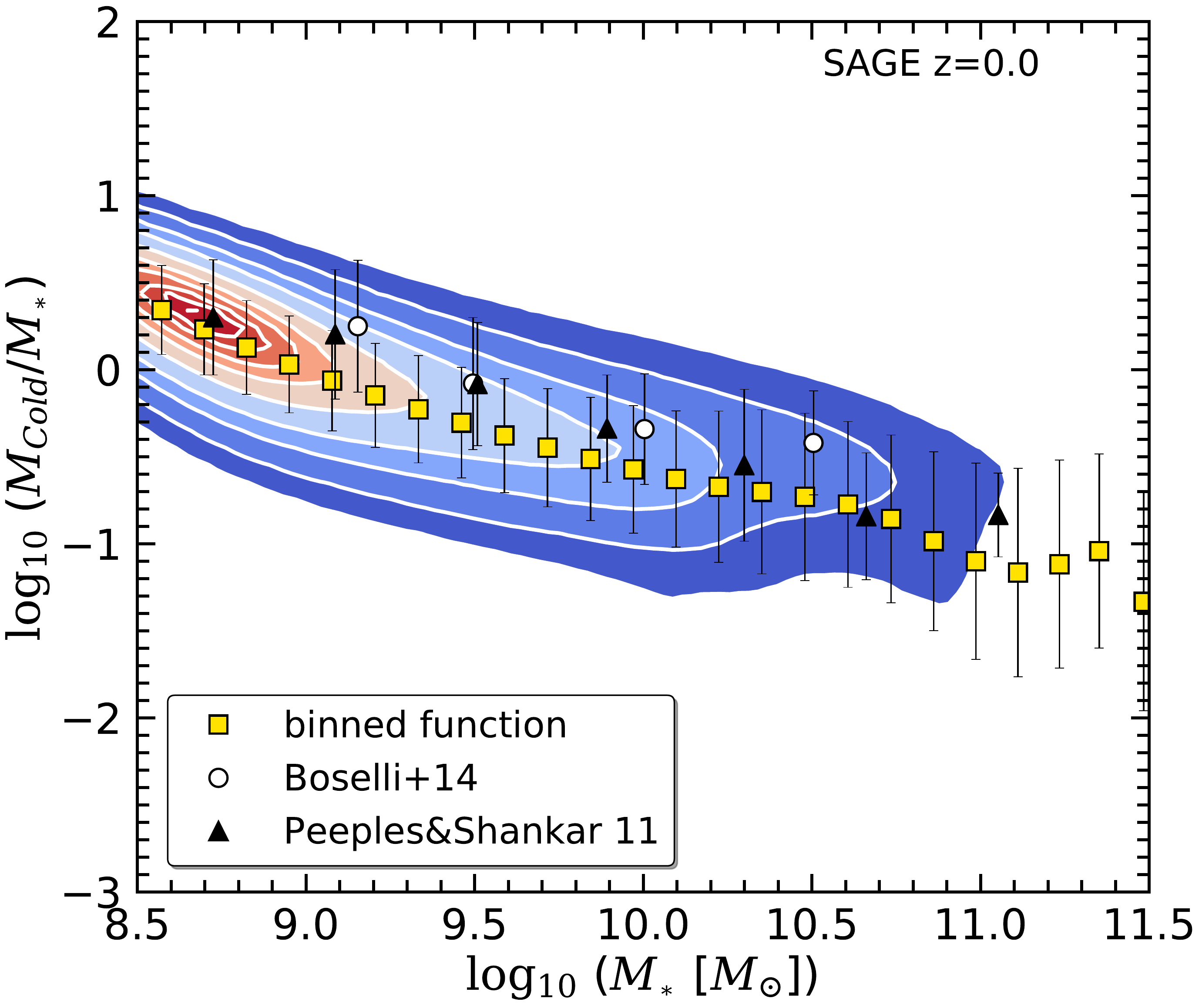}
 	\end{subfigure}
   	\caption{Fraction of cold gas compared to stellar mass as a function of stellar mass at redshift $z=0$ compared to observations from \citep[their Fig.5a]{Boselli14} (open circles) and \citet[Table 2]{Peeples11} (black triangles). The yellow squares represent the binned data points of the same relation for a certain model.}
	 \label{fig:mcold2mstar} \vspace{-0.3cm}
  \end{figure}
 
An important tracer for star formation, age and metallicity is the fraction of cold gas to stellar mass. We therefore show in \hyperref[fig:mcold2mstar]{\Fig{fig:mcold2mstar}} the \CGF vs. stellar mass for \galacticus (\textit{top panel}), \sag (\textit{middle panel}) and \sage (\textit{bottom panel}) at redshift $z=0$ as coloured contours and binned data points (yellow squares). We report that \sag and \sage are in excellent agreement with the observational data points from \citet[Fig.5a; open circles]{Boselli14}. This also applies to considering HI and CO detected late-type objects \citep[compilation Table 2; black triangles]{Peeples11} as well as considering HI from 21cm and HI+HII detected star-forming objects. Every data point of the binned function of \sage and \sag is located within the error bars of at least one of the observations, with the exception of \sag for $\log_{10}(M_{*}~[\Msun])~>11.0$. But note that \sag\ has been calibrated to the \citet{Boselli14} data for which there are no data points beyond $\log_{10}(M_{*}~[\Msun])~>10.5$. \galacticus' \CGF drops rapidly between $10.0~<~\log_{10}(M_{*}~[\Msun])~<~11.0$. This is related to the current model of AGN feedback in \galacticus\ which is quite extreme, and dramatically reduces gas cooling above this scale. However, star formation rates remain high in these galaxies after AGN feedback kicks in, so they rapidly deplete their gas supply. However, we note that all our models are consistent with standard theories of star formation where massive, red galaxies either already used up their gas reservoir or (cold) gas became unavailable due to feedback mechanisms. Or -- compared to the total stellar mass -- they simply contain too small cold gas fractions \citep{Lagos14}. All of this explains the low \CGF\ for the high-$M_{*}$ galaxies see in this figure. 

%Although a more detailed analysis would be beyond the scope of our paper and will be presented in accompanying studies, we also discuss the total gas-phase metallicity as a function of stellar mass in the next section.

%%%%%%%%%%%%%%%%%%%%%%%%%%%%%%%%%%%%%%%%%%%%%%%%%%%%%%%%%%%%%%%%%%%%%%%%%%%
\subsection{The Mass-Metallicity Relation} \label{sec:zgas2mstar}
%%%%%%%%%%%%%%%%%%%%%%%%%%%%%%%%%%%%%%%%%%%%%%%%%%%%%%%%%%%%%%%%%%%%%%%%%%%

 \begin{figure}
  \vspace{-0.3cm}
 	\begin{subfigure}{\columnwidth}
 	  \centering
 	  \includegraphics[width=\columnwidth]{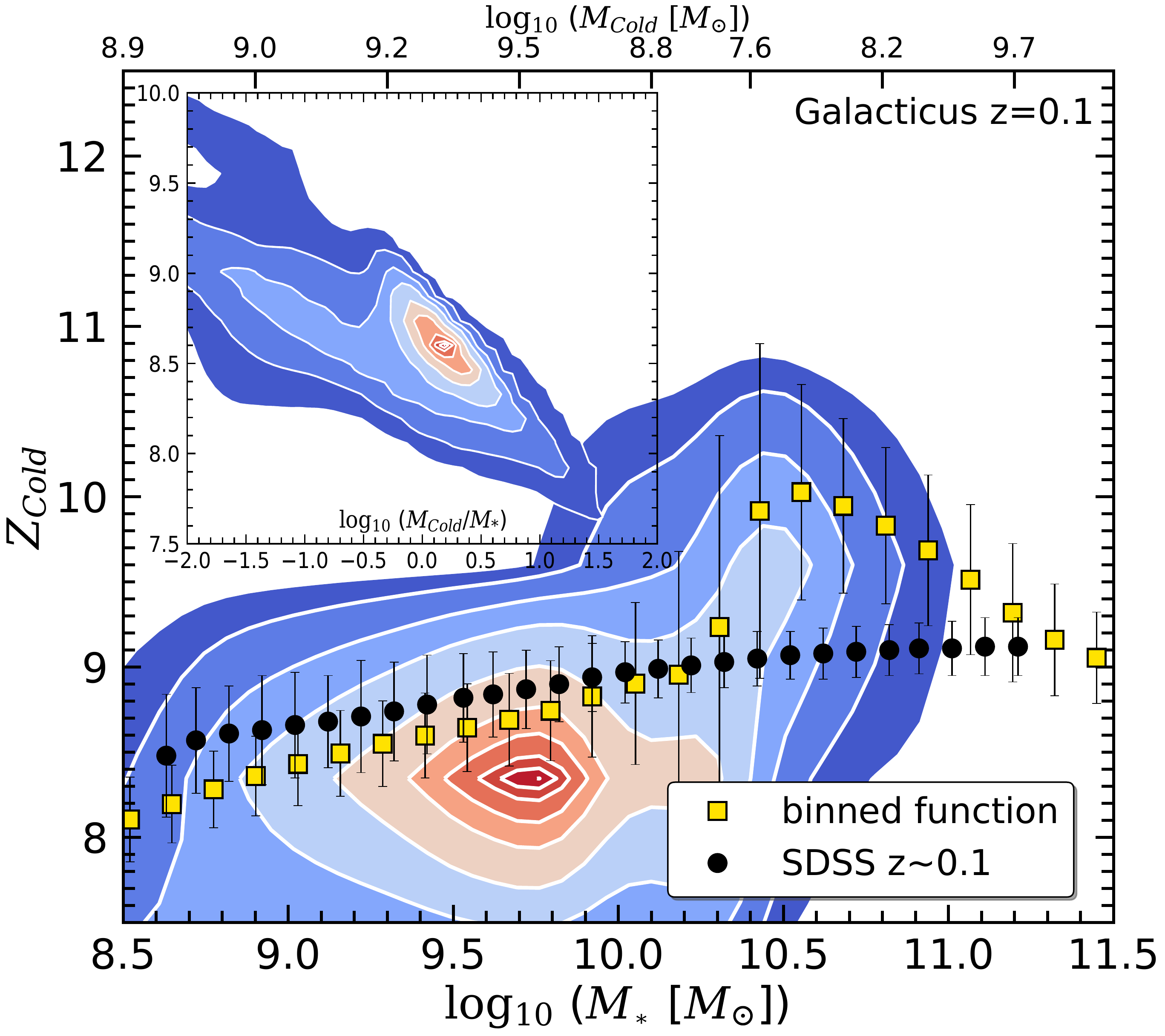}\vspace{0.2cm}
 	\end{subfigure}
 	\begin{subfigure}{\columnwidth}
 	  \centering
 	  \includegraphics[width=\columnwidth]{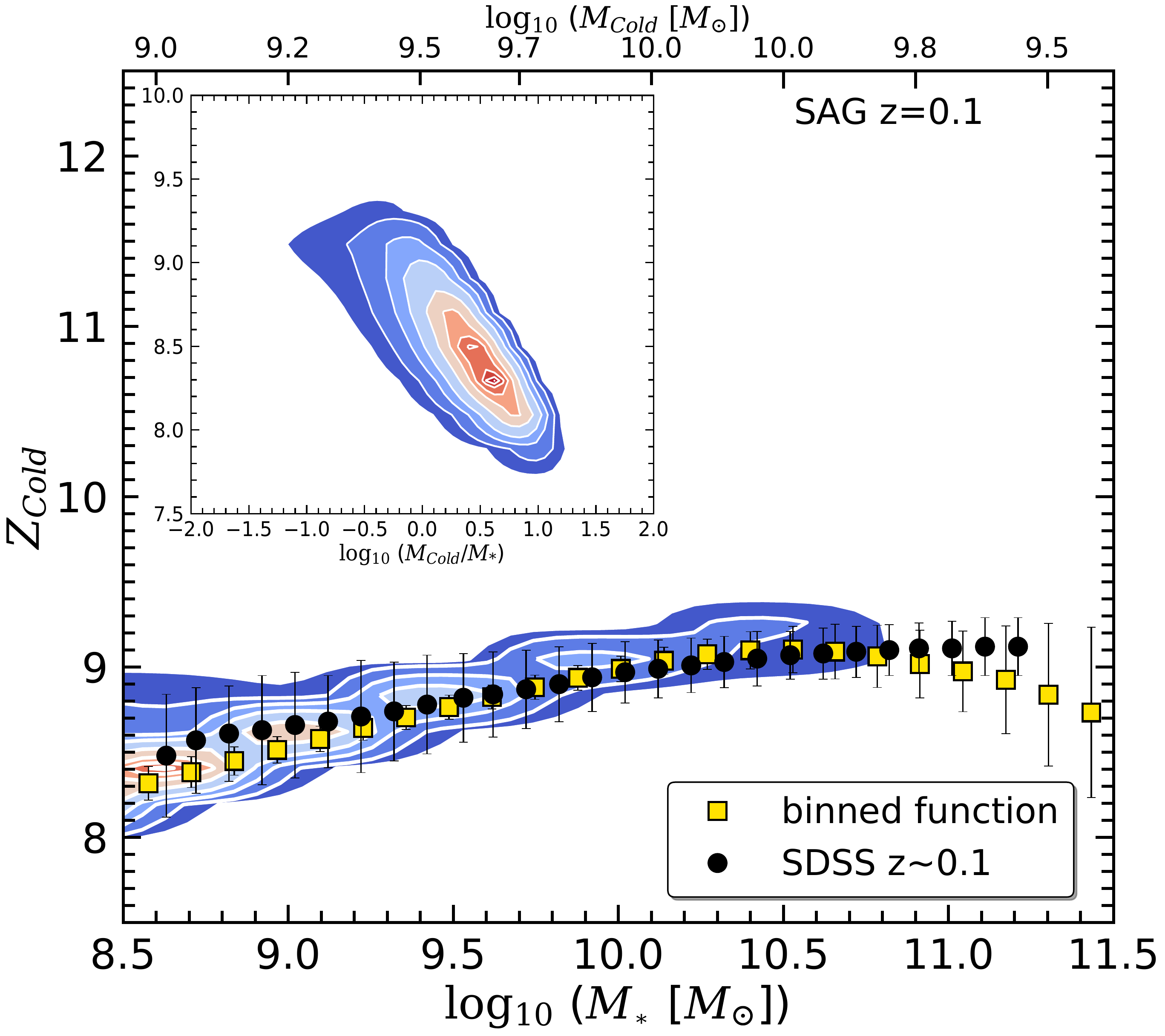}\vspace{0.2cm}
 	\end{subfigure}
 	\begin{subfigure}{\columnwidth}
 	  \centering
 	  \includegraphics[width=\columnwidth]{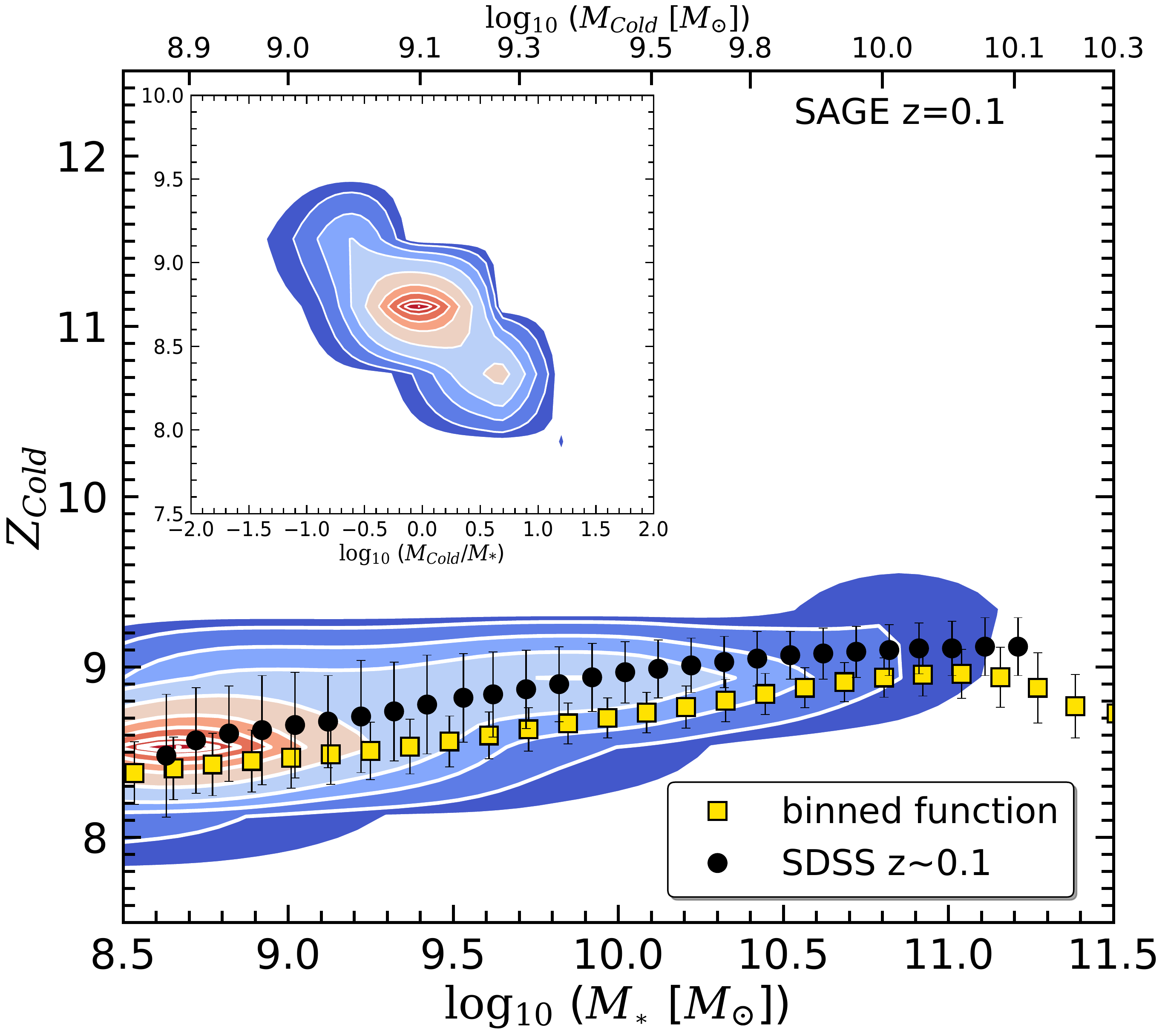}
 	\end{subfigure}
   	\caption{The total gas-phase metallicity to stellar mass relation at redshift $z=0.1$ compared to observations from \citet{Tremonti04} (black dots error bars represent the 2.5/97.5 percentile of the distribution). The yellow squares represent the binned data points of the same relation for a certain model. The inset plot shows the same gas-phase metallicity as in the outer plot, but now compared to \Mcold/\Mstar.}
	 \label{fig:zgas2mstar} \vspace{-0.3cm}
 \end{figure}

Metals in galaxies are produced in stars and released into the inter-stellar and inter-galactic medium when stars let go of their gaseous envelopes or explode as supernovae. And as metals act as cooling agents in the process of star formation, their distribution throughout the galaxy also influences the (distribution of the) next generation of stars providing a link between metallicities and galaxy morphology \citep{Lara-Lopez09,Lara-Lopez10,Yates2013}. They are further strongly linked to stellar mass and star formation, leading to pronounced correlations with luminosities, and circular velocities as well. 

We now verify such a relation between metallicity and stellar mass in the models by considering the total gas-phase abundance as a function of stellar mass using 
\begin{equation}
 	Z_{\rm cold} =8.69+\log_{10}(M_{Z, \rm cold}/M_{\rm cold})-\log_{10}(Z_{\odot}),
\end{equation}

where $M_{Z,\rm cold}$ is the mass of metals in the cold gas-phase and \Mcold is the total gas mass. $Z_{\rm cold}$ is normalized by the metallicity of the Sun $Z_{\odot}=0.0134$ \citep{Asplund09}, while the factor $8.69$ \citep{Allende01} corresponds to its oxygen abundance. Note that $Z_{\rm cold} $ as defined here is a conversion of cold gas metalicity to the oxygen abundance. Displaying metallicities this way is a commonly used approach in the literature, and hence it is also adopted here. Note that for \sage the total gas mass is given by the cold gas disc mass and that the other two models additionally provide a cold gas component for the bulge.

We present the results for the total gas-phase metallicity to stellar mass relation in \hyperref[fig:zgas2mstar]{\Fig{fig:zgas2mstar}}. We find that the SAMs in general are in good agreement with the observational data from \citet{Tremonti04}. Compared to \hyperref[fig:mcold2mstar]{\Fig{fig:mcold2mstar}} where the \Mcold/\Mstar\ ratio is decreasing, here the metallicity $Z_{\rm cold}$ is increasing with mass. That means that more massive galaxies tend to have a smaller cold gas reservoir and higher metallicity. The larger extent of the cold gas fraction seen in \hyperref[fig:mcold2mstar]{\Fig{fig:mcold2mstar}} for \galacticus\ is mirrored here again: for a fixed \Mstar\ value, the spread in predicted metallicity is largest for \galacticus further hinting at a similar bimodality as for the cold gas fraction. The peak seen for this model at $\log_{10}(M_{*}) \approx 10.5$ again relates to the depletion of gas due to the AGN feedback implementation in \galacticus: these galaxies have almost no inflow of pristine gas, and rapidly consume their gas supply. As expected from simple chemical evolution models the metallicity of the cold gas is driven up to the effective yield in this case. The other two models \sag\ and \sage\ show excellent agreement with the observational data -- noting that this relation has been used during the parameter calibration for \sage. And the marginal offset seen for that model is simply due to the conversion from metal fraction to $Z_{\rm gas}$ being different for this plot versus the \sage\ calibration plot. To relate metallicity with cold gas mass we also include upper (approximated) tick marks representing the total cold gas mass \Mcold. Recent studies of the \Mstar-$Z_{\rm cold}$ relation suggest that there is an additional dependence of this relation on \SFR \citep{Ellison08,Mannucci10,Lara10a,Lara10b,Yates12}. Additional projections are used by various authors in their works including \SFR and \CGF\ to investigate the parameter space of these properties in more detail. The picture drawn by these works clearly corresponds to our current knowledge about galaxy formation. However, they report a `turnover' towards low metallicities at low-\Mcold/\Mstar\ \citep[see Fig.6 in][]{Yates12} for galaxies with stellar masses $\log_{10}{(\Mstar[\Msun])}\geq10.5$. Since cold gas is the fuel for star formation -- and metals are the required coolant -- the gas-to-stellar mass ratio \Mcold/\Mstar, or equally $Z_{\rm cold}$, should correlate with the enrichment of the inter-stellar medium, hence metallicity of the gas. \citet{Yates12} concluded that galaxies with low \sSFR contribute to that turnover occurring at higher masses, in the sense that they tend to have lower ratio than other galaxies of a similar mass, caused by a gradual dilution of the gas phase in some galaxies. This is triggered by a gas-rich merger which shuts down subsequent star formation without impeding further cooling. They also drew a link between this `turnover' and the black hole mass where they claim that these `turned-over' galaxies also exhibit a larger central black hole mass. A detailed study inspired by \citet{Yates12} would be interesting but beyond the scope of this paper. However, we here tested a few relations presented in their paper and can report a similar behaviour for our SAMs (see inset plots of \hyperref[fig:zgas2mstar]{\Fig{fig:zgas2mstar}}).
Furthermore, currently there is only limited observational data available to study this relation as well as its dependence on \SFR, meaning that modelling metallicities will remain a very important tool and challenging task until sufficient data has been collected.

%%%%%%%%%%%%%%%%%%%%%%%%%%%%%%%%%%%%%%%%%%%%%%%%%%%
\subsection{Stellar-to-Halo Mass Fraction (\SHMF)} \label{sec:SHMF}
%%%%%%%%%%%%%%%%%%%%%%%%%%%%%%%%%%%%%%%%%%%%%%%%%%%
\begin{figure}
	\includegraphics[width=\linewidth]{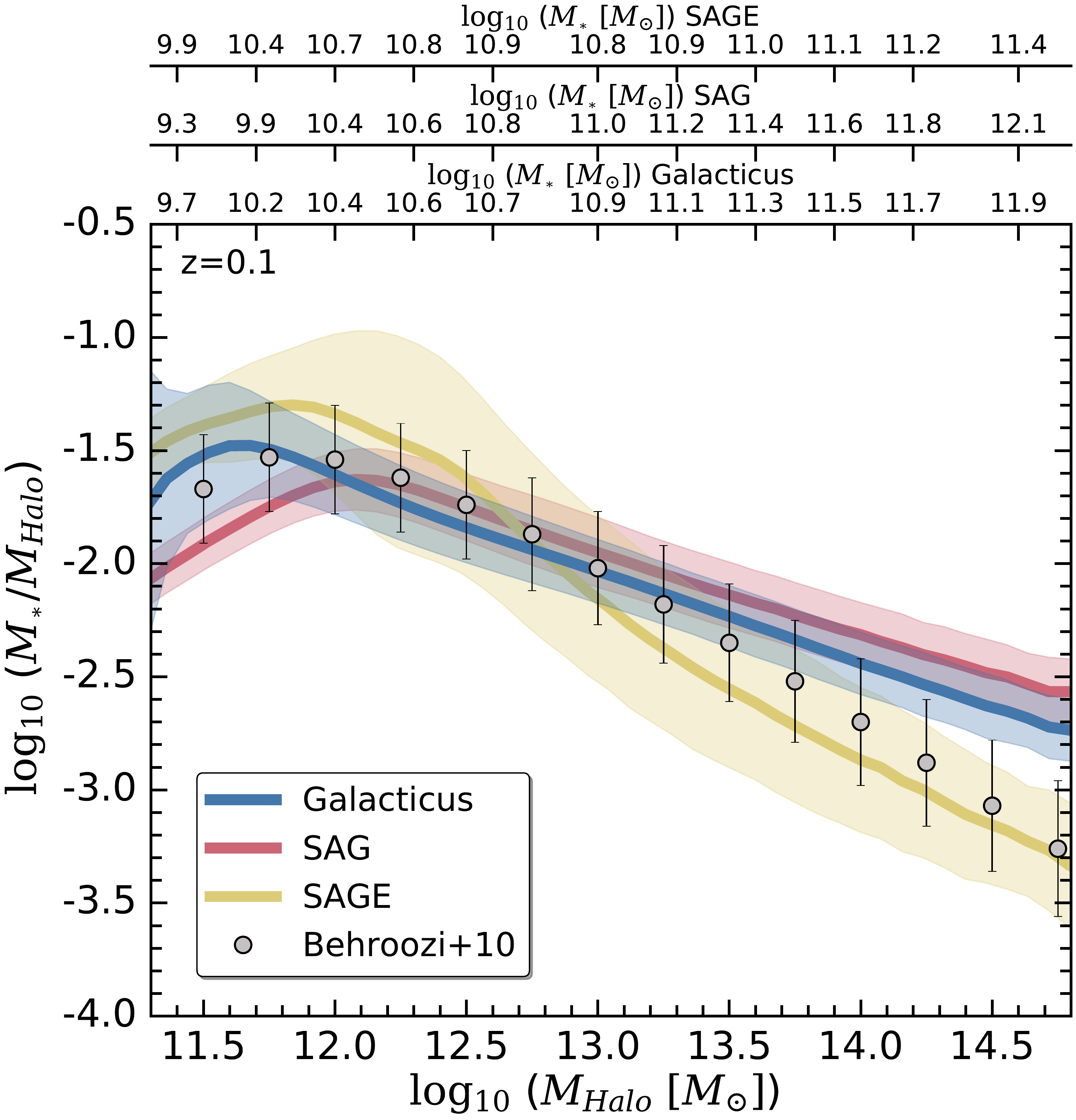}\vspace{-0.2cm}
	\caption{Stellar-to-halo mass ratio as a function of halo mass compared to the models of \citet{Behroozi10} for \textit{non-orphan galaxies} at $z=0.1$.}
	\label{fig:SHMF}
	\vspace{-0.3cm}
\end{figure}

The previous sub-sections only dealt with the stellar and gas content of the galaxies (and its related properties). Here we draw a link to the dark-matter haloes they reside in. For this purpose we show in \hyperref[fig:SHMF]{\Fig{fig:SHMF}} the stellar-to-halo mass fraction $M_{\ast}/M_{\rm~Halo}$ (\SHMF) as a function of dark-matter host mass $M_{\rm Halo}$. Note that we excluded orphan galaxies from this plot as they do not have an associated dark matter (sub-)halo any more by definition. Further, for satellite galaxies we assign the mass of their actual (sub-)halo to them and not the halo mass at the time of accretion to the encompassing dark matter host halo. Note that these two halo masses will be different as dark matter will be tidally stripped when orbiting within the overall host. We are aware that this will introduce a bias towards larger \Mstar/\Mhalo\ values for satellite galaxies.

We compare our SAMs' \SHMF to the abundance matching model of \cite{Behroozi10} at $z=0.1$. We report that our SAMs show a distinct peak around $\log_{10}{(\Mhalo[\Msun])}\thicksim~12$, but slightly shifted either vertically or horizontally from the Behroozi data. The location of this peak as well as the slope of the \SHMF provide deep insight into the physics of our models; the peak marks the halo mass for which the suppression of star formation changes from being controlled by AGN (higher halo mass) to domination of stellar feedback (lower halo masses). This peak should roughly coincide with the knee of the stellar mass function. To allow for such a comparison we provide in \hyperref[fig:SHMF]{\Fig{fig:SHMF}} as upper tickmarks an approximate conversion from halo to stellar mass, which is derived from a convolution of the \SMF as presented in \hyperref[fig:SMF]{\Fig{fig:SMF}} with the stellar-to-halo mass ratio presented here. We note that all our models also agree with this expectation. The marginal excess at the high-\Mhalo\ end for \galacticus\ and \sag\ is yet another reflection of the increased stellar masses at the high-mass end of the \SMF: those galaxies -- residing in the same dark-matter haloes as for \sage\ -- have higher stellar masses than the corresponding galaxies in \sage.

%%%%%%%%%%%%%%%%%%%%%%%%%%%%%%%%%%%%%%%%%%%%%%%%%%%
\subsection{Luminosity Functions (\LF)} \label{sec:lf}
%%%%%%%%%%%%%%%%%%%%%%%%%%%%%%%%%%%%%%%%%%%%%%%%%%%
\begin{figure*}
	  \hspace{-0.0cm}\includegraphics[width=6.9cm]{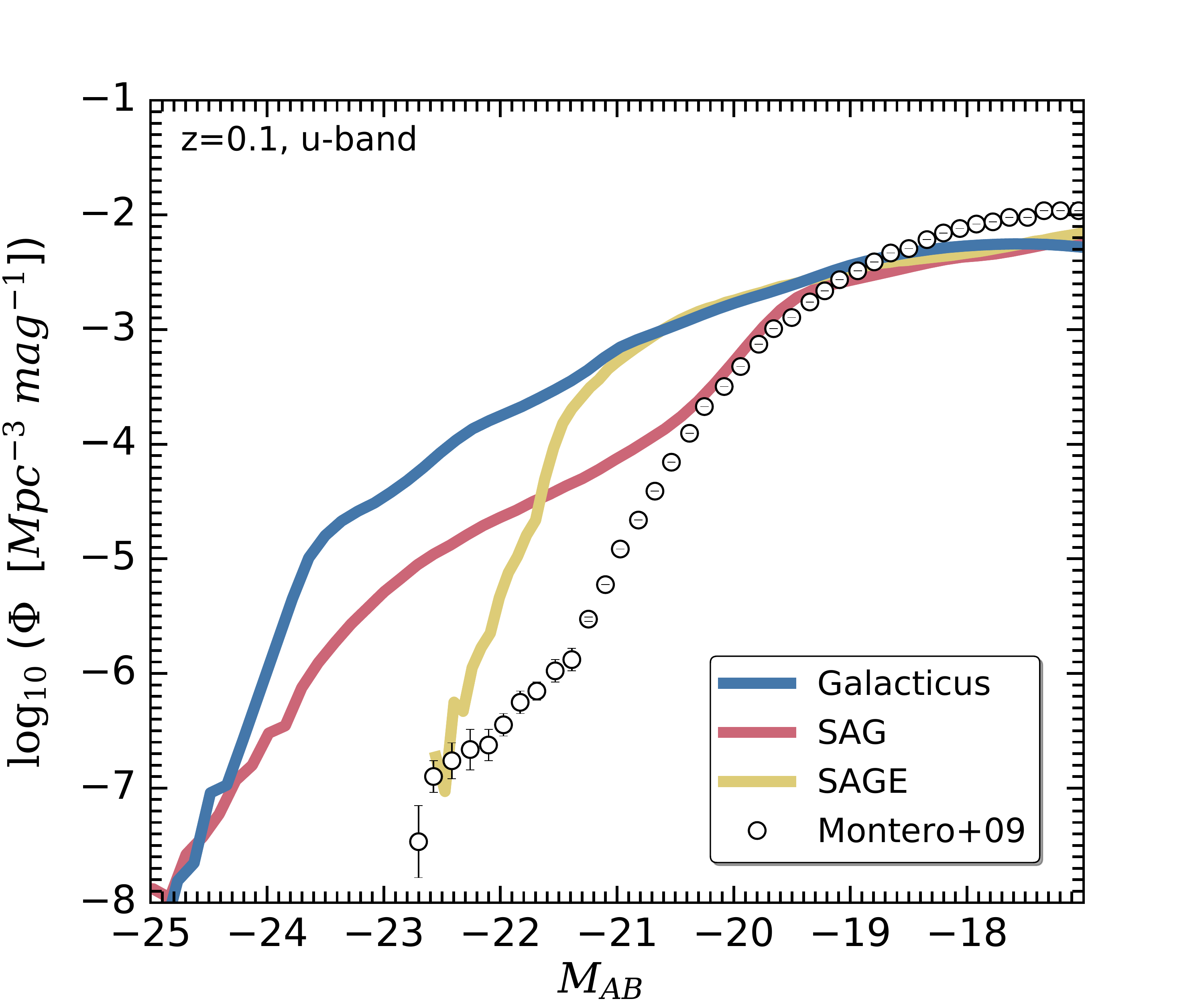}%
	  \hspace{-1.5cm}\includegraphics[width=6.9cm]{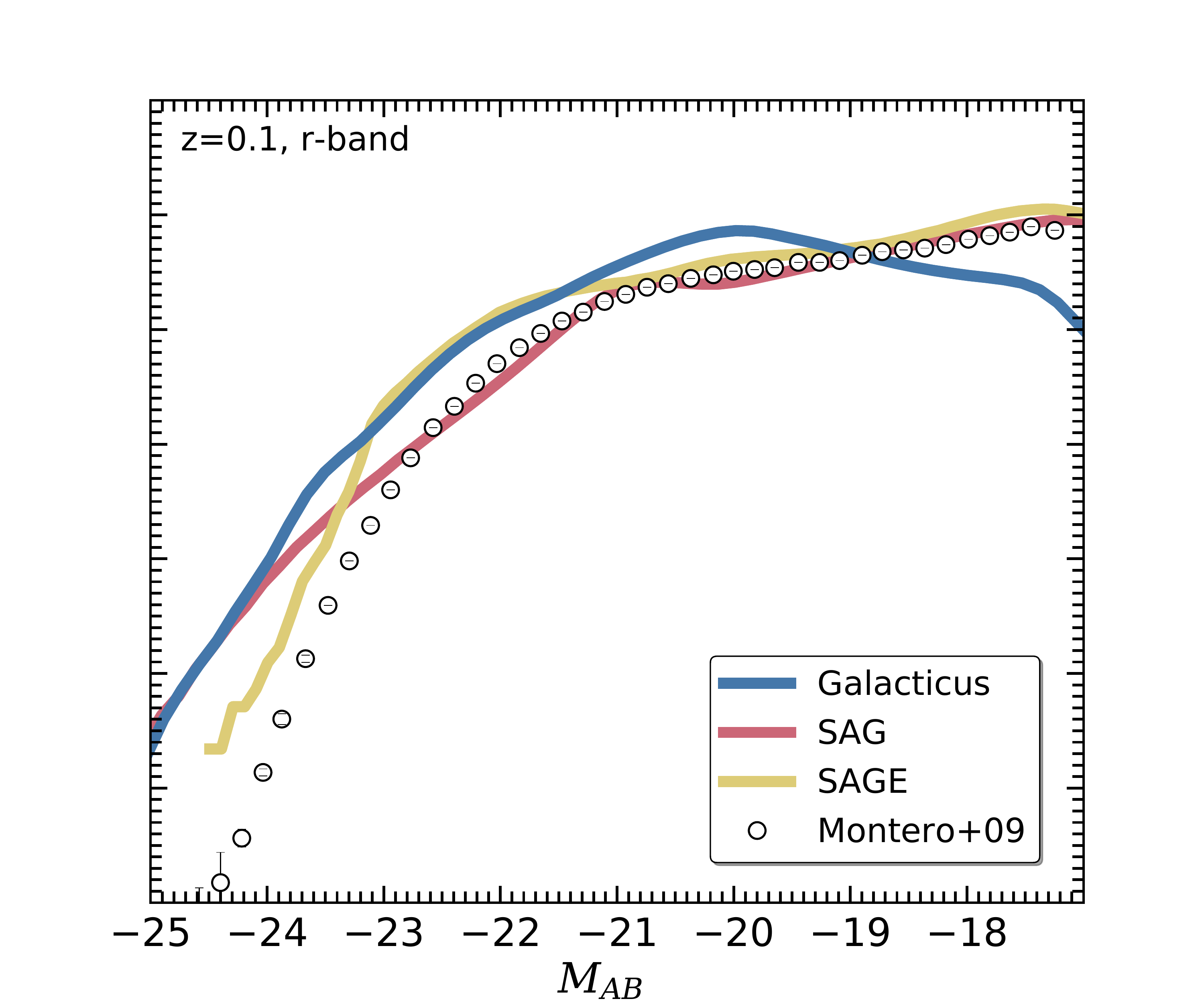}%
	  \hspace{-1.5cm}\includegraphics[width=6.9cm]{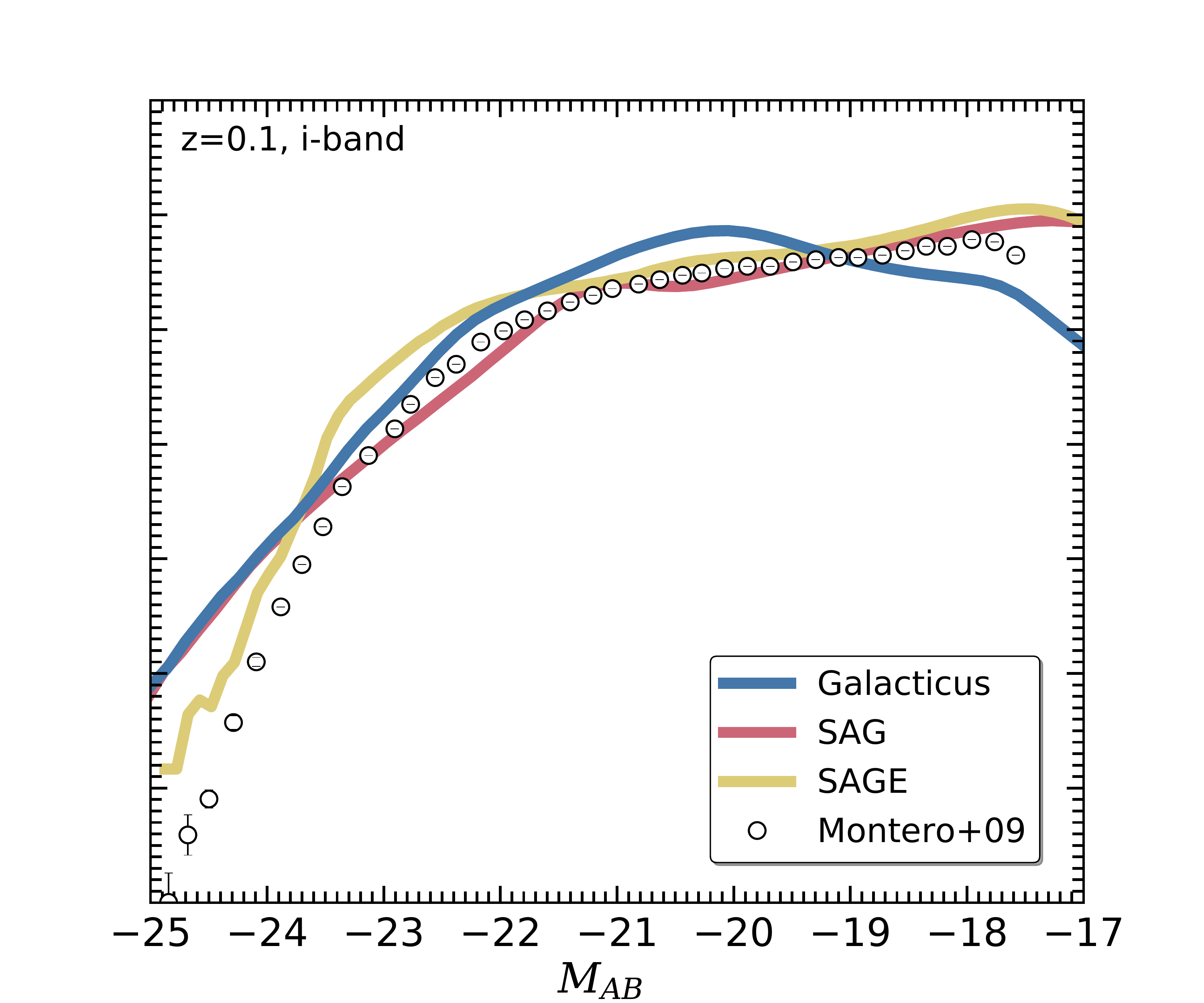}%
	\caption{Luminosity function (rest-frame magnitudes) for the SDSS bands $u,r,i$ for the SAM models compared to observations from SDSS DR6 \citep{Montero09}.}
	\label{fig:LF}
	\vspace{-0.3cm}
\end{figure*}

\begin{table*}
	\caption{\textit{Stellar population synthesis (SPS)} and dust (extinction) models applied to the SAMs in order to generate luminosities.}
	\label{tab:SED}
	\begin{center} 
		\begin{tabular}{llll}
		\hline
		\textsc{Model}	& \textsc{SPS model} & \textsc{Dust Model} & \textsc{Provided Properties}\\
		\hline
		 \galacticus	& \citet{Conroy09}	 		& \citet{Ferrara99} & total luminosities\\
		 \sag			& \citet{bruzual_tab_2007} 	& observational constraints from \citet{Wang96} & rest frame magnitudes AB-system\\
		 \sage		& \citet{bruzual_tab_2003} 	& \textit{Calzetti} extinction curve \citep{Calzetti97, Calzetti01}    & rest frame magnitudes AB-system \\
		 \hline
		\end{tabular}
	\end{center}
\end{table*}

We close the general presentation of the properties of our SAM galaxies with a closer look at luminosities. However, not all of the three models have returned luminosity-based properties as they introduce another layer of modelling, i.e. the employed stellar population synthesis (\SPS) and dust model. In particular, the \sage\ model has not provided luminosities \textit{ab initio} and they were modelled in post-processing via the \tao\footnote{\url{https://tao.asvo.org.au/tao/}\label{ftn:TAO}} \citep[TAO,][]{Bernyk16}. This approach complies with the viewpoint of the \sage team: the majority of the computing time is spent on the construction of the primary galaxy catalogues, and the additional layer of \SPS and dust is preferentially kept modular and separate from the rest of the SAM. The other two models directly returned either luminosities (\galacticus) or magnitudes (\sag) that have been uploaded to the database, whereas the reader should use TAO to generate \sage's luminosities. An overview of their applied \SPS used to create luminosities and dust extinction models can be found in \hyperref[tab:SED]{\Tab{tab:SED}}. In what follows we describe how to unify the provided output and obtain rest-frame magnitudes for them, respectively.

\galacticus\ provides luminosities $L$ as an output (with the bandpass shifted to the emission rest-frame) which can be readily converted into flux densities $f$

\begin{equation}
 f = L / 4 \pi D_L^2 \ ,
\end{equation}

where $D_L$ is the luminosity distance in $cm$. The resulting units of the flux density are $[erg s^{-1} cm^{-2} Hz^{-1}]$ and the zero-point flux density of the AB-System is given by $1~\text{ Jy}~= [10^{-23}erg s^{-1} cm^{-2} Hz^{-1}]$ \citep{OkeGunn1983AB}. We have to further apply a redshift correction factor %of $(1+z)$
to the flux to gain the correct fluxes in the frame of the filter. Using the standard equation to convert flux density into magnitudes in the AB-system and to calculate the magnitudes in the different SDSS \textit{ugriz} bands we hence arrive at

\begin{equation}
 m_{AB} = -2.5 \log_{10} (f/3631 \text{Jy}) -2.5 \log_{10} (1+z) .
\end{equation}

\noindent
$m_{AB}$ are the magnitudes in the filter bands \textit{ugriz} in the observed-frame. Note that these magnitudes correspond to the total galaxy luminosity and that also a dust correction has been applied (see \hyperref[tab:SED]{\Tab{tab:SED}}). In order to calculate the absolute magnitudes in the rest-frame we have to substract the distance modulus and the K-correction to these magnitudes

\begin{equation}
 M_{AB} = m_{AB} - DM - K_{cor}
 \end{equation}
 
 \noindent
where $DM = 5\log_{10}(D_L / 10 \text{ pc})$ and $K_{cor}$ is the K-correction. The latter is calculated using the publicly available `K-corrections calculator'\footnote{\url{http://kcor.sai.msu.ru/}} \citep{Chilingarian2010, Chilingarian2012}. 

The \sag model provides dust corrected absolute magnitudes in the rest-frame, therefore we do not need to apply any conversion.

As mentioned before, \sage's magnitudes were calculated with \TAO. This tool is a highly flexible and allows to select from a huge sample of filter band, \SPS and dust extinction models to create magnitudes and colours for galaxies as a post-processing step separated from the actual simulation and galaxy creation. To generate magnitudes with \TAO we used a subsample with 350 \hMpc\ side-length and applied Chabrier IMF and the SPS and the dust extinction models presented in \hyperref[tab:SED]{\Tab{tab:SED}}; we further only considered galaxies with stellar mass $M_{*}>1.46 \times 10^{8}$.

We present the resulting luminosity function for the three SAMs in \hyperref[fig:LF]{\Fig{fig:LF}}. The figure shows $LF$s in SDSS $u,r,i$ bands at $z=0.1$ compared to the observational data from \citet{Montero09}. Note that the observational data has been corrected to also give rest-frame luminosities allowing for an adequate comparison to our SAM data. While we find reasonable agreement at low-luminosities there are systematically too many bright galaxies for all three models, especially when considering the $u$-band. However, in the case of \galacticus and \sag this phenomenon is readily explained by the fact that for these two SAM models the \SMF\ also shows an excess of high-mass galaxies: they contain too many stars, giving rise to too much light.

\begin{figure*}
	 \hspace{-0.0cm}\includegraphics[width=6.9cm]{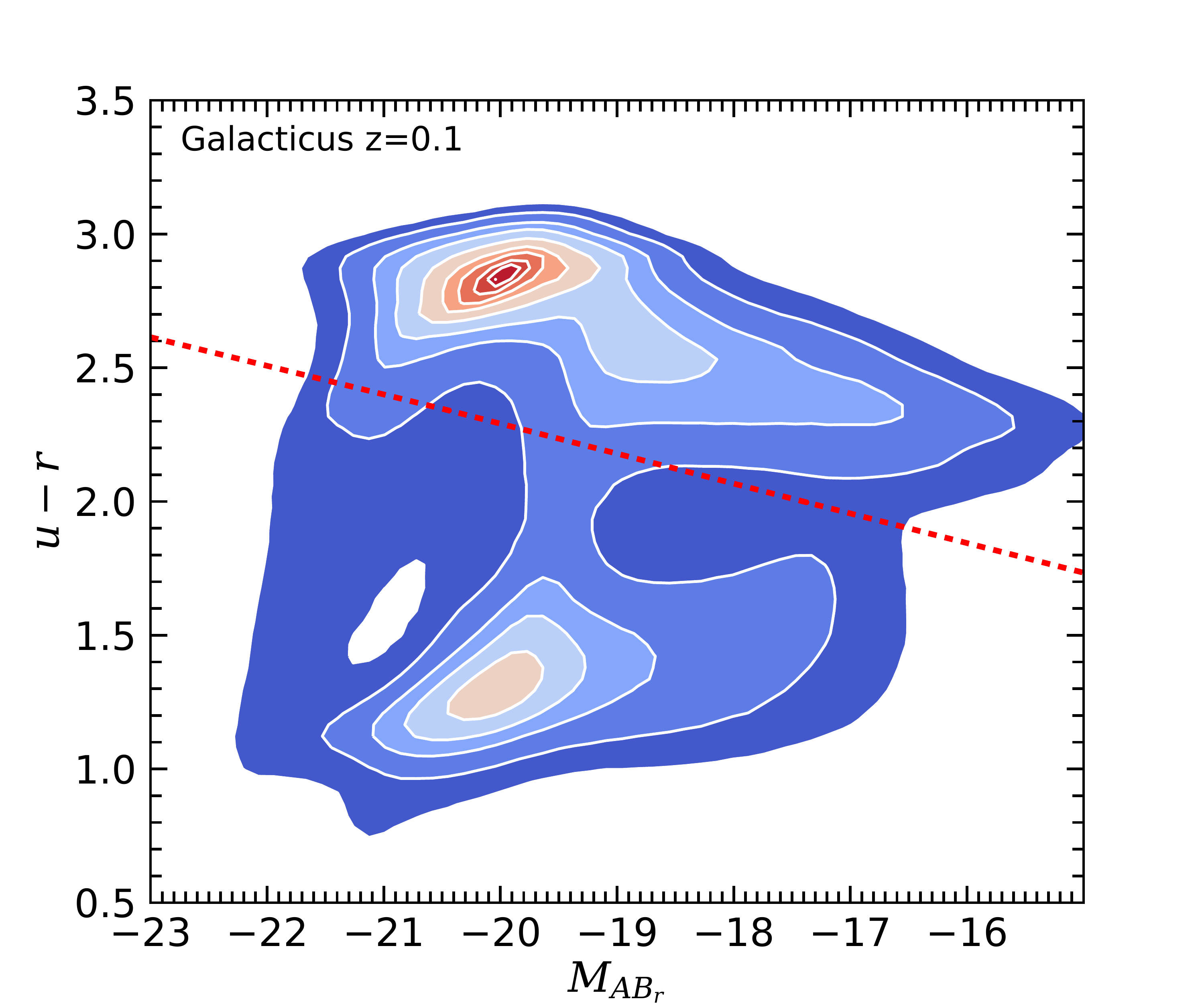}\vspace{-0.5cm}%
	 \hspace{-1.5cm}\includegraphics[width=6.9cm]{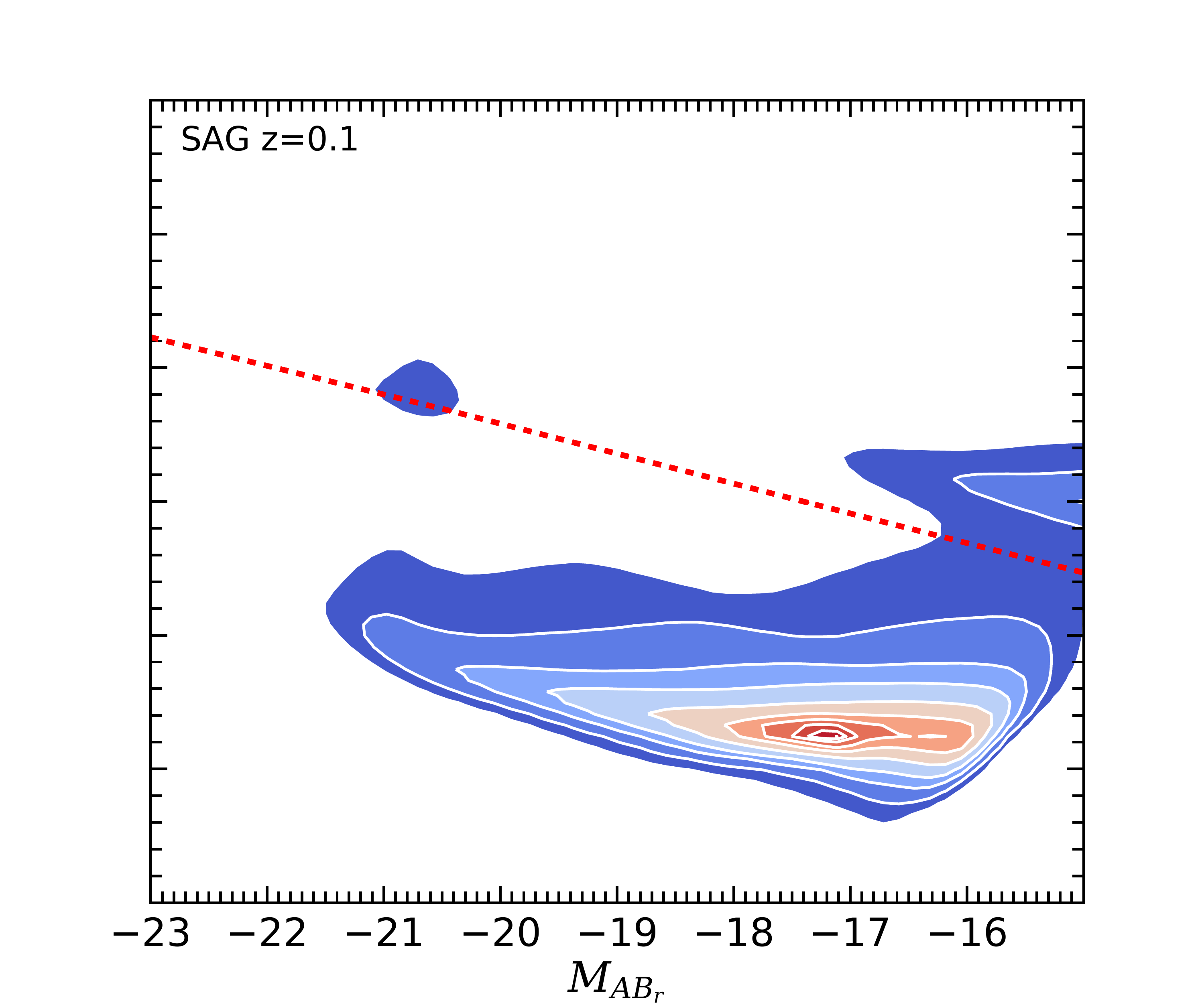}%
	 \hspace{-1.5cm}\includegraphics[width=6.9cm]{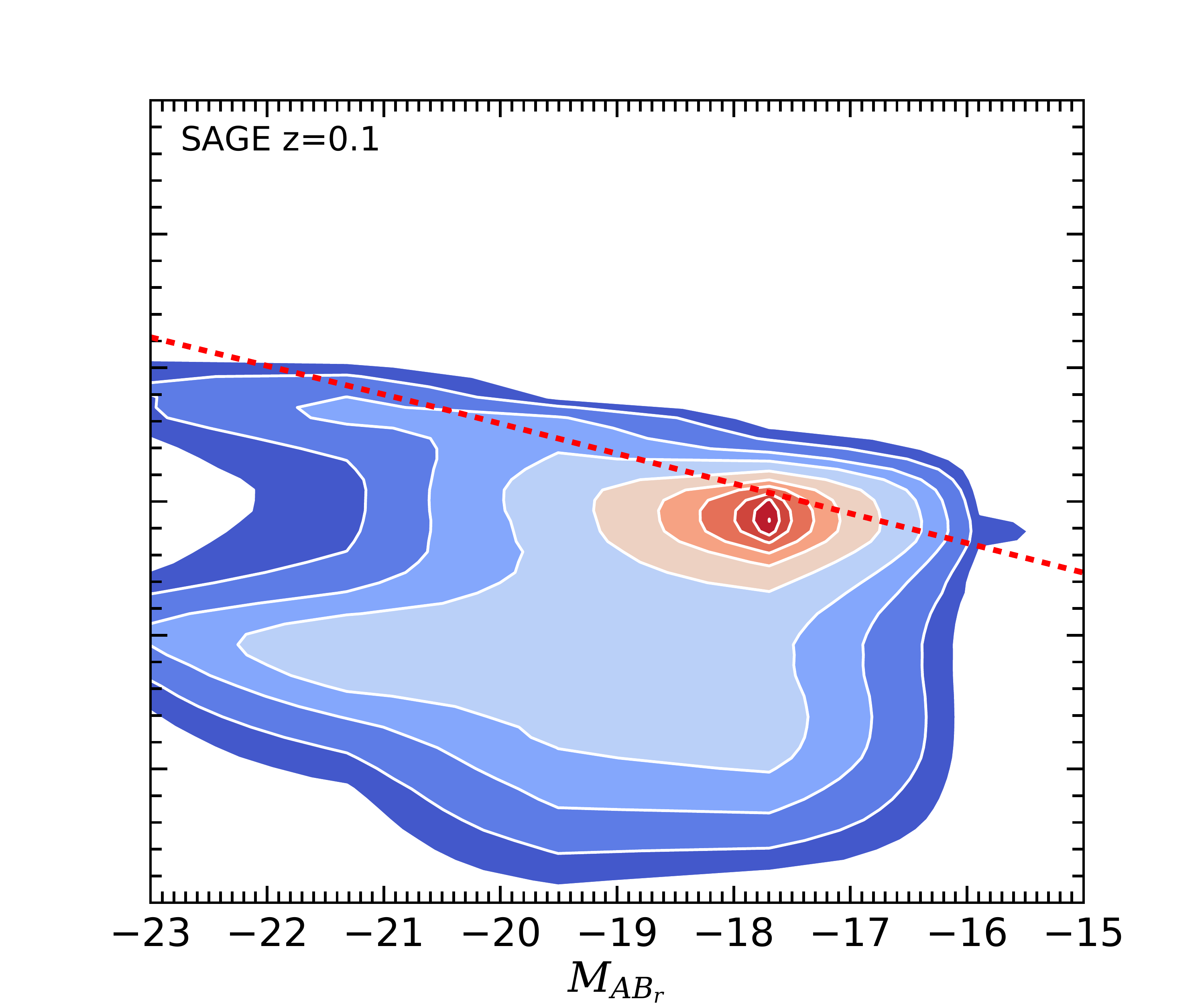}%
	 
	 \hspace{-0.0cm}\includegraphics[width=6.9cm]{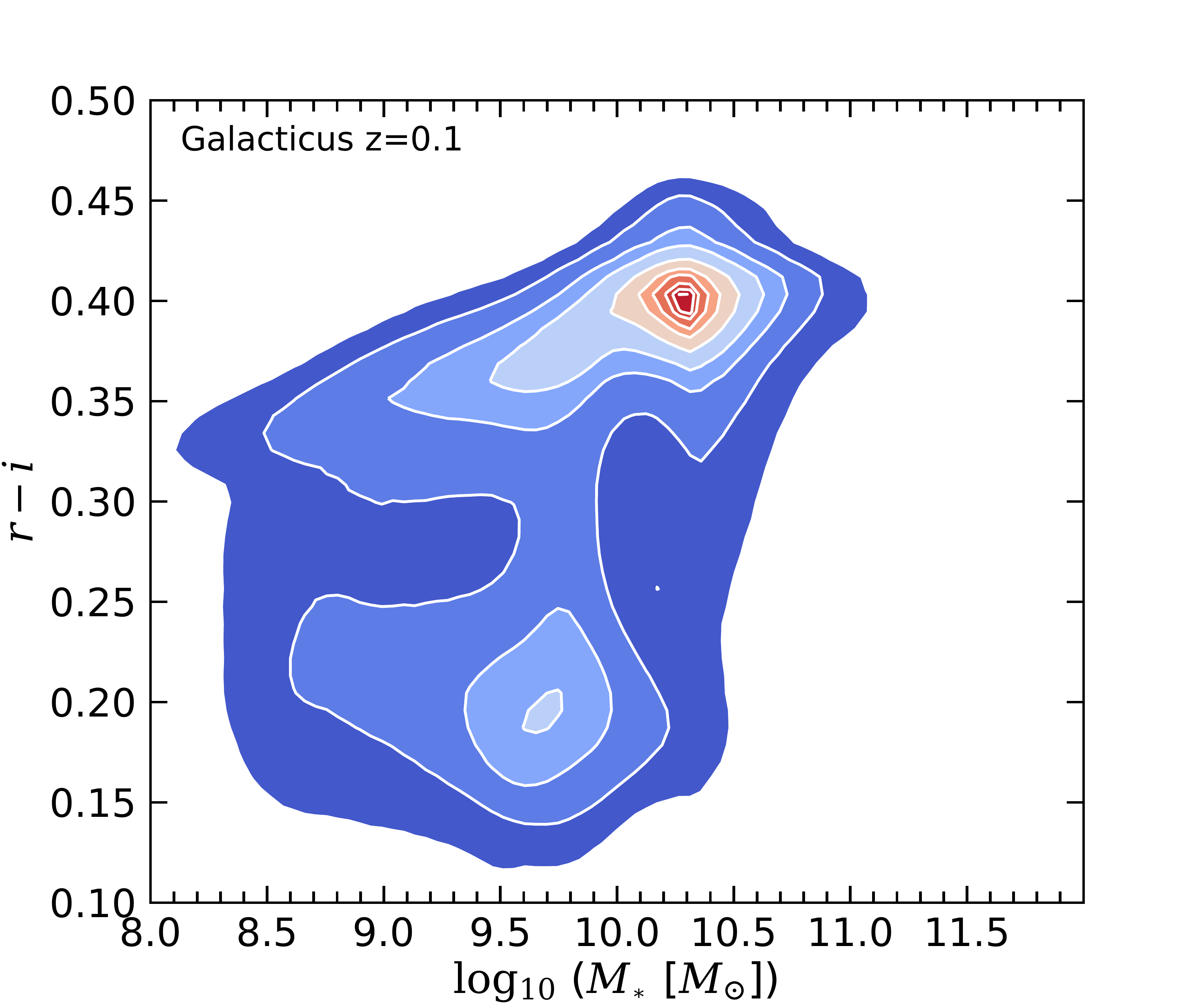}%
	 \hspace{-1.5cm}\includegraphics[width=6.9cm]{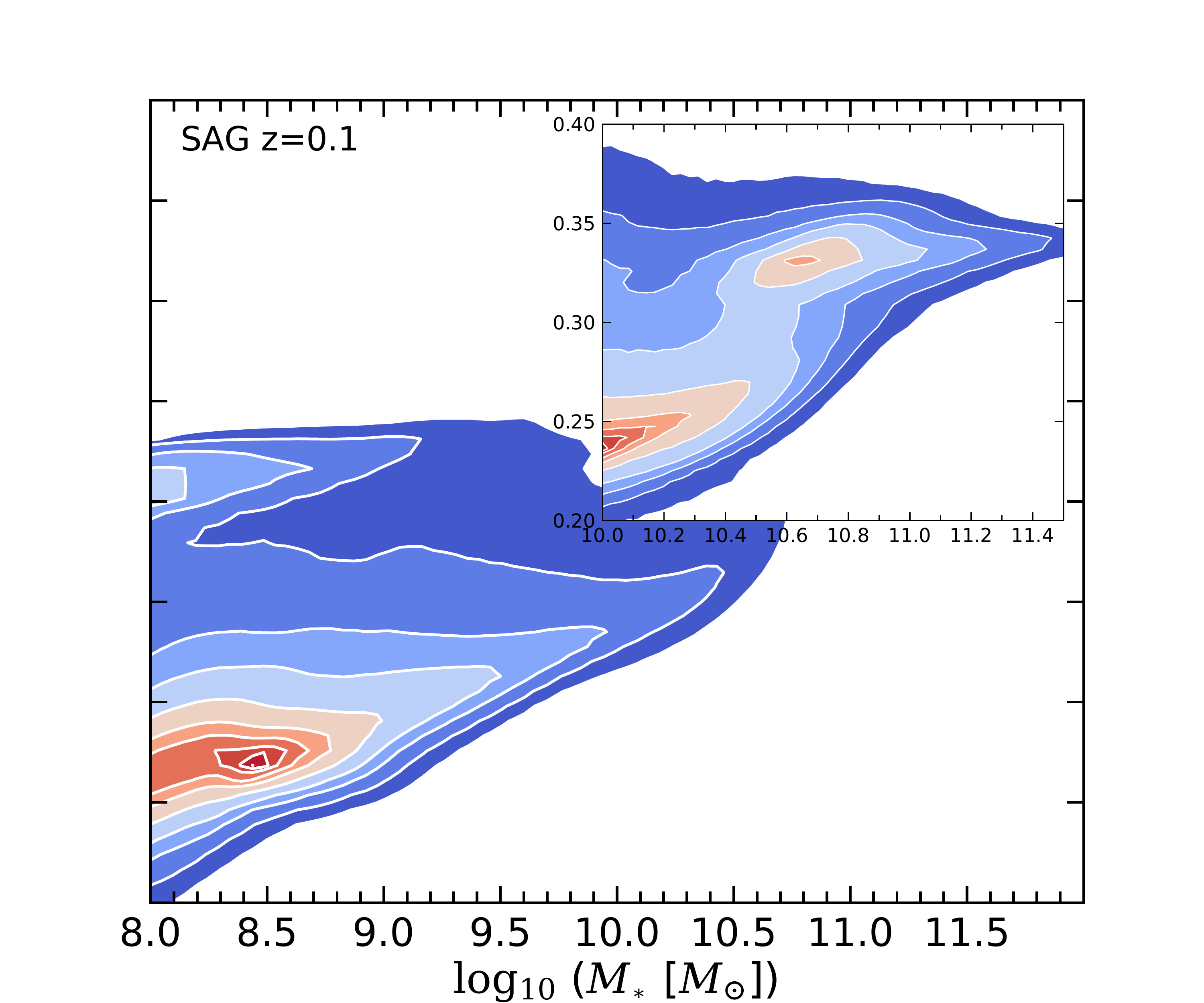}%
	 \hspace{-1.5cm}\includegraphics[width=6.9cm]{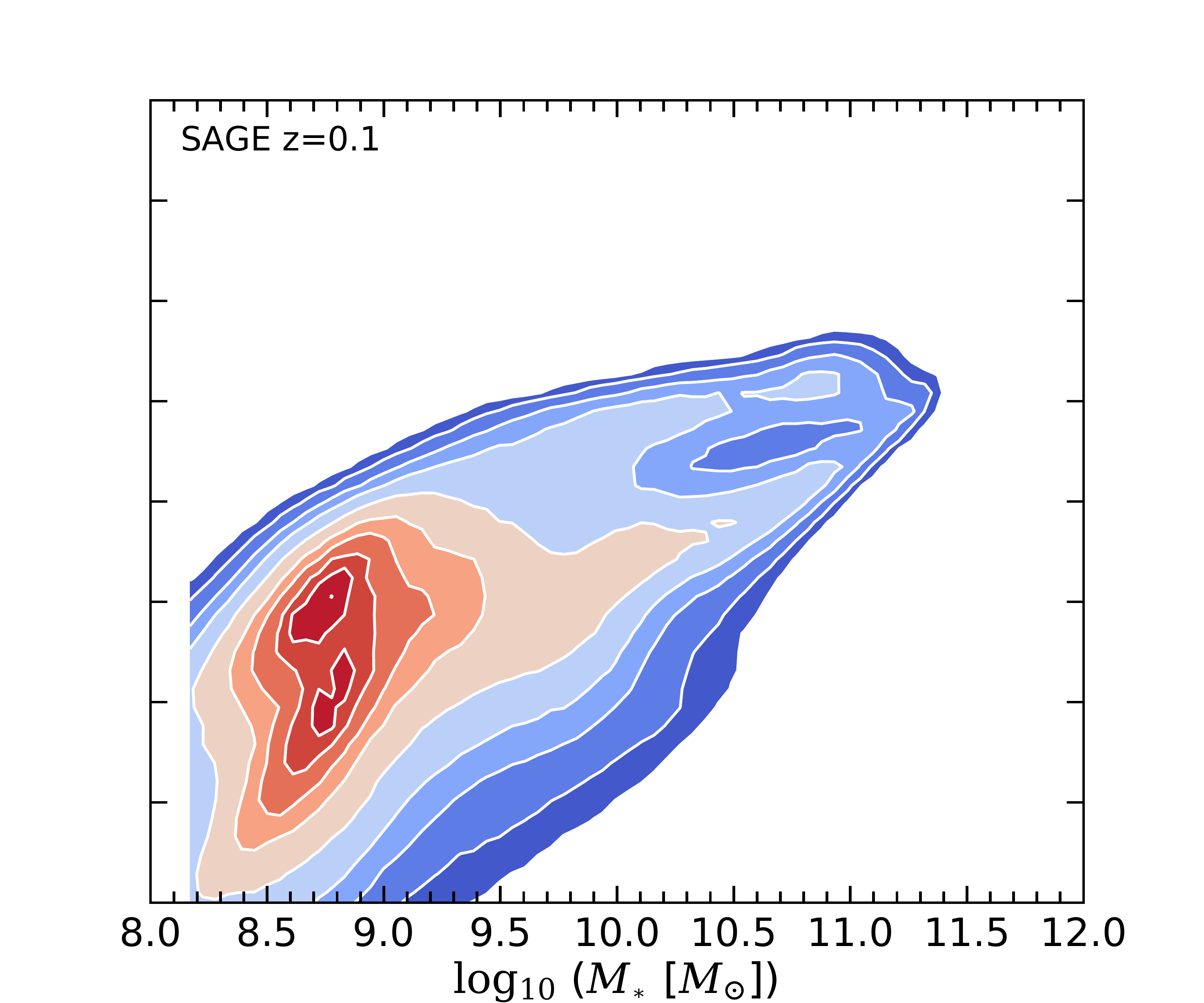}%
			
	\caption{Colour-magnitude or colour-stellar mass diagrams in the rest-frame, respectively at redshift $z=0.1$ for \galacticus (\textit{left column}), \sag (\textit{middle column}) \sage (\textit{right column}). \textit{Top}: SDSS $u-r$ band to $r$-band relation. The dashed red line corresponds to the commonly used separation of red and blue galaxies \citep{Strateva01}. \textit{Bottom}: SDSS $r-i$ band to stellar mass relation.}
	\label{fig:colours}
	\vspace{-0.3cm}
\end{figure*}

To gain more insight into this we present in \hyperref[fig:colours]{\Fig{fig:colours}} typical colour-magnitude and colour-stellar mass combinations at redshift $z=0.1$ for \galacticus (left column), \sag (middle column), and \sage\ (right column). In the \textit{top panel} the SDSS rest-frame $u-r$ to the $r$-band relation is shown. The red dashed line corresponds to the commonly used separation of red and blue galaxies \citep{Strateva01}. In the \textit{bottom panel} we present the SDSS rest-frame $r-i$ to stellar mass relation.

\galacticus also shows a clear separation between a red and blue population as indicated by the reference line (dashed red) in the top panel as well as a reasonable colour-to-stellar mass relation. For \sag the top panel confirms what we already showed in \hyperref[fig:SFRdensity]{\Fig{fig:SFRdensity}}; \sag shows a higher star formation rate than \galacticus for redshifts $z<1$, and hence the majority of galaxies is blue. However, this does not necessarily translate into a negligible red fraction: \sag also provides a reasonable red population, but due to the amount of lower mass blue galaxies (cf. bottom panel), its redder population is not resolved very well in these contour plots. We therefore include an inset panel for \sag when showing $r-i$ vs. stellar mass: in this representation -- for which the same contour levels have been used, but the number density of galaxies is different due to applying a cut in stellar mass on the $x$-axis -- the red population is clearly visible, too.

%%%%%%%%%%%%%%%%%%%%%%%%%%%%%%%%%%%%%%%%%%%%%%%%%%%
\section{Galaxy Clustering}\label{sec:clustering}
%%%%%%%%%%%%%%%%%%%%%%%%%%%%%%%%%%%%%%%%%%%%%%%%%%%

\begin{figure*}
	  \hspace{-0.0cm}\includegraphics[width=6.9cm]{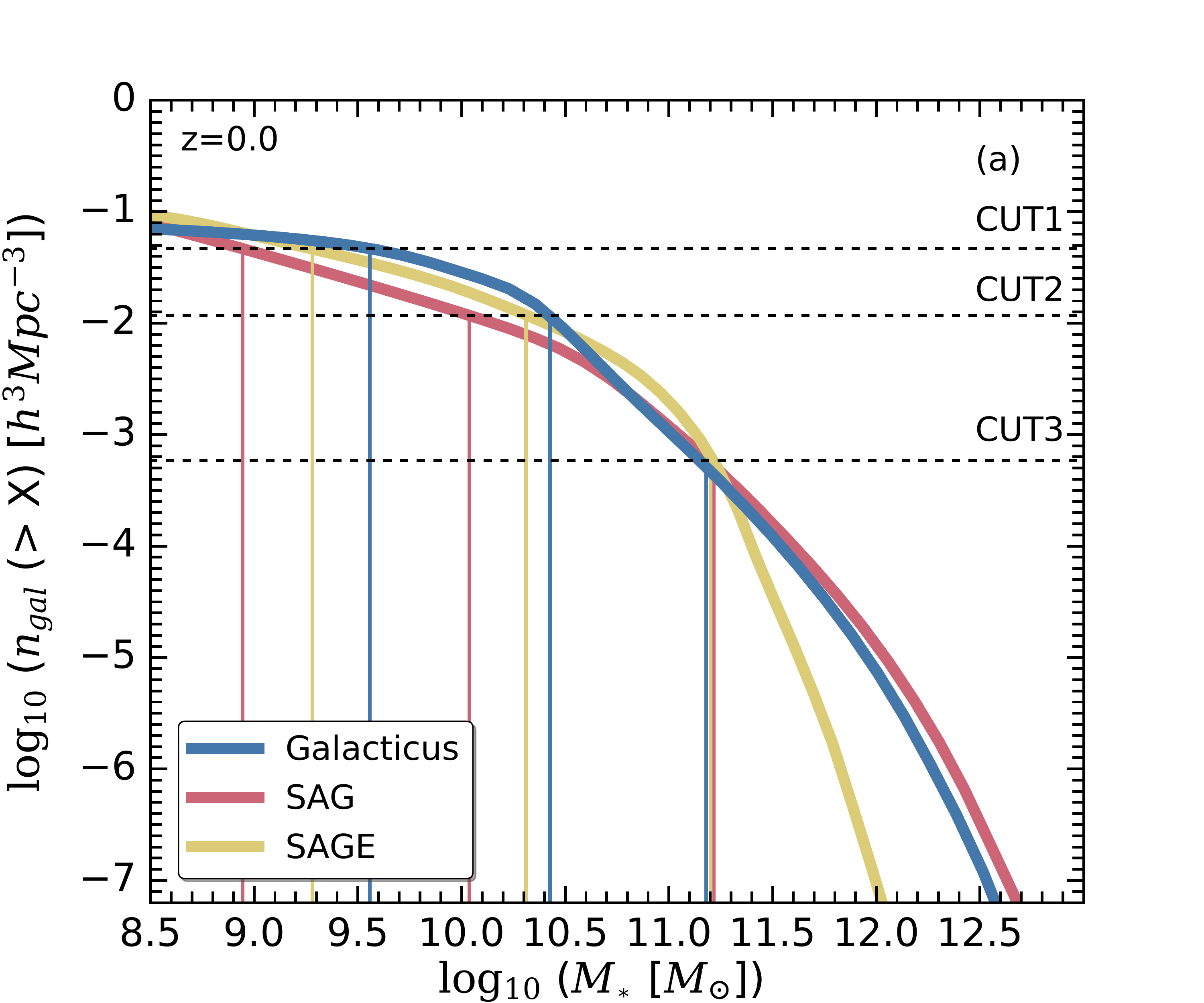}%
	  \hspace{-1.5cm}\includegraphics[width=6.9cm]{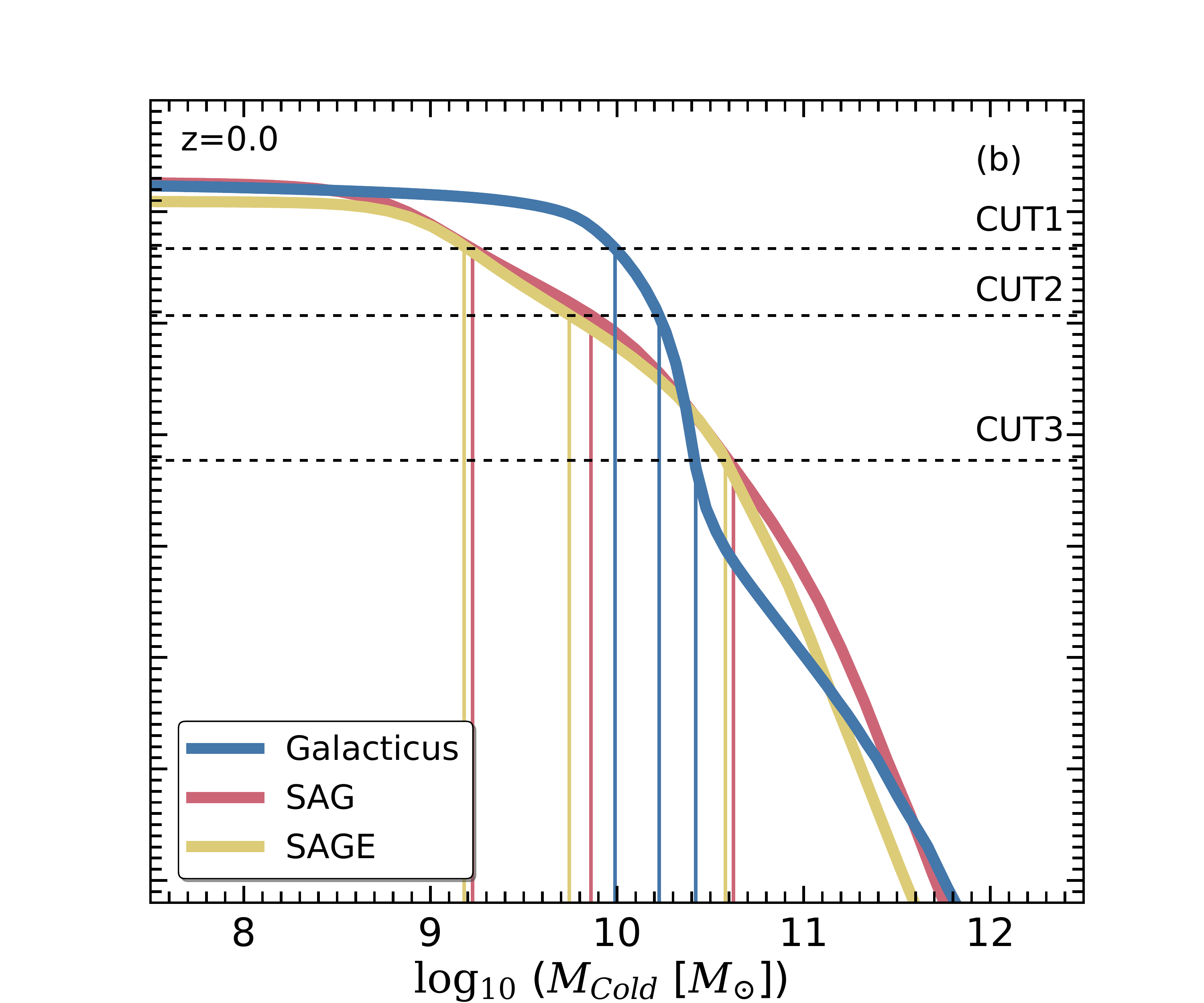}%
	  \hspace{-1.5cm}\includegraphics[width=6.9cm]{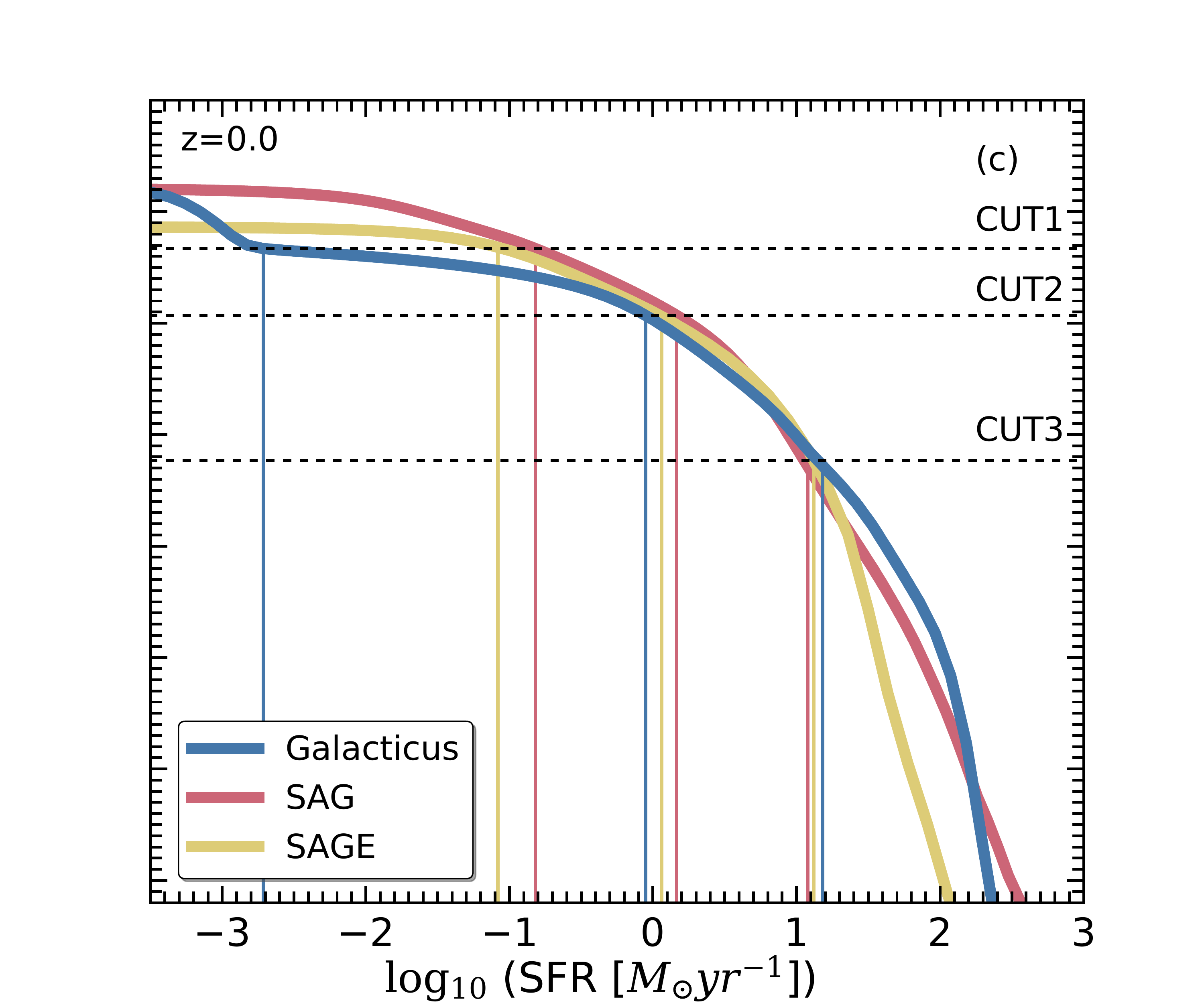}%
	  \caption{The cumulative abundance of SAM galaxies ranked by (a) \Mstar, (b) \Mcold and (c) \SFR. The vertical lines indicate how the applied number density cut translates into a lower limit for the respective galaxy property (see \Tab{tab:ncut} for the actual values).}
	\label{fig:CF}
	\vspace{-0.3cm}
\end{figure*} 

\begin{table}
	\caption{Number density cuts for the selection of galaxy samples for the 2PCF calculation. The column labelled $X$ gives the translation of the number density cut into the corresponding cut in the respective galaxy property (i.e. $X$ can be \Mstar, \Mcold, or \SFR). }
	\label{tab:ncut}\vspace{-0.2cm}
	\begin{center} 
		\begin{tabular}{cccccc}
		\hline
		\textsc{CUT}				& $n_{\rm CUT}$ 					&  $X$	& \multicolumn{3}{c}{$\log_{10}(X_{\rm cut})$}\\
								& [(\hMpc)$^3$]					&  		& \galacticus\ 	& \sag 	& \sage\\
		\hline
		\hline
		\multirow{3}{*}{\textsc{\#1}} 	&\multirow{3}{*}{46.75$\times10^{-3}$}	& \Mstar	 &9.56		&8.94	&9.28\\
		                                  		& 								& \Mcold	 &9.99		&9.23	&9.18\\
		                                  		& 								& \SFR	 &-2.71		&-0.82	&-1.08\\
		\hline
		\multirow{3}{*}{\textsc{\#2}} 	&\multirow{3}{*}{11.77$\times10^{-3}$}	& \Mstar	 &10.43		&10.04	&10.31\\
		                                  		& 								& \Mcold	 &10.23		&9.86	&9.74\\
		                                  		& 								& \SFR	&-0.05		&0.16	&0.06\\
		\hline
		\multirow{3}{*}{\textsc{\#3}} 	&\multirow{3}{*}{0.53$\times10^{-3}$}	& \Mstar	 &11.24		&11.22	&11.20\\
		                                  		& 								& \Mcold	 &10.42		&10.62	&10.58\\
		                                  		& 								& \SFR	 &1.18		&1.08	&1.12\\
		\hline
		\vspace{-0.6cm}
		\end{tabular}
	\end{center}
\end{table}

\begin{figure*}
	\begin{subfigure}{\linewidth}
	  \includegraphics[width=7.0cm]{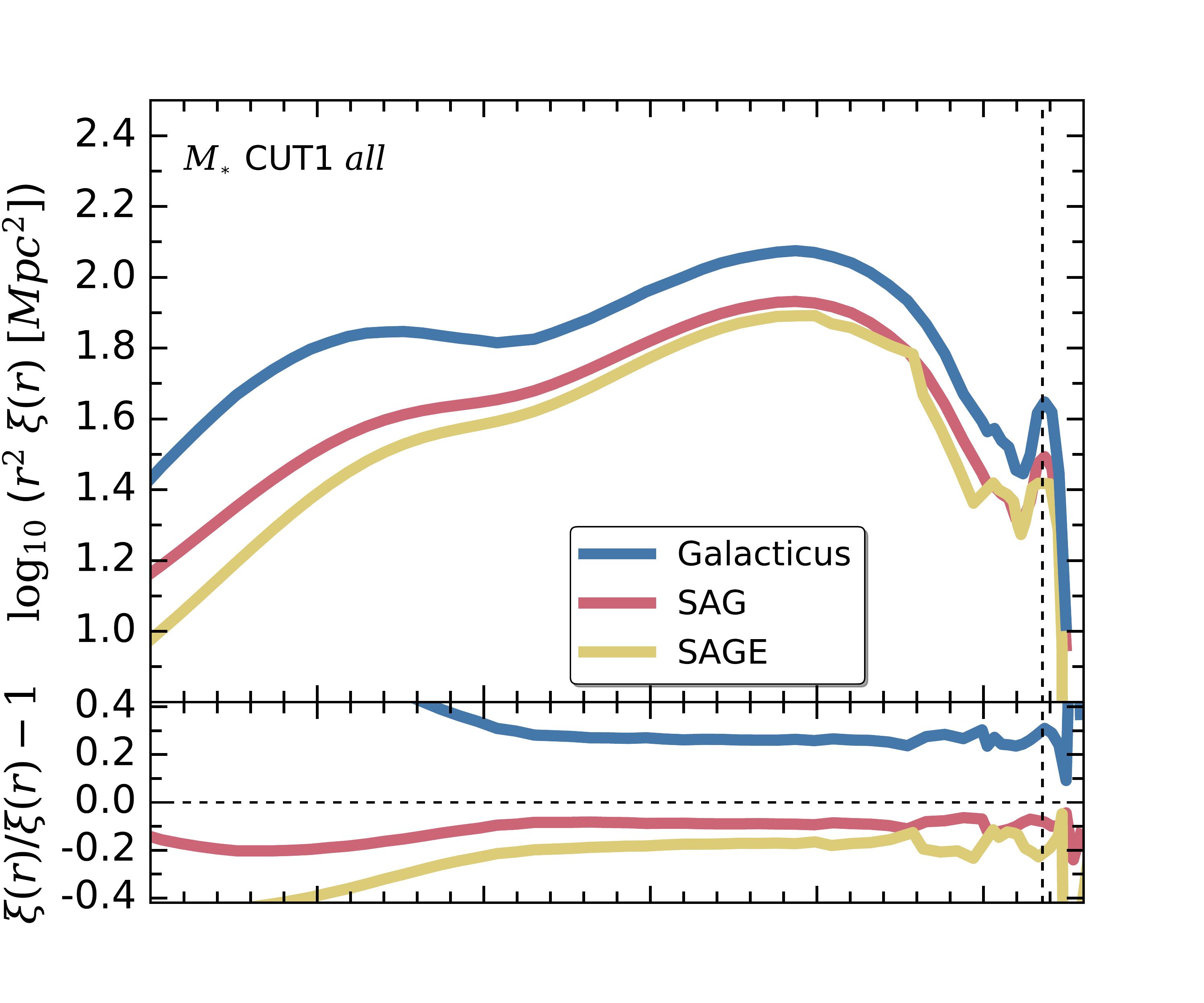}\hspace{-1.0cm}%
	  \includegraphics[width=7.0cm]{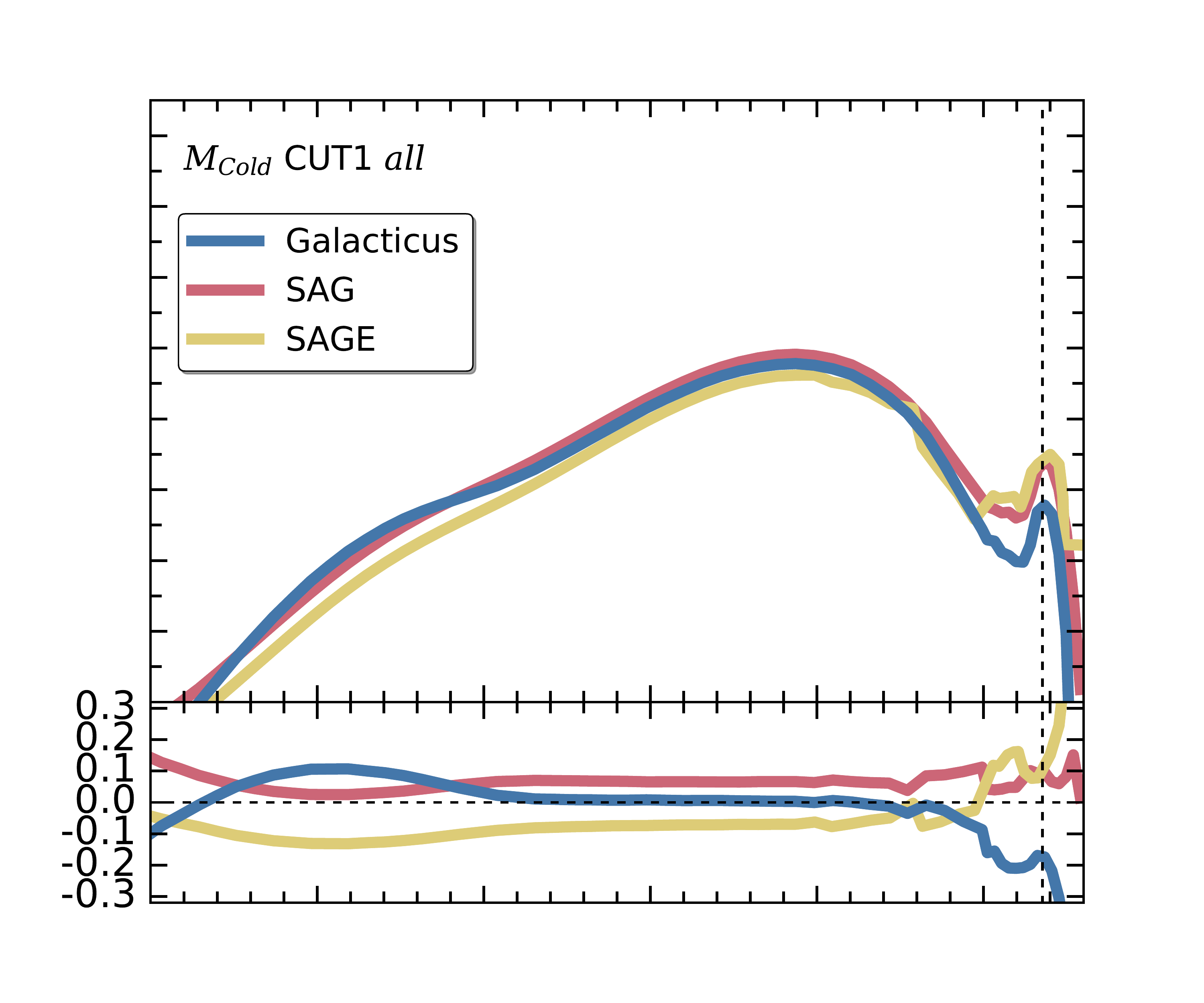}\hspace{-1.0cm}%
	  \includegraphics[width=7.0cm]{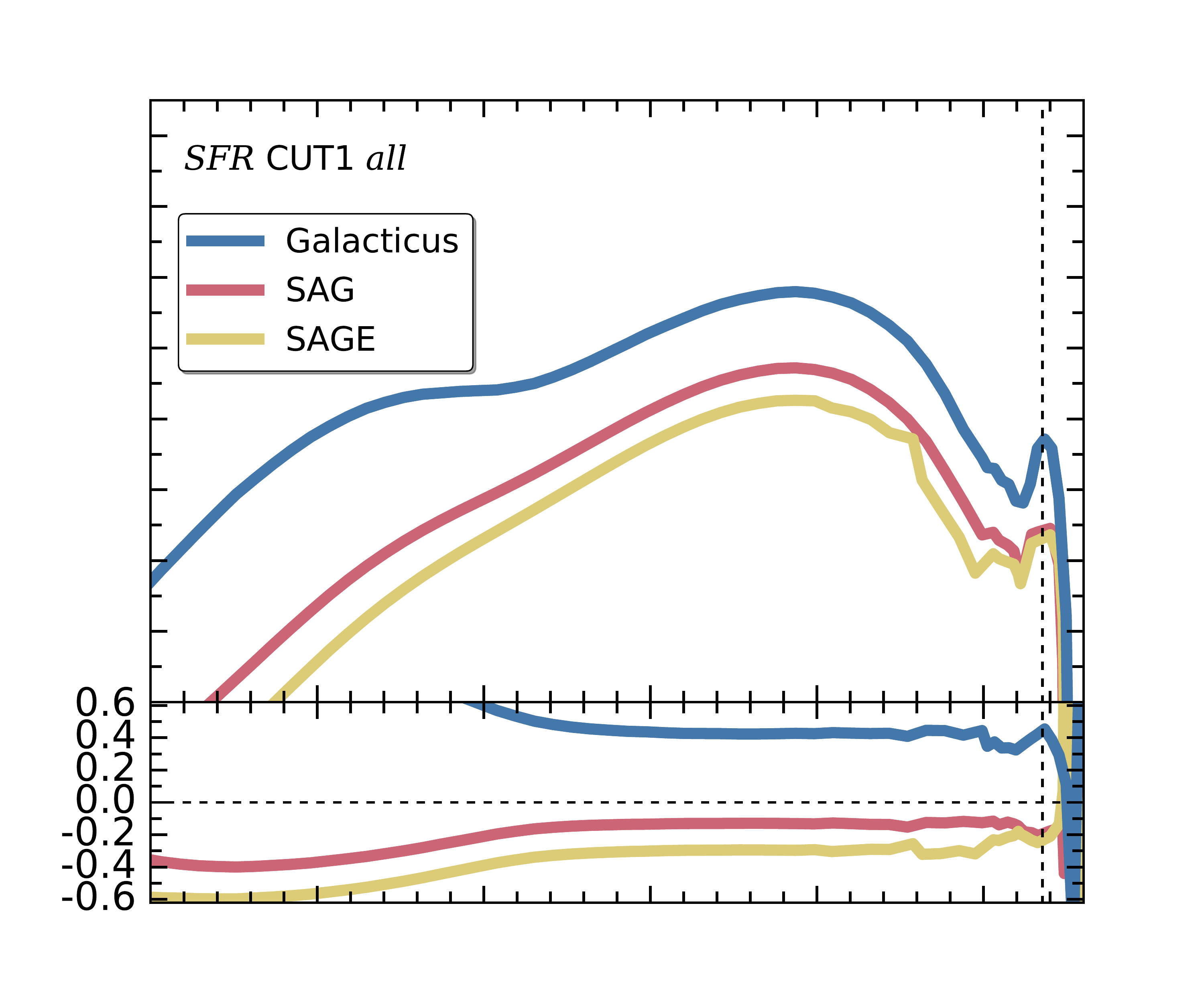}\hspace{-1.0cm}\vspace{-1.4cm}%
	\end{subfigure}
	\begin{subfigure}{\linewidth}
	  \includegraphics[width=7.0cm]{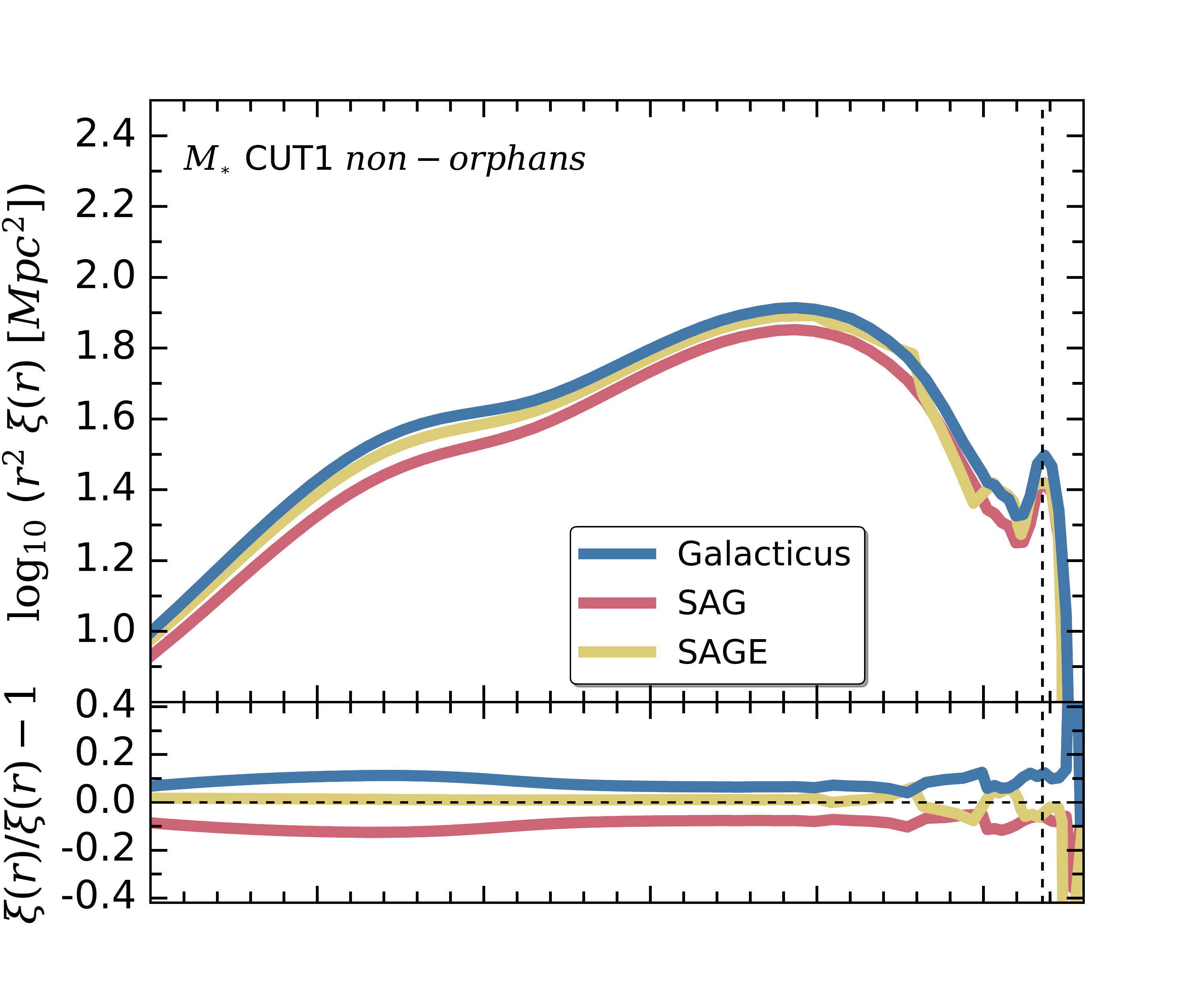}\hspace{-1.0cm}%
	  \includegraphics[width=7.0cm]{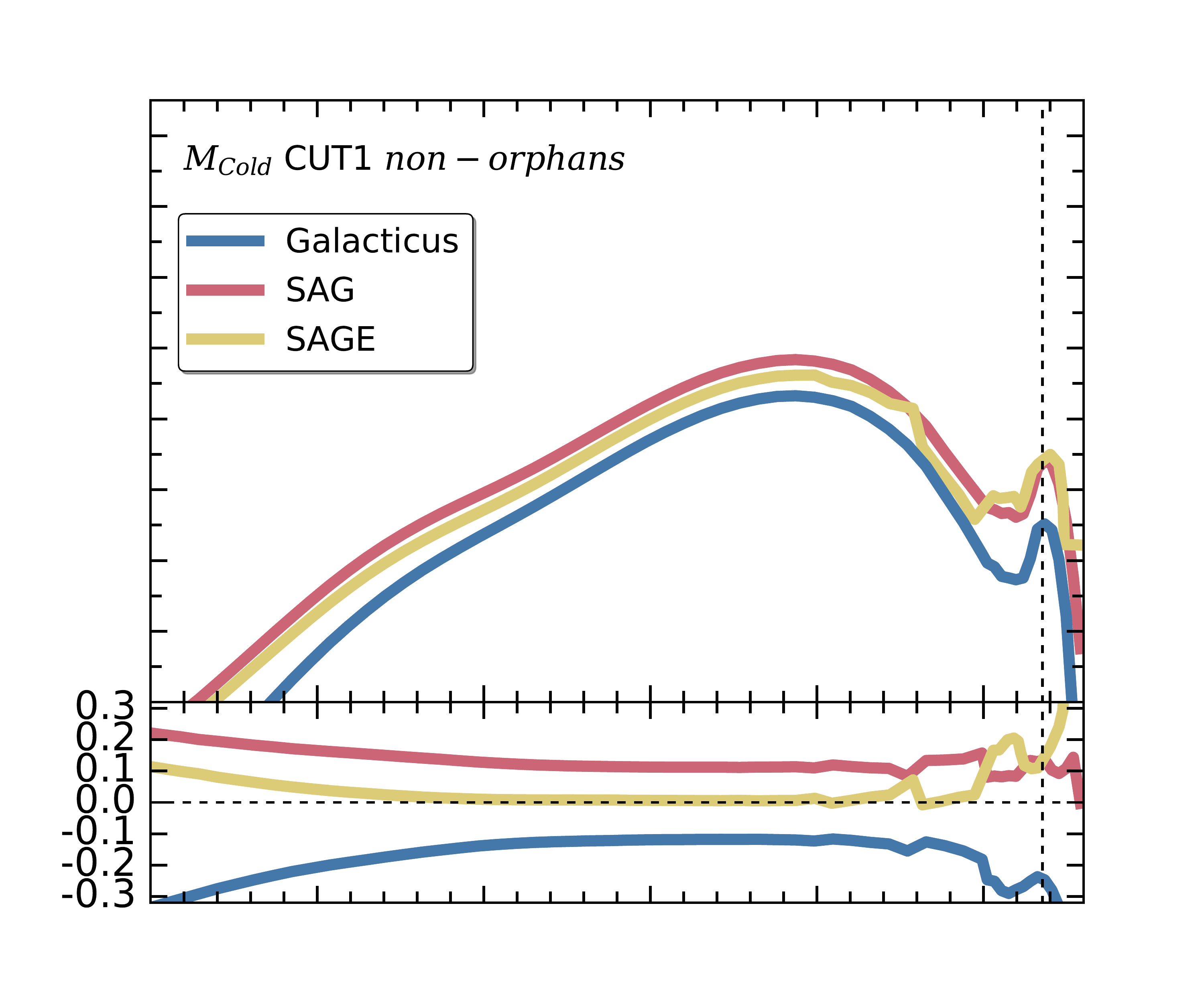}\hspace{-1.0cm}%
	  \includegraphics[width=7.0cm]{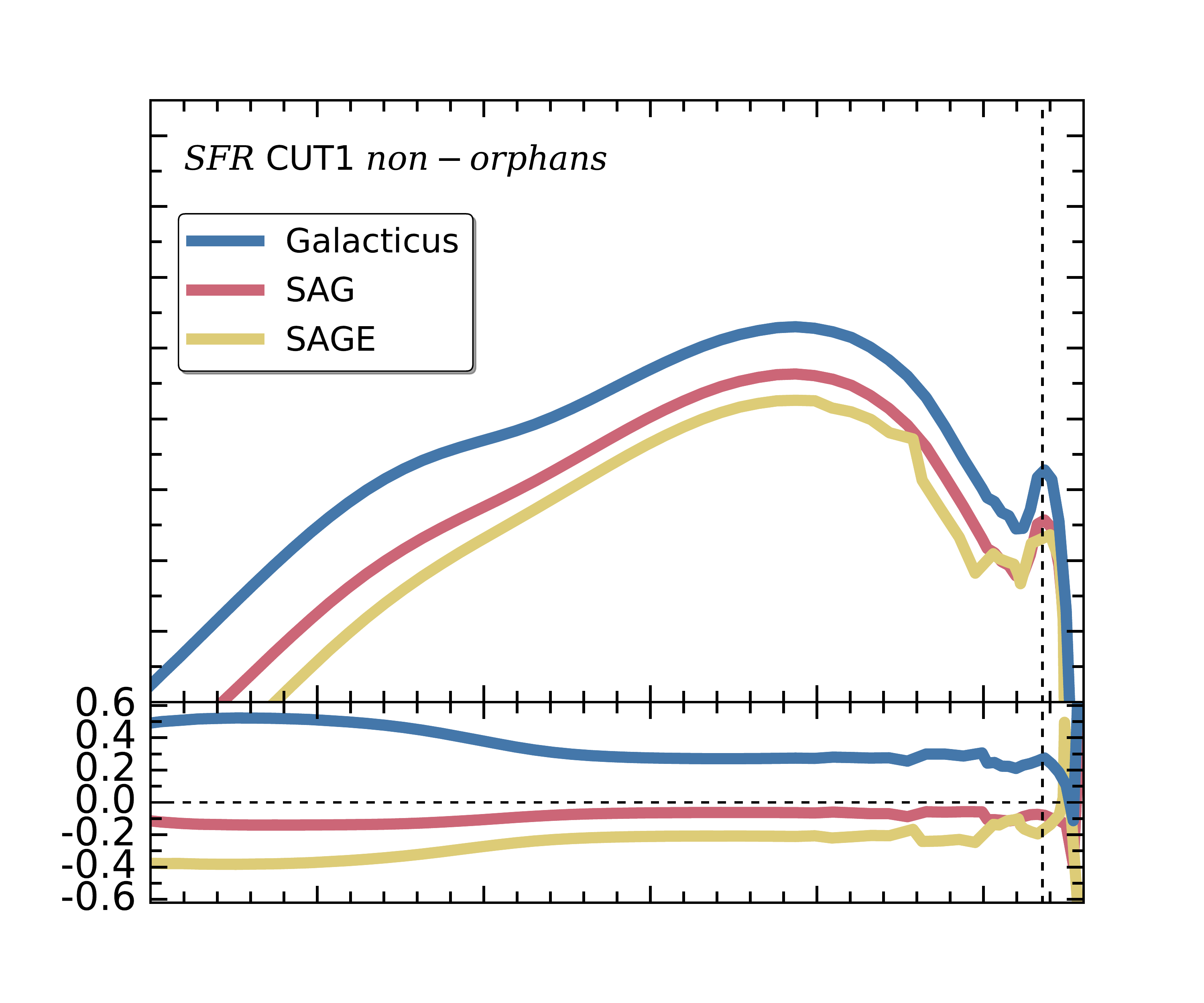}\hspace{-1.0cm}\vspace{-1.4cm}%
	\end{subfigure}
	\begin{subfigure}{\linewidth}
	  \includegraphics[width=7.0cm]{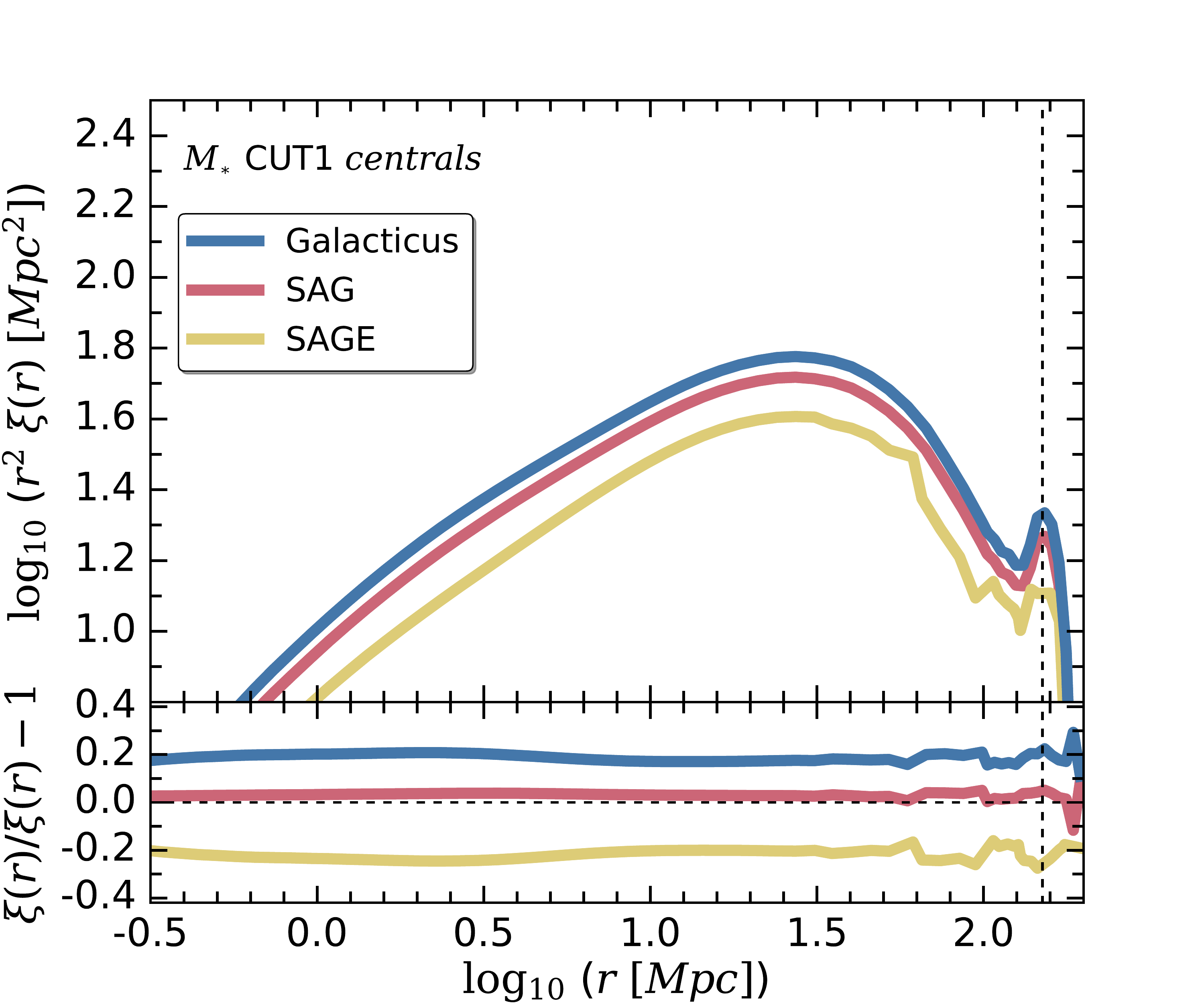}\hspace{-1.0cm}%
	  \includegraphics[width=7.0cm]{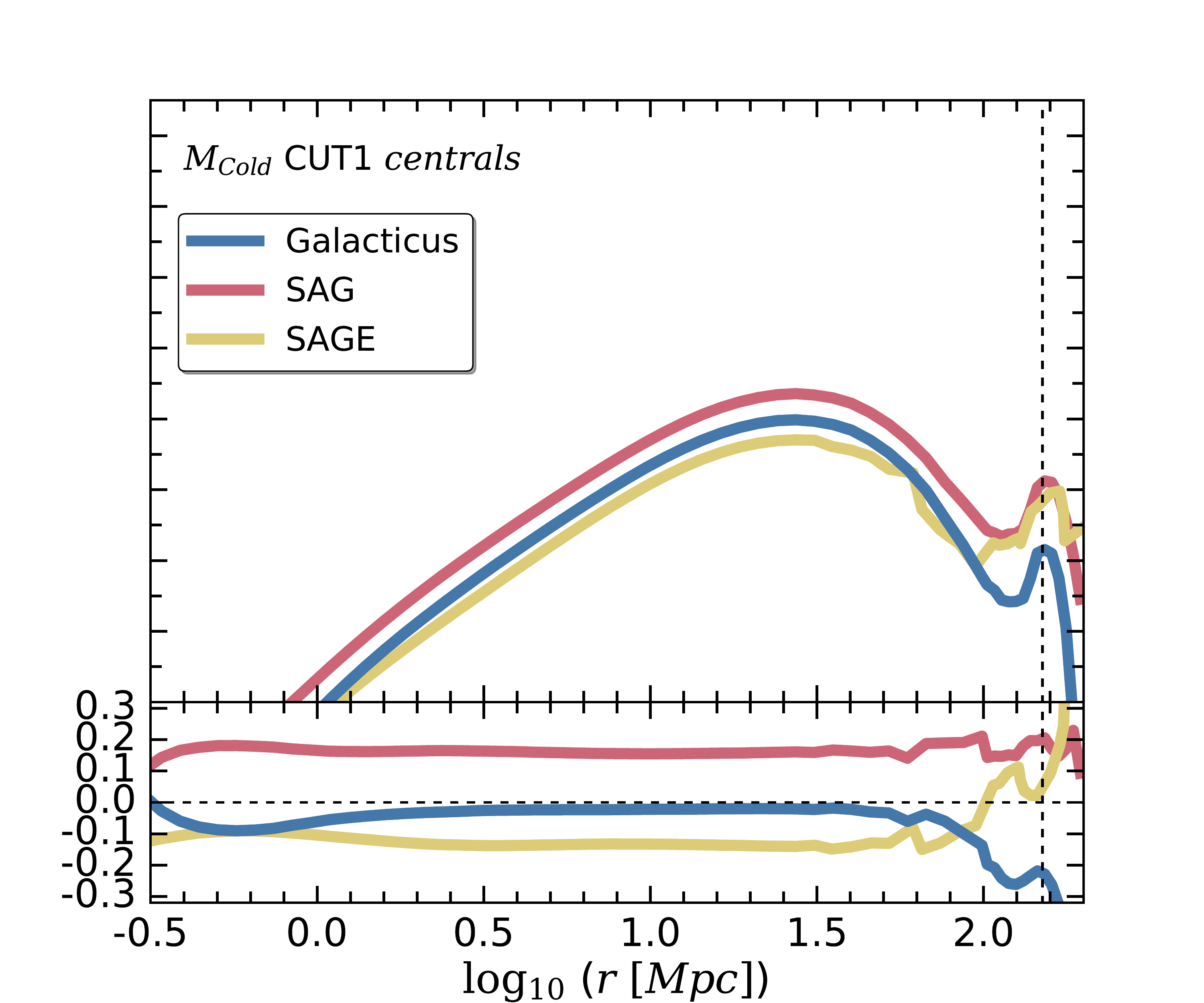}\hspace{-1.0cm}%
	  \includegraphics[width=7.0cm]{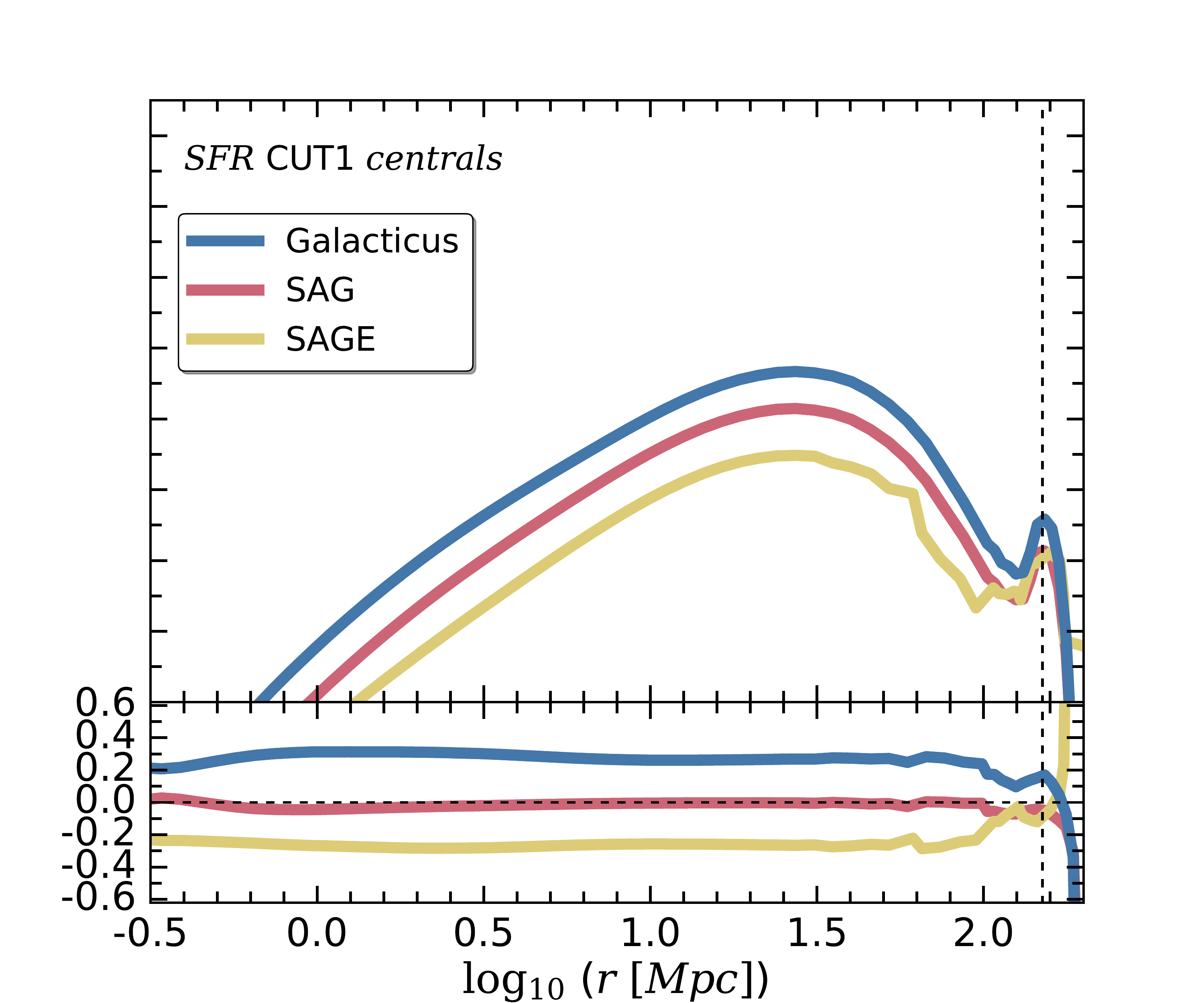}\hspace{-1.0cm}\vspace{-0.4cm}%
	\end{subfigure}
	\caption{The real-space two-point correlation function at redshift $z=0.0$ for the density $CUT1$ ranked by the abundances of the following galaxy properties form left to right: \Mstar (left column), \Mcold (middle column) and \SFR (right column) and from top to bottom: `all', `non-orphan' and `central' galaxies. The lower panel in each sub-plot shows the fractional difference with respects to the mean correlation function $\bar{\xi}(r)$. The vertical line indicates the position of the Baryonic Acoustic Oscillations peak. As \galacticus\ does not integrate the orbits of orphans, the positions of them correspond to the position of the central galaxy they orbit for that model. \sage\ does not feature orphans at all and hence the `all' and `non-orphan' curves are the same.}
	\label{fig:2PCF_CUT1}
	\vspace{-0.3cm}
\end{figure*}

\begin{figure*}
	\begin{subfigure}{\linewidth}
	  \includegraphics[width=7.0cm]{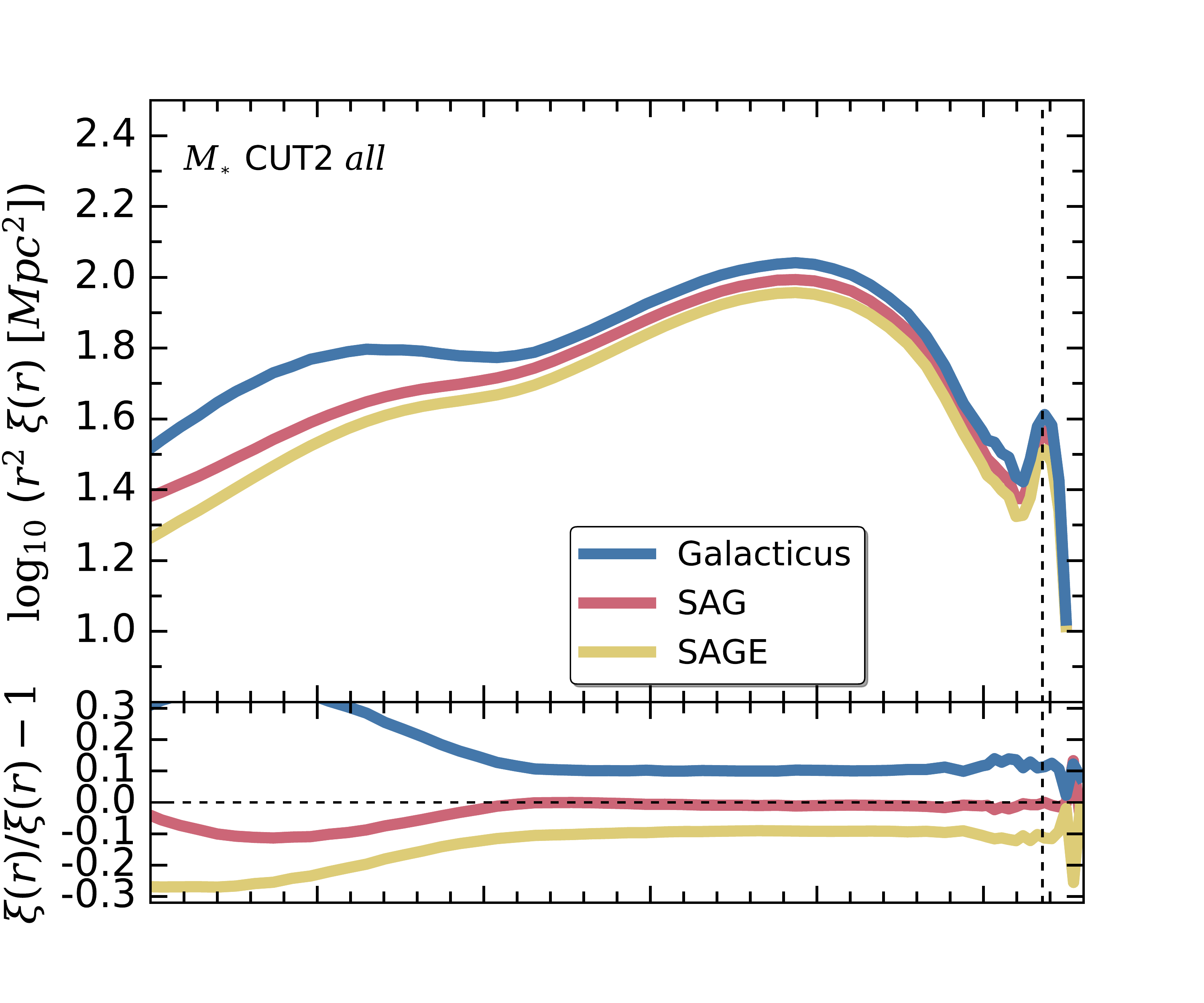}\hspace{-1.0cm}%
	  \includegraphics[width=7.0cm]{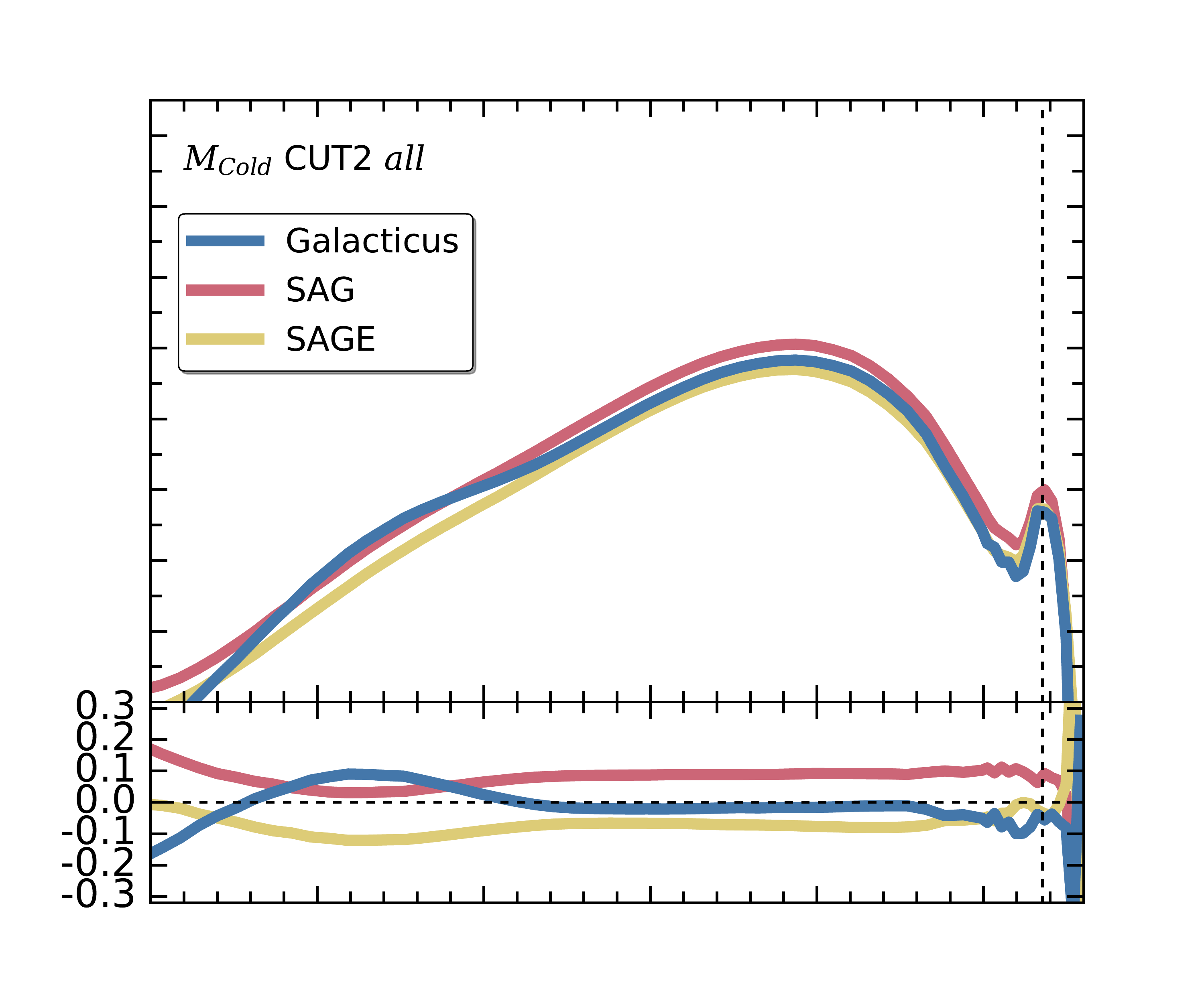}\hspace{-1.0cm}%
	  \includegraphics[width=7.0cm]{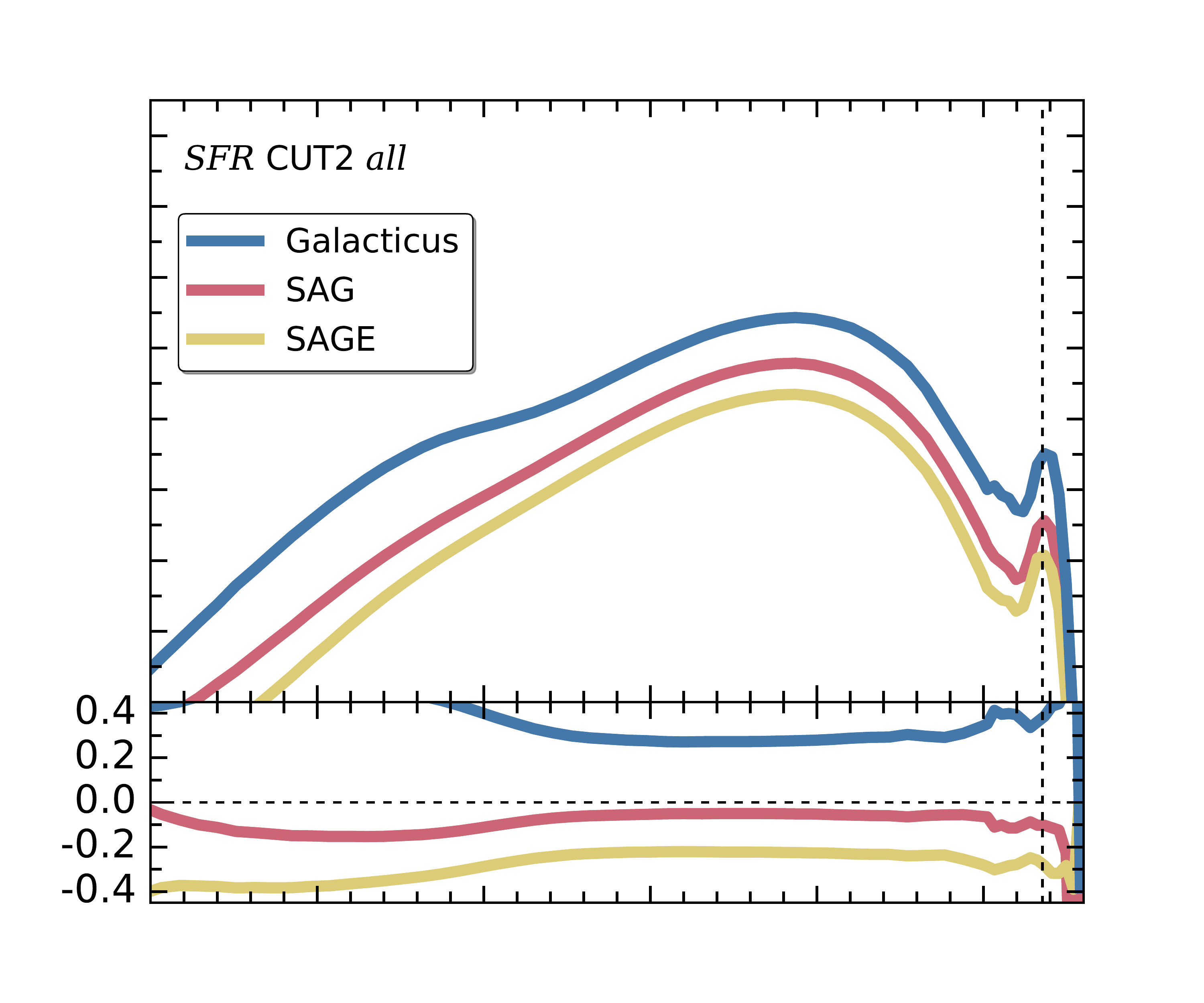}\hspace{-1.0cm}\vspace{-1.4cm}%
	\end{subfigure}
	\begin{subfigure}{\linewidth}
	  \includegraphics[width=7.0cm]{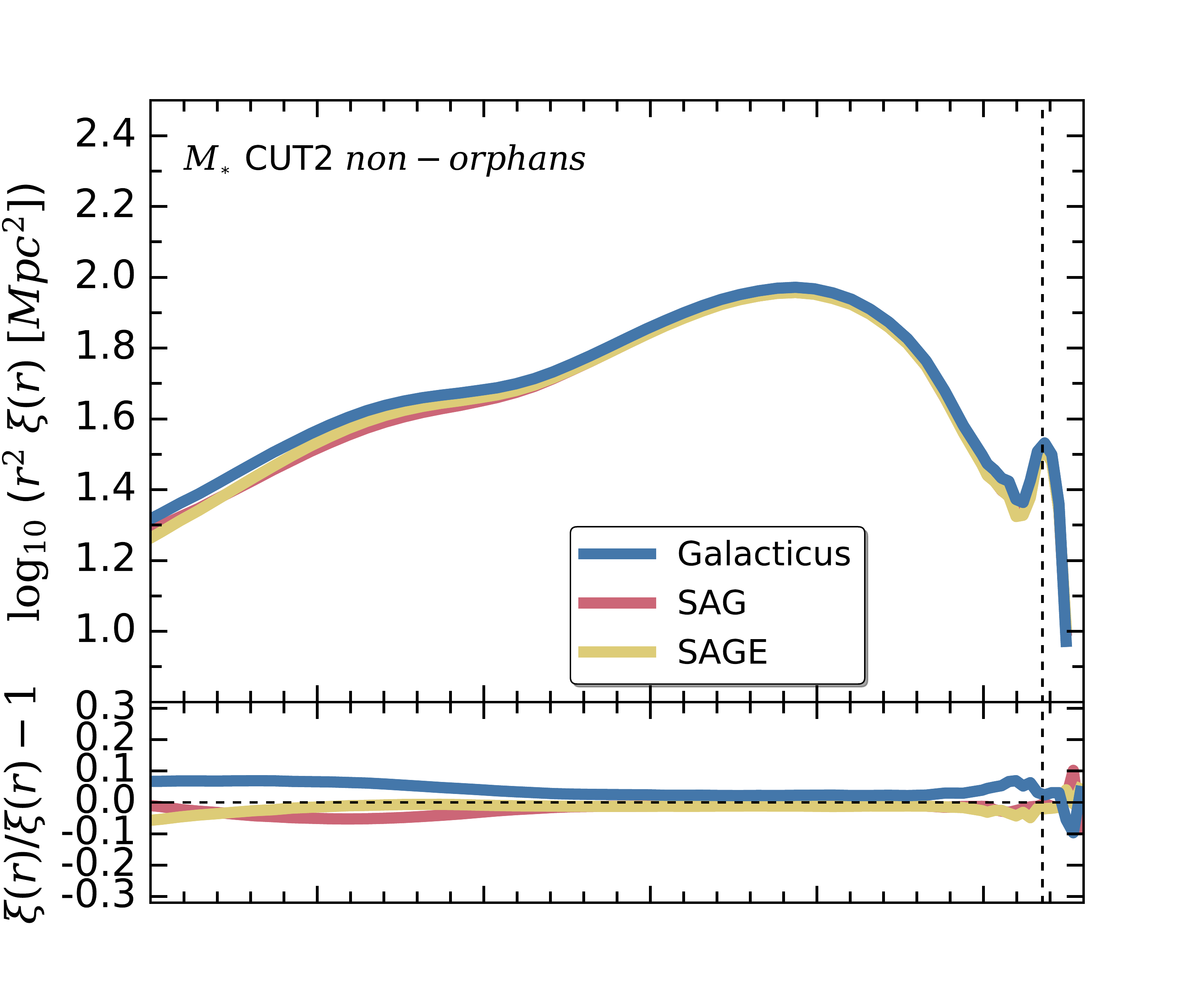}\hspace{-1.0cm}%
	  \includegraphics[width=7.0cm]{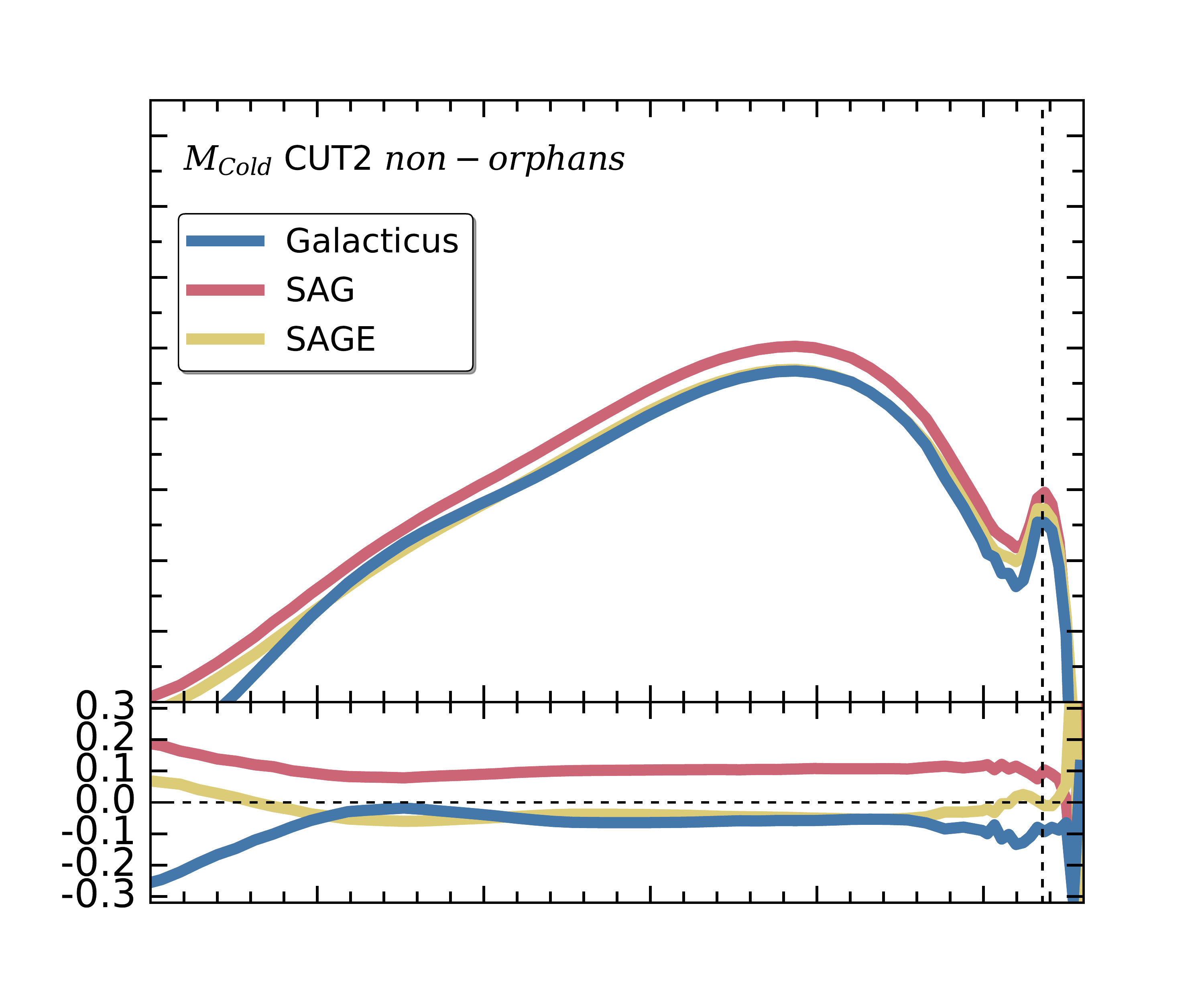}\hspace{-1.0cm}%
	  \includegraphics[width=7.0cm]{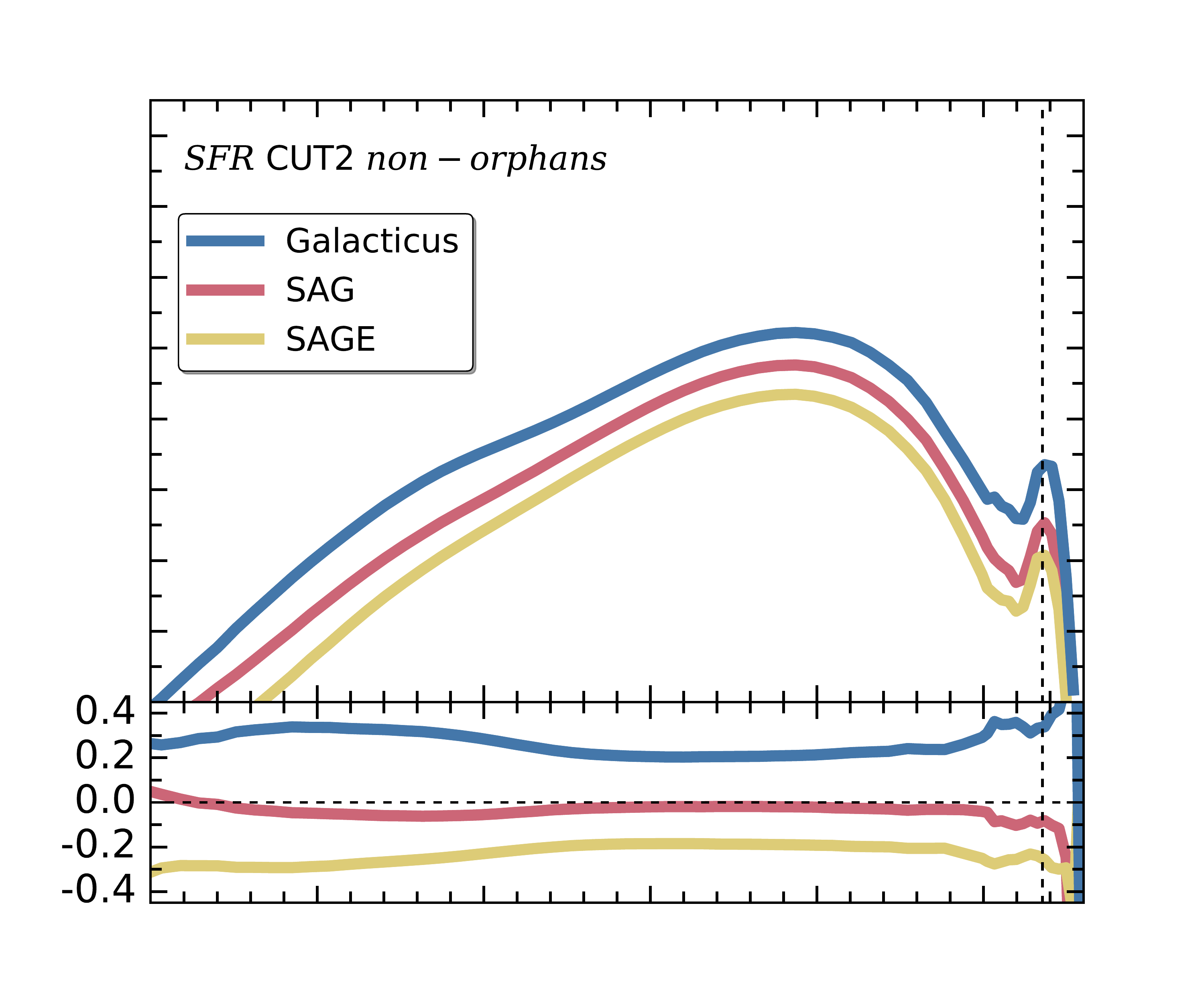}\hspace{-1.0cm}\vspace{-1.4cm}%
	\end{subfigure}
	\begin{subfigure}{\linewidth}
	  \includegraphics[width=7.0cm]{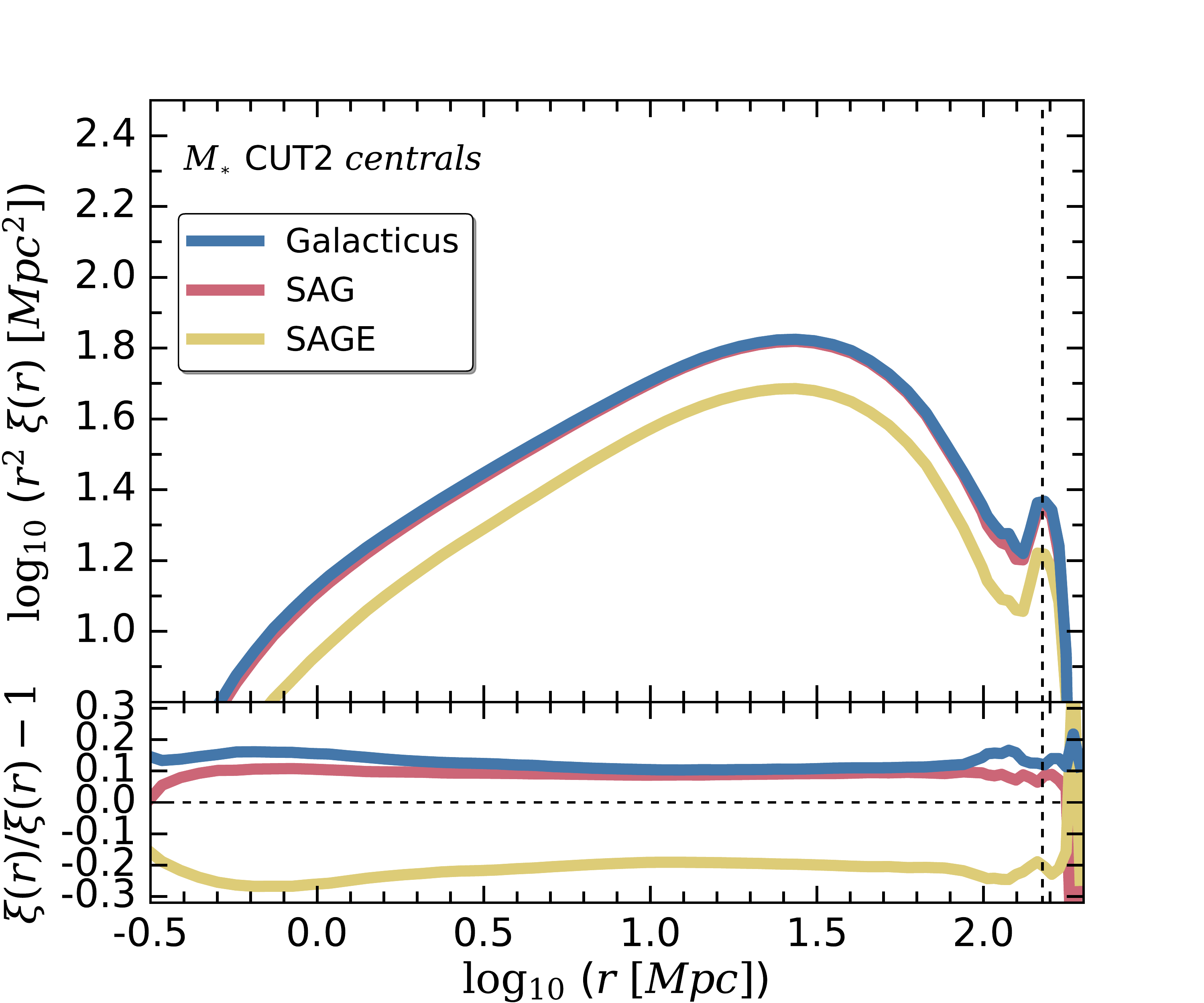}\hspace{-1.0cm}%
	  \includegraphics[width=7.0cm]{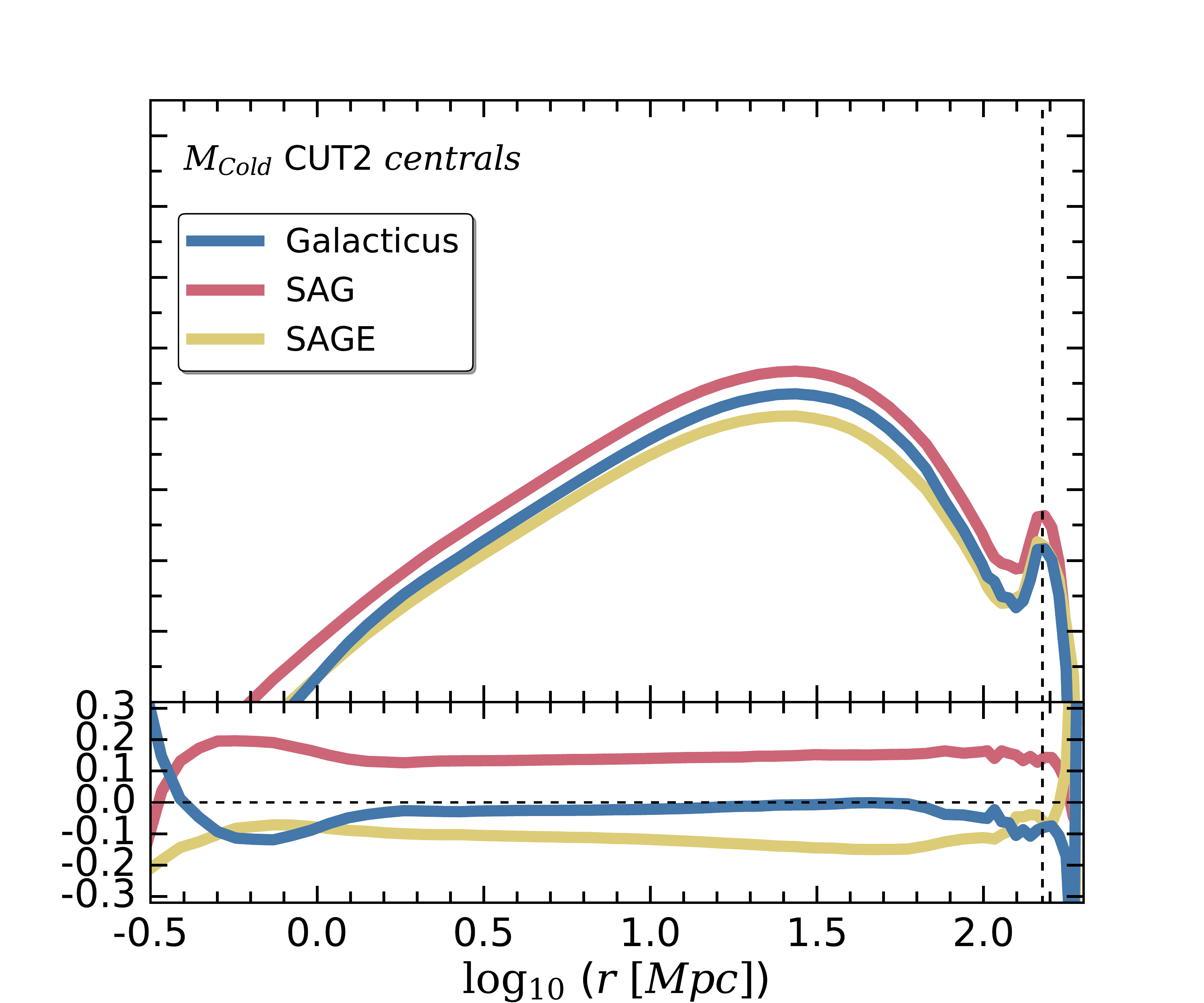}\hspace{-1.0cm}%
	  \includegraphics[width=7.0cm]{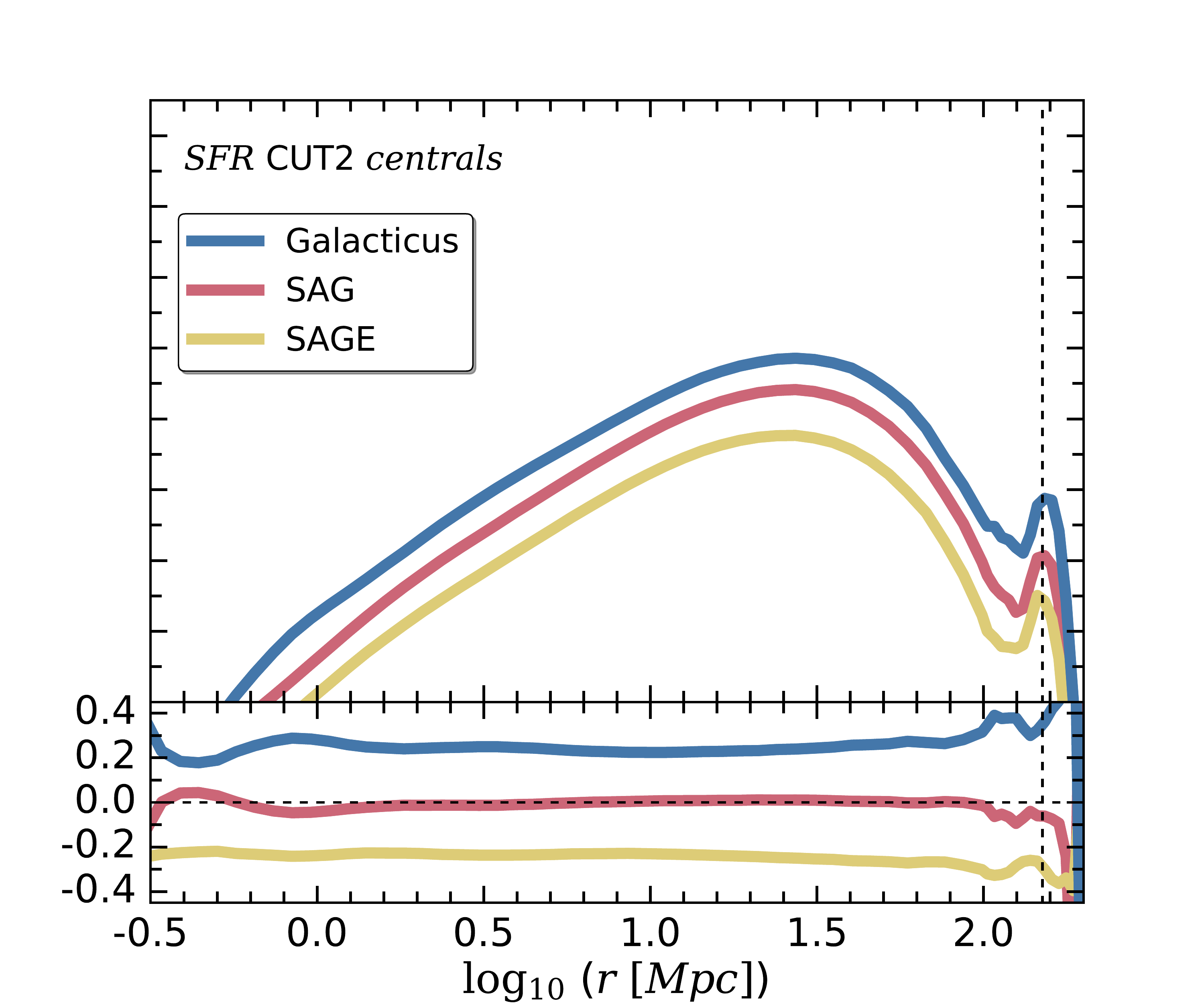}\hspace{-1.0cm}\vspace{-0.4cm}%
	\end{subfigure}
	\caption{Same as \hyperref[fig:2PCF_CUT1]{\Fig{fig:2PCF_CUT1}}, but for CUT2.}
	\label{fig:2PCF_CUT2}
	\vspace{-0.3cm}
\end{figure*}

\begin{figure*}
	\begin{subfigure}{\linewidth}
	  \includegraphics[width=7.0cm]{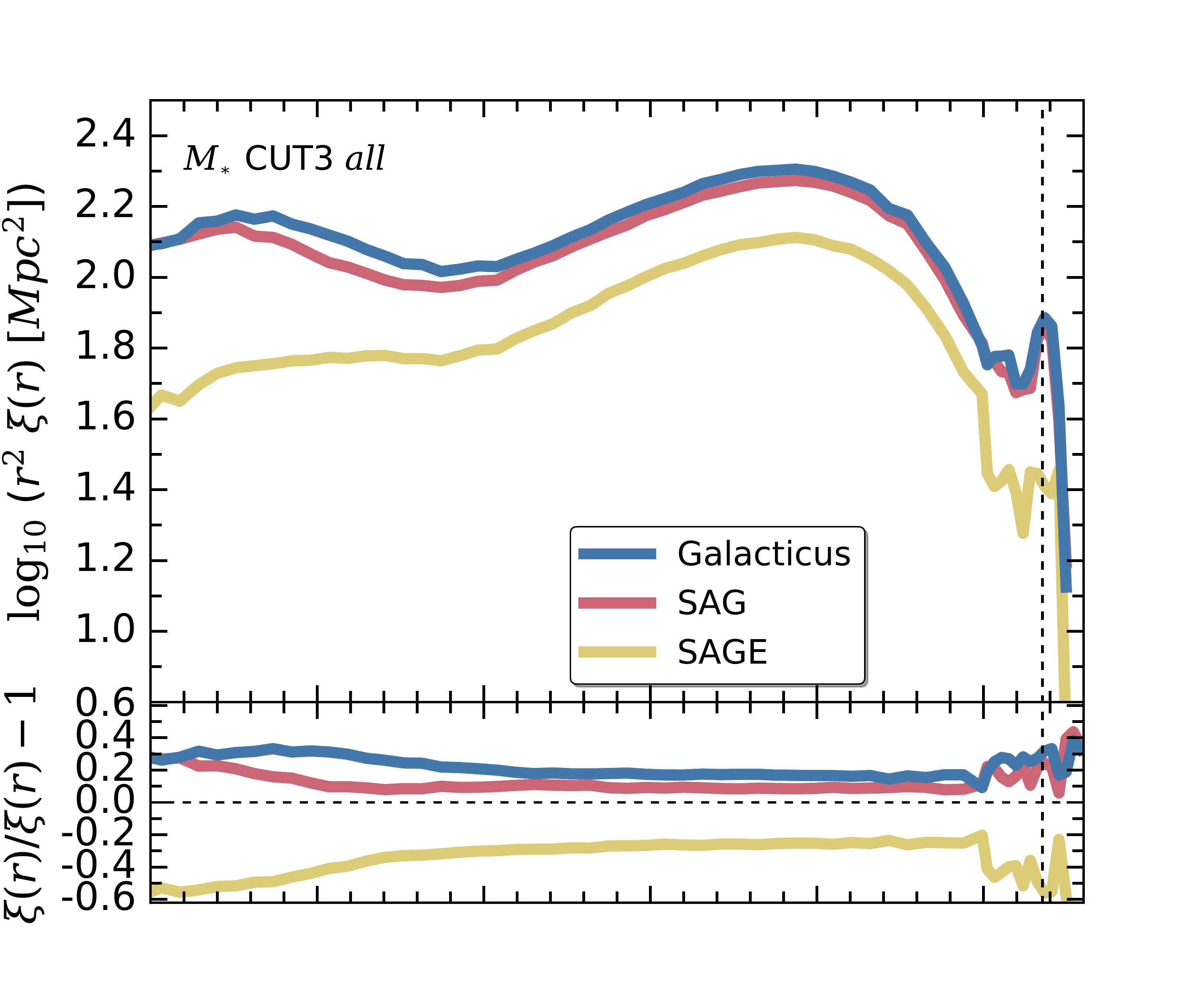}\hspace{-1.0cm}%
	  \includegraphics[width=7.0cm]{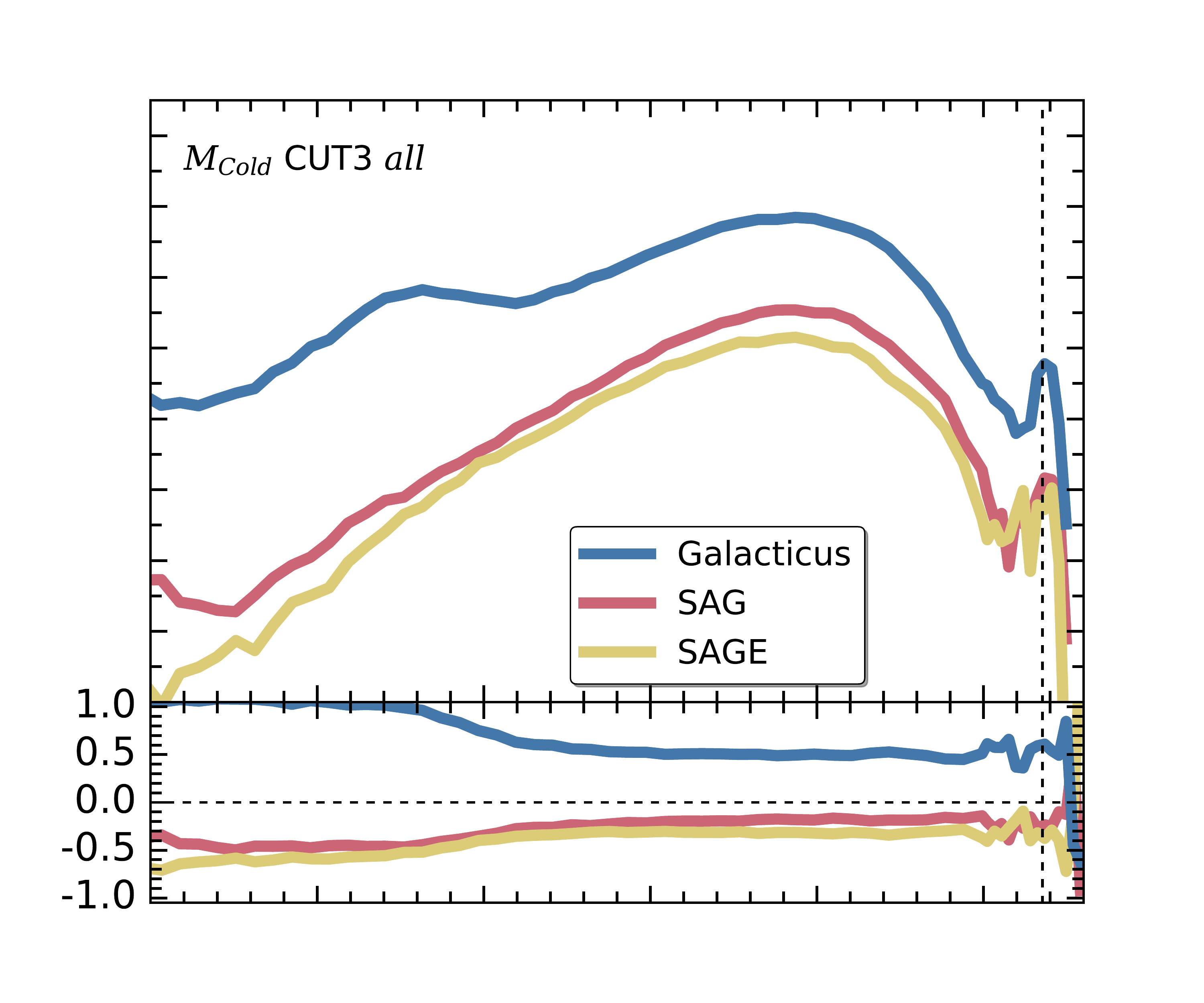}\hspace{-1.0cm}%
	  \includegraphics[width=7.0cm]{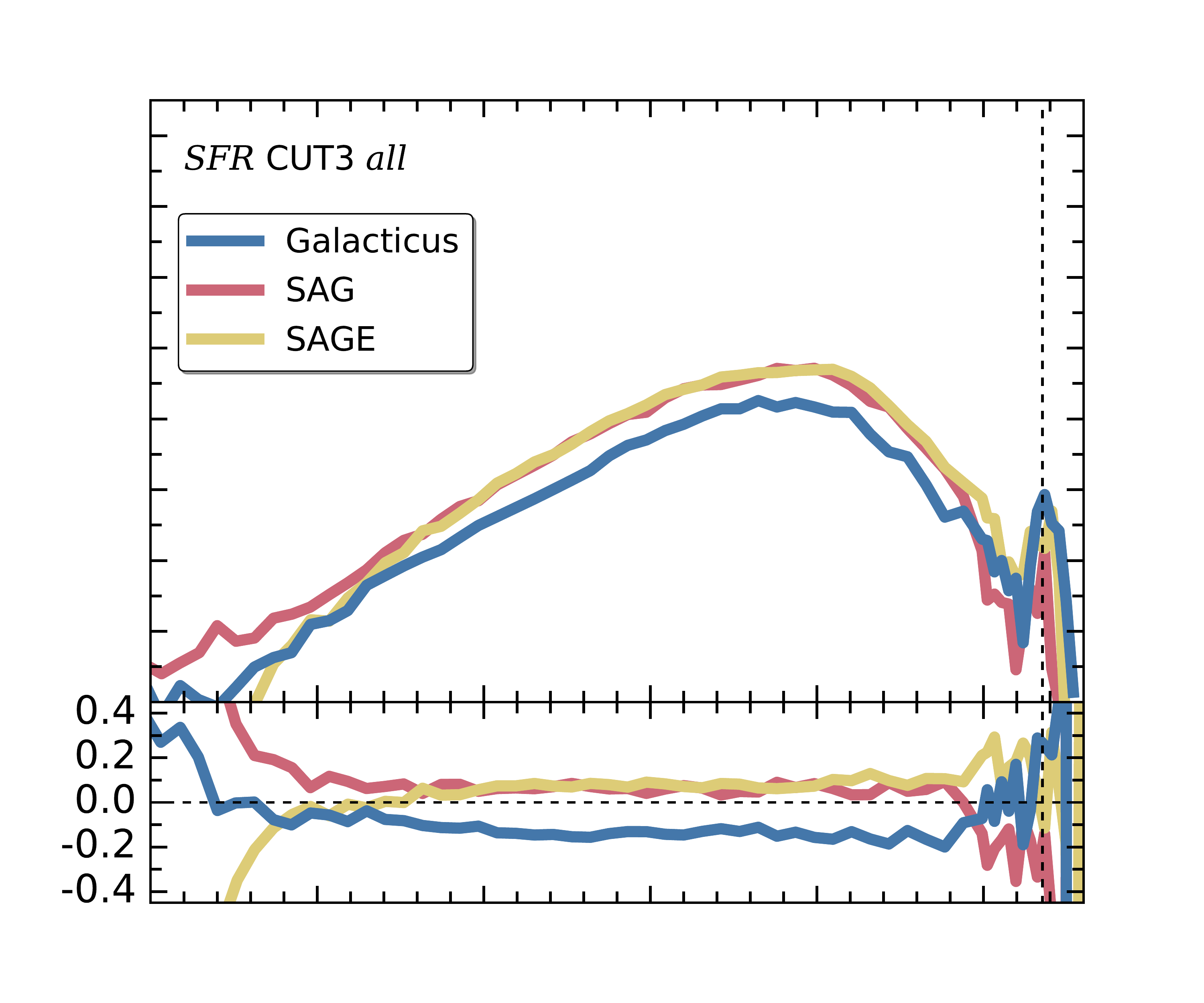}\hspace{-1.0cm}\vspace{-1.4cm}%
	\end{subfigure}
	\begin{subfigure}{\linewidth}
	  \includegraphics[width=7.0cm]{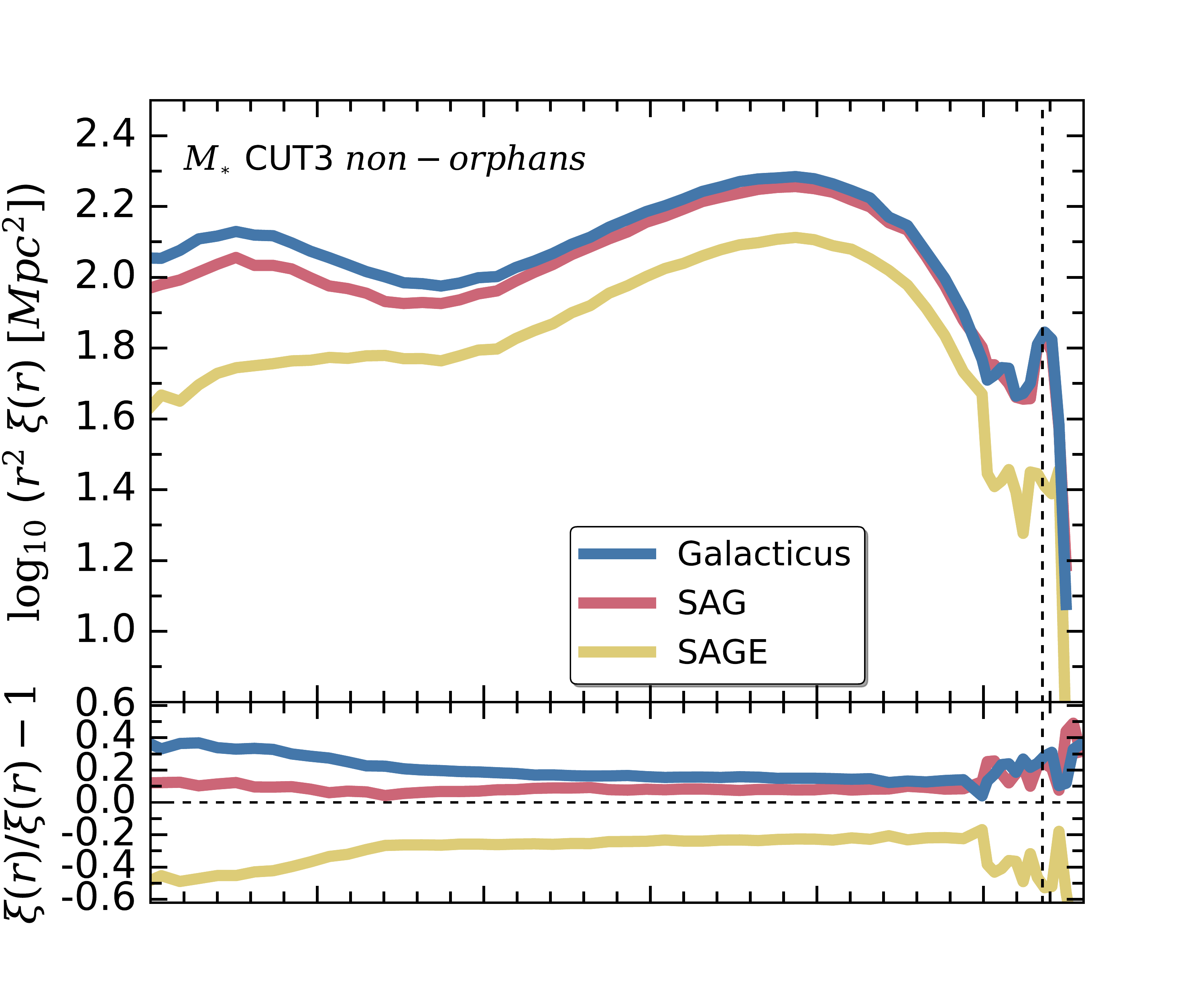}\hspace{-1.0cm}%
	  \includegraphics[width=7.0cm]{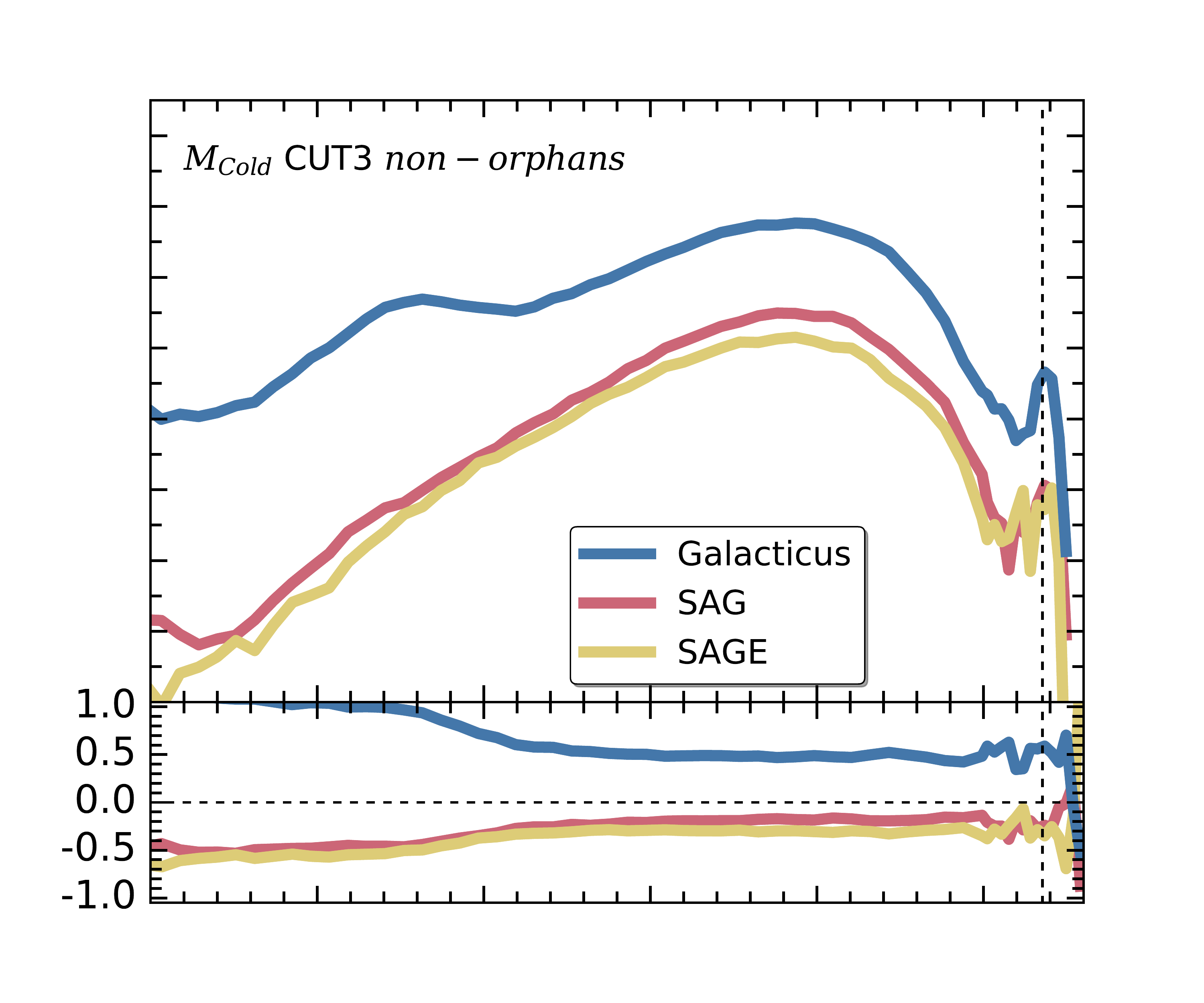}\hspace{-1.0cm}%
	  \includegraphics[width=7.0cm]{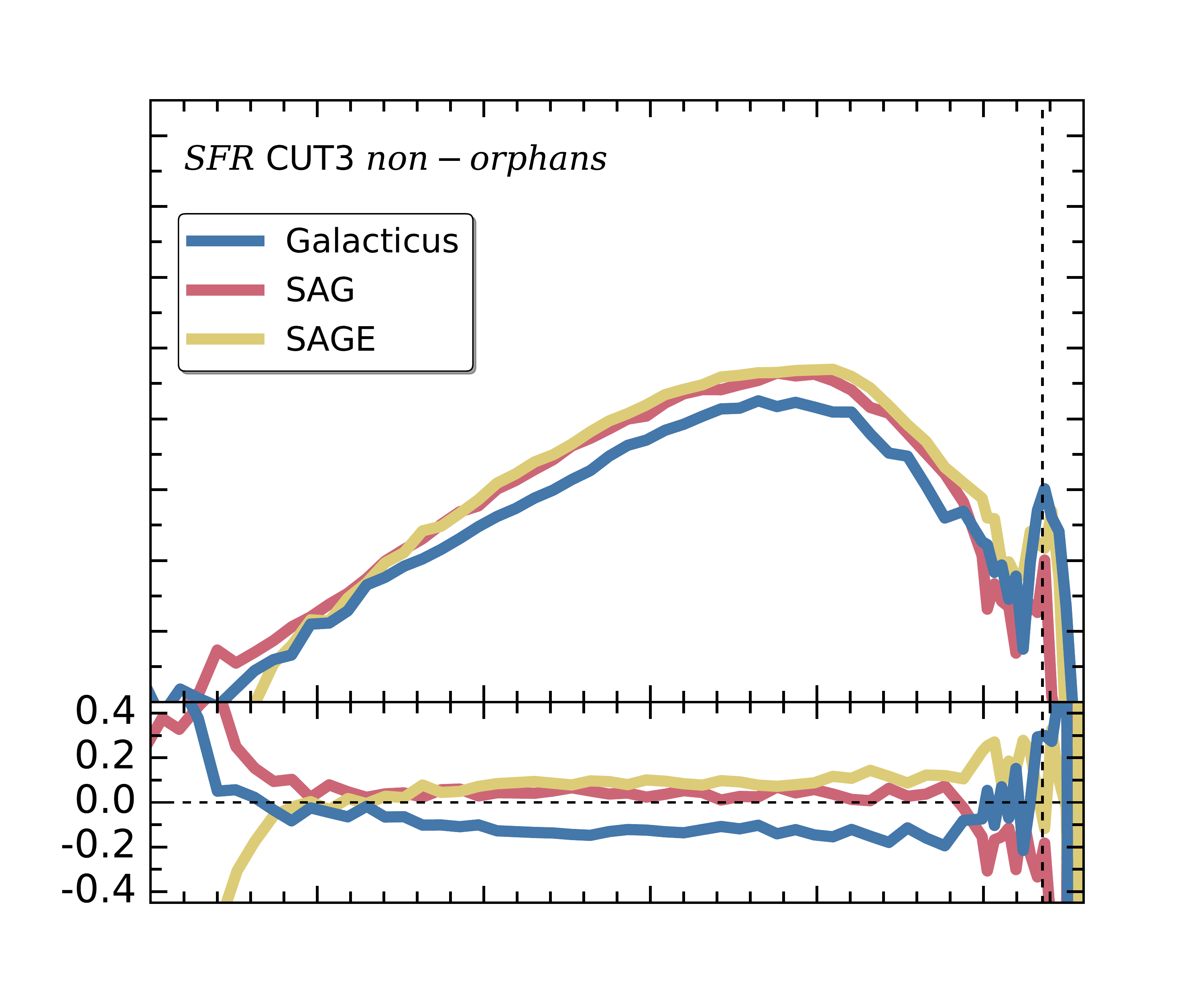}\hspace{-1.0cm}\vspace{-1.4cm}%
	\end{subfigure}
	\begin{subfigure}{\linewidth}
	  \includegraphics[width=7.0cm]{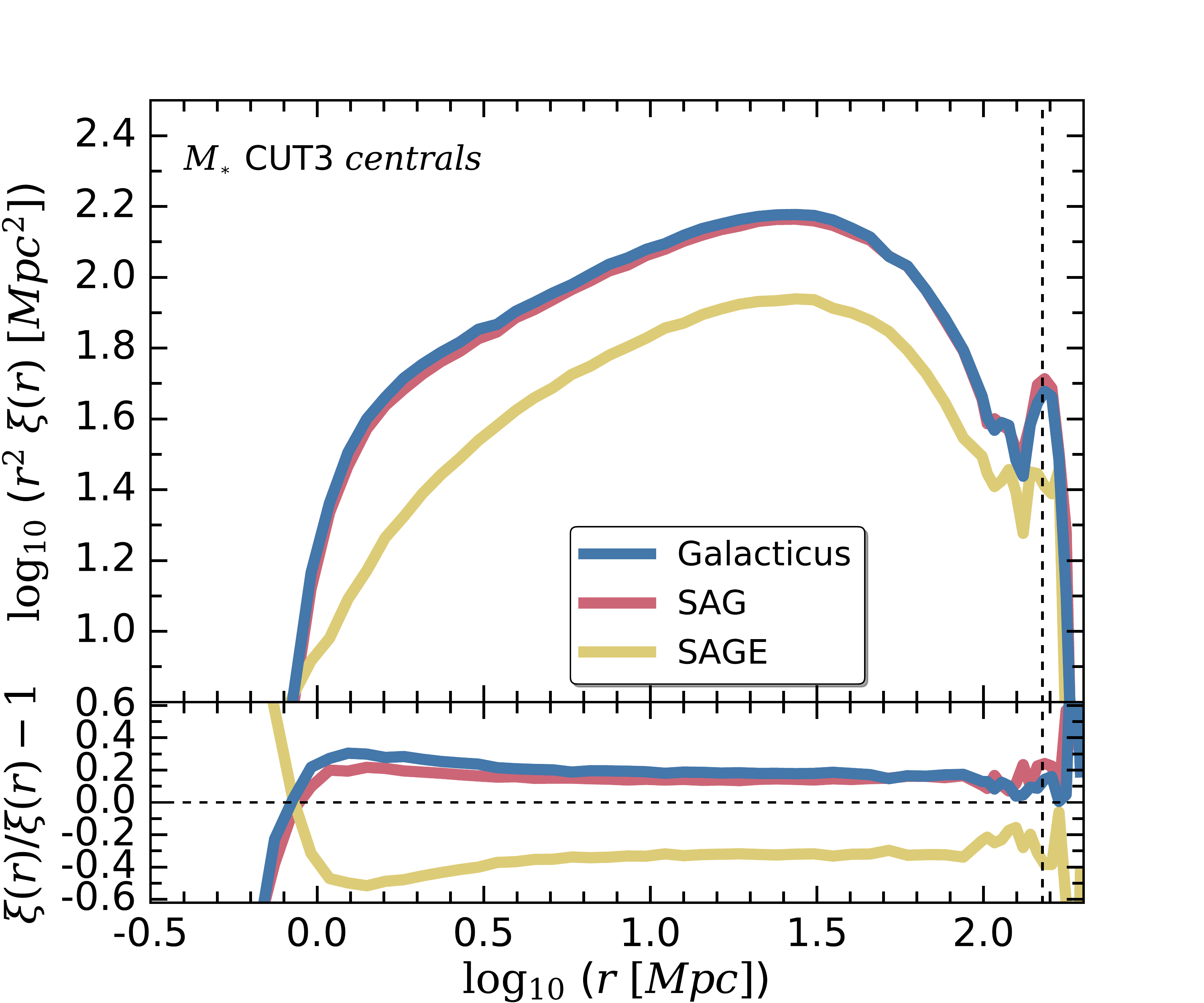}\hspace{-1.0cm}%
	  \includegraphics[width=7.0cm]{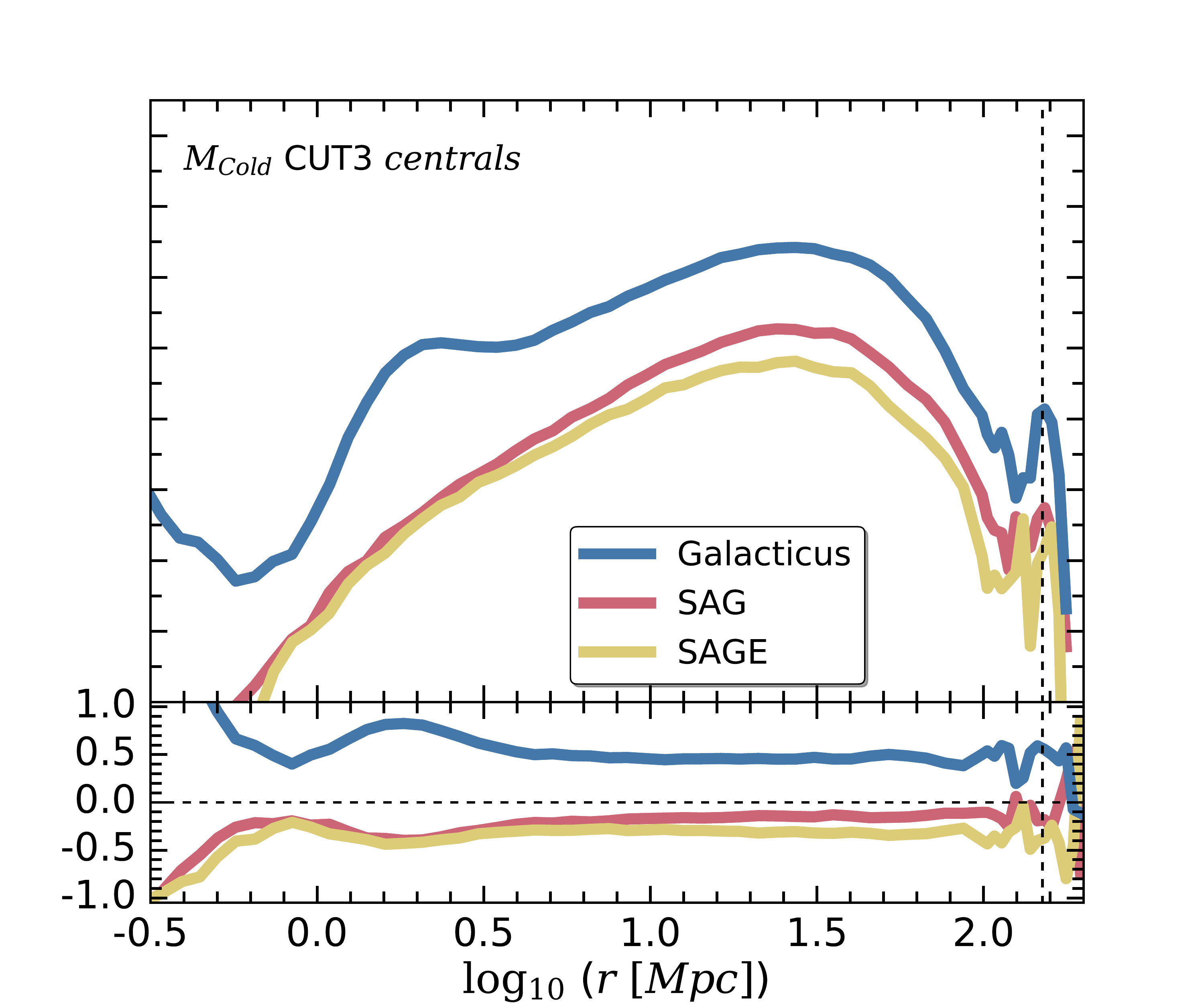}\hspace{-1.0cm}%
	  \includegraphics[width=7.0cm]{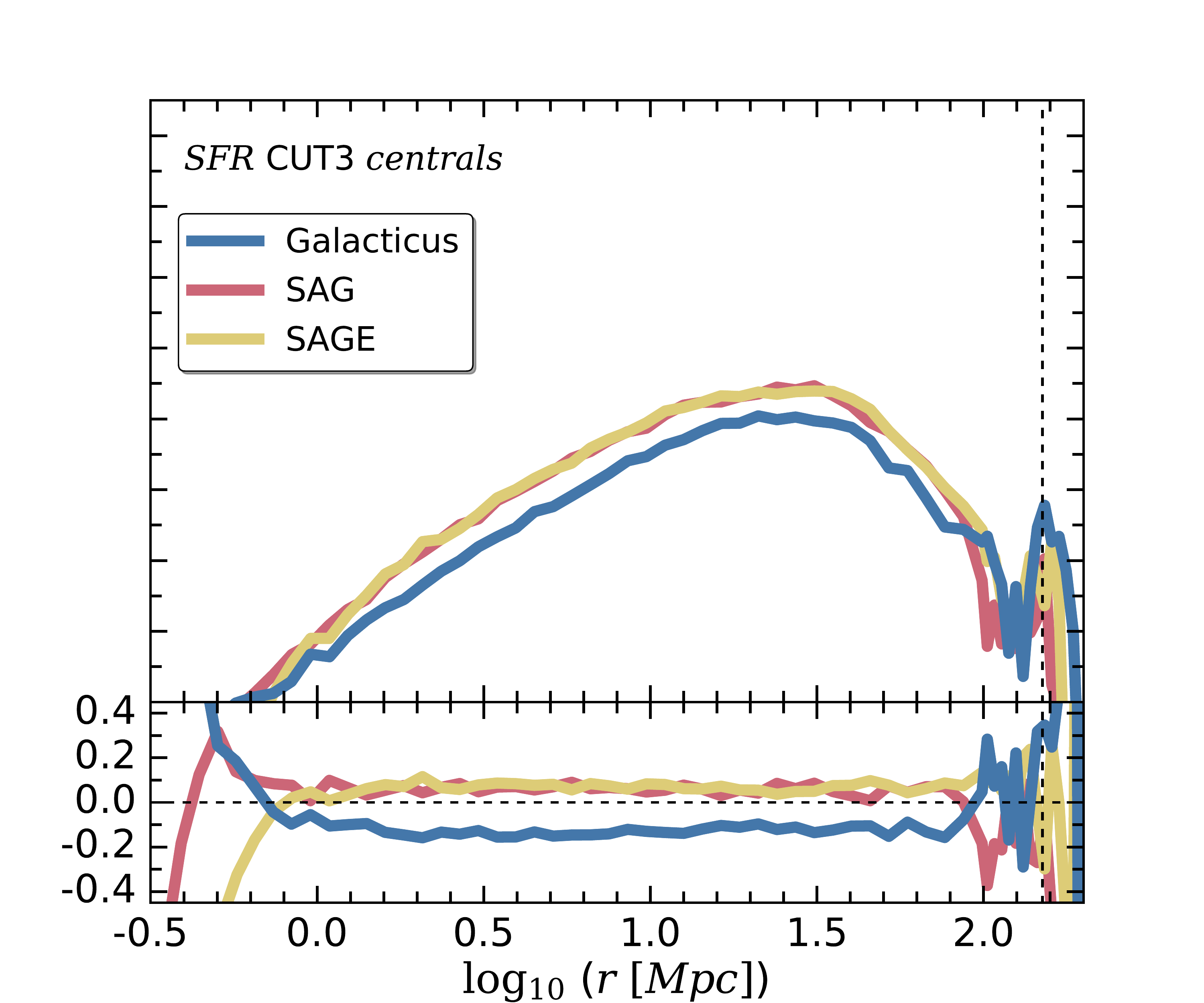}\hspace{-1.0cm}\vspace{-0.4cm}%
	\end{subfigure}
	\caption{Same as \hyperref[fig:2PCF_CUT1]{\Fig{fig:2PCF_CUT1}}, but for CUT3.}
	\label{fig:2PCF_CUT3}
	\vspace{-0.3cm}
\end{figure*}

The spatial distribution of galaxies and their clustering properties in the matter density field carries an extensive amount of information, especially about cosmological parameters. Because of this, we are witnessing an ever-growing demand for mapping the three-dimensional distribution of galaxies across the sky and throughout the Universe through either ground-based (e.g. eBOSS, J-PAS, DES, HETDEX, DESI) or space-born (e.g. Euclid, WFIRST) missions.
%and determinations of the two-point correlation function from it \citep{Zehavi05,Zehavi:11}.
But the interpretation of those (redshift or photometric) galaxy surveys requires exquisite theoretical modelling. First and foremost, galaxies only serve as tracers of the underlying (dark) matter density field. And while galaxies do form within the potential wells of dark matter \citep[e.g.][]{White78}, their clustering amplitude cannot straightforwardly be related to the clustering amplitude of the matter density field due to the uncertainties in the bias relation. Further, individual surveys target only certain galaxies which introduces another level of bias and complexity.

In a recent work carried out as part of the `nIFTy Cosmology' program\footnote{\url{http://popia.ft.uam.es/nIFTyCosmology}} we have presented a clustering comparison of 12 galaxy formation models, including variants of the SAM models presented here \citep{Pujol17}. Like in the present study, all models were applied to the same halo catalogues and merger trees, but the sidelength of the cosmological box was only 62.5\hMpc\ and hence probing galaxy clustering on much smaller scales. \citet{Contreras13}, on the other hand, used two different SAM models to study the two-point correlation function in the Millenium simulation. While both works found that the models generally agree in their clustering predictions, the observed differences for small scales reported in \citet{Pujol17} can be attributed to orphan galaxies. Here we extend such a study by investigating the clustering properties on much larger scales. We will nevertheless put a focus on one of the prime differences between our three SAM models, i.e. the treatment of orphan galaxies. For the calculation of the 2PCFs we used the \textsc{corrfunc} software package\footnote{\url{http://corrfunc.readthedocs.io/en/master/index.html}\label{ftn:corrfunc}}. \textsc{corrfunc} is a set of high-performance routines to measure clustering statistics in a simulation box or on a mock catalogue \citep{Sinha17}. To calculate the correlation functions we are using always 60 log-spaced bins in the range of $0.1 < r_p < 200$ \Mpc\ and in case of calculating the projected correlation function we integrate up to $\pi_{max}=60$ \Mpc.

For the calculation of the two-point correlation function (in real-space) we divide our galaxy catalogues into distinct galaxy sub-samples following the ideas of \citet{Contreras13} by applying various cuts in number density. This initial idea of comparing catalogues from galaxy formation models at a fixed number density was developed by \citet{Berlind03} and \citet{Zheng05} within their analysis of a Halo-Occupation-Distribution (HOD) from hydrodynamical and semi-analytic models. By comparing the models at a fixed abundance, the authors were able to single out common features in their models. We are now choosing the same density cuts as given in \citet{Contreras13} -- who applied the same procedure -- and listed here again in \hyperref[tab:ncut]{\Tab{tab:ncut}}. Those cuts\footnote{We like to remark that the applied cuts in $M_{*}$ select galaxies more massive than $10^{9}$\Msun. However, the cuts in \Mcold\ and \SFR\ will give galaxies with much lower stellar mass in the respective sample.} are applied to all our SAMs by using the 
\begin{itemize}
  \item[a)] cumulative stellar mass function,
  \item[b)] cumulative cold gas mass function, and
  \item[c)] cumulative star formation rate.
\end{itemize}
The respective distributions are shown in \hyperref[fig:CF]{\Fig{fig:CF}} and our applied cuts are illustrated as dashed lines. We like to remark that fixing the number density results in selecting galaxies for the three models with different cuts in the respective galaxy property. To better understand how the constant number density cut translates into the corresponding lower limit for the property in each model we show in \hyperref[fig:CF]{\Fig{fig:CF}} (as vertical lines) the intersection of the cumulative property distribution function with the applied number density cut. The resulting lower limits are additionally listed in \hyperref[tab:ncut]{\Tab{tab:ncut}}.

Before calculating the two-point correlation functions we further sub-divided the `3 models $\times$ 3 CUTs' roster of catalogues into three different galaxy populations: `all' referring to the whole sample, `centrals' restricting the calculation to central galaxies (i.e. galaxies residing at the centre of their main host halo, see definition in \hyperref[fig:halopointers]{\Fig{fig:halopointers}}), and `non-orphans' (i.e. galaxies with a host subhalo). Note, \sage\ does not feature orphans and hence the `all' and `non-orphans' sample are identical for this model. Further, \galacticus\ does not integrate the orbits of orphan galaxies but rather stores the position of dark matter halo at the time it was last found in the merger tree. While this makes their positions not suitable, in order not to loose the orphan galaxies and their contribution to at least the two-halo term\footnote{The one-halo term measures clustering on scales smaller than the typical size of haloes, i.e. correlations of substructure -- whereas the two-halo term quantifies the clustering of distinct haloes. But please note that substructure also contributes to the two-halo term, i.e. subhaloes in different distinct haloes are adding to the large-scale clustering signal.} of the correlation function we assign to them the position of the central galaxies of the halo they orbit in.

The clustering results for the three CUT samples is shown in \hyperref[fig:2PCF_CUT1]{\Fig{fig:2PCF_CUT1}} (CUT1), \hyperref[fig:2PCF_CUT2]{\Fig{fig:2PCF_CUT2}} (CUT2), and \hyperref[fig:2PCF_CUT3]{\Fig{fig:2PCF_CUT3}} (CUT3) as a 3$\times$3 grid on which the rows refer to `all' (upper), `non-orphan' (middle), and `central' (lower) galaxies and the columns to cuts in \Mstar (left), \Mcold (middle), and \SFR (right). Each individual panel is further sub-divided into an upper part where we show the actual correlation function (multiplied by $r^2$ for clarity) and a lower part showing the fractional difference to the mean curve $\bar{\xi}(r)=\sum_{i=1}^{3}\xi_{i}(r)/3$ (summing over the three models). The vertical line indicates the position of the Baryonic Acoustic Oscillations peak \citep{Beutler11}. In the following sub-section those figures will be discussed in the context of 

\begin{itemize}
  \item variations in number density, i.e. CUT1 vs. CUT2 vs. CUT3,
  \item changing galaxy property to define the sample, i.e. \Mstar\ vs. \Mcold\ vs. \SFR,
  \item different galaxy populations, i.e. `all' vs. `centrals' vs. `non-orphans'
  \item model-to-model variations, i.e. \galacticus\ vs. \sag\ vs. \sage
\end{itemize}

%%%%%%%%%%%%%%%%%%%%%%%%%%%%%%%%%%%%%%%%%%%%%%%%%%%%%
\subsection{Number Density Influence}\label{sec:CUTEnumberdensity}
%%%%%%%%%%%%%%%%%%%%%%%%%%%%%%%%%%%%%%%%%%%%%%%%%%%%%
As expected, we clearly observe that the correlation functions become more noisy when lowering the number density cut -- especially on large scales. We also find that this introduces more disparity between the different models. For instance, the variations between \galacticus\ and \sag/\sage\ for \Mcold\ non-orphans is minimal for CUT1/2, whereas it rises to 50 per cent when considering the CUT3 sample. As a matter of fact, the clustering continuously decreases for \SFR-selected galaxies in \galacticus\ when lowering the threshold -- whereas it remains rather constant for the other two models. For galaxies selected via a \Mstar-cut, we find that lowering the threshold increases the correlation on small scales. This is primarily driven by non-central galaxies for which the clustering on small scales naturally declines (see discussion in \hyperref[sec:CUTEgalaxypopulation]{\Sec{sec:CUTEgalaxypopulation}} below). The number density cuts have the smallest effect on \sag\ and \sage\ as well as galaxies selected via a \SFR-cut: here we only observe a general increase of the noise level.

%%%%%%%%%%%%%%%%%%%%%%%%%%%%%%%%%%%%%%%%%%%%%%%%%%%%%
\subsection{Galaxy Property Influence}\label{sec:CUTEgalaxyproperty}
%%%%%%%%%%%%%%%%%%%%%%%%%%%%%%%%%%%%%%%%%%%%%%%%%%%%%
We remind the reader that lowering the number density cuts for the \Mstar\ selection basically means restricting the analysis to  more massive galaxies, lowering in \Mcold\ selects those with huge reservoirs of cold gas (i.e. galaxies with lower stellar mass according to \hyperref[fig:mcold2mstar]{\Fig{fig:mcold2mstar}}), and lowering \SFR\ corresponds to preferring star forming galaxies.

We observe that preferring star forming galaxies primarily affects the two-point correlation function due to a change in number density: the overall shape is preserved -- at least on scales $r\gtrsim 1$\Mpc. The largest effect is found when changing the \Mstar\ number density cut. But this can be explained by the fact that more massive galaxies tend to be centrals and hence restricting the analysis to them will wash out any clustering signal on scales $r\lesssim 1-2$\Mpc, which is where the effect is observed to be strongest.

%%%%%%%%%%%%%%%%%%%%%%%%%%%%%%%%%%%%%%%%%%%%%%%%%%%%%
\subsection{Galaxy Population Influence}\label{sec:CUTEgalaxypopulation}
%%%%%%%%%%%%%%%%%%%%%%%%%%%%%%%%%%%%%%%%%%%%%%%%%%%%%
The difference between the three populations is that `centrals' limit the analysis to those galaxies that reside at the centre of a distinct dark-matter host halo, i.e. a halo that itself is not a subhalo of any larger object. For this sample, we do not expect a strong clustering signal on scale $r$\ltsima 1--2\Mpc\ which corresponds to the size of these objects. The `non-orphans' are a class of galaxies that do have a dark-matter host (sub-)halo which itself could be a distinct halo or a subhalo. Restricting the analysis to such objects comes closest to methods where dark matter halo catalogues are populated with galaxies by means of, for instance, halo abundance matching (HAM) as presented in the recent study by \citet{Rodriguez-Torres16} for the BOSS galaxy clustering. The `all' sample now covers all galaxies for which positional information is available, and that might include orphan galaxies (only for \sag\ though).

The main observation for changes in the galaxy population is the division of the clustering signal into a contribution from scales larger than the typical size of dark matter haloes and correlations inside those haloes, i.e. the decomposition into the so-called two- and one-halo term. We find that `non-orphans' show correlations below $r$\ltsima 1\Mpc\ whereas this is suppressed for `centrals', especially for the CUT3 sample.

%%%%%%%%%%%%%%%%%%%%%%%%%%%%%%%%%%%%%%%%%%%%%%%%%%%%%
\subsection{SAM Model Influence}\label{sec:CUTEsammodel}
%%%%%%%%%%%%%%%%%%%%%%%%%%%%%%%%%%%%%%%%%%%%%%%%%%%%%
We like to restate that one of the obvious differences between the models is the treatment of orphan galaxies: \galacticus\ provides physical properties for orphan galaxies (like masses, star formation rates, luminosities, etc.), but does not integrate the orbits; we therefore assigned the position of the central galaxy they orbit to them. \sag\ follows the trajectories of orphans after their dark matter halo disappeared and hence gives full information; \sage\ does not provide any information on orphans at all. 

We observe that differences between models only become apparent when lowering the threshold for the CUT. While the clustering signal in general follows the same shape with differences in the amplitude of order less than 20 per cent, it rises above that for CUT3. But the model-to-model variations also depend on the galaxy property used in the CUT-selection. For instance, the largest model-to-model variations are found for galaxies selected via the \SFR-cut. Here we observe deviations larger than 20 per cent across all CUT samples. And the increase in model differences when lowering the \Mstar-threshold is just a reflection of the differences seen in the stellar mass function in \hyperref[fig:SMF]{\Fig{fig:SMF}}. \galacticus\ and \sag\ have a very similar high-mass end of the \SMF\ and also show comparable clustering properties for these objects (also see  \hyperref[fig:CF]{\Fig{fig:CF}}). Similar arguments can be used to explain the similarities and differences seen across the other CUT properties \Mcold\ and \SFR: models showing correspondence in these (distributions of) properties are also alike when it comes to the clustering signal. 

A lot of the differences seen in the 2PCF across models for various CUTs can also be attributed to the fact that keeping the number density constant leads to differing cuts in the respective galaxy property. This is readily verified in \hyperref[fig:CF]{\Fig{fig:CF}} where it can be seen that, for instance, CUT1 selects galaxies from the \galacticus\ catalogue with \Mstar$>10^{10}\Msun$ (\Mcold$>10^{10}\Msun$) whereas this mass limit is \Mstar$>10^{9}\Msun$ (\Mcold$>10^{9.3}\Msun$) for \sag. But we conclude that the shape of the 2PCF remains largely the same for the models and hence appears to be independent of the implementation of the physical processes. 
%However, its amplitude (and thus any measurement of the galaxy bias) are affected. For instance, star forming galaxies in \sage\ are less clustered than for the other two models, in particular for the more complete CUT1 and CUT2. This might to be due to the larger spread in the \SHMF\ as seen in \hyperref[fig:SHMF]{\Fig{fig:SHMF}}.

%%%%%%%%%%%%%%%%%%%%%%%%%%%%%%%%%%%%%%%%%%%%%%%%%%%%%
\subsection{Comparison to SDSS main galaxies}\label{sec:CUTEp2CF}
%%%%%%%%%%%%%%%%%%%%%%%%%%%%%%%%%%%%%%%%%%%%%%%%%%%%%
We close the presentation of the clustering statistics with a comparison of our model projected 2PCF (p2PCF) to different samples drawn from the SDSS DR7 main galaxy sample \citep{Strauss02}. To this extent, we selected SAM galaxy samples within the following four absolute $r$-band magnitude bins

\begin{itemize}
  \item[(a)] $M_r \in [-19,-18]$,
  \item[(b)] $M_r \in [-20,-19]$,
  \item[(c)] $M_r \in [-21,-20]$, and
  \item[(d)] $M_r \in [-22,-21]$.
\end{itemize}

Note that the samples are only selected by $r$-band magnitude and no additional cuts have been made.

As for the real-space correlation function  we used the \textsc{corrfunc} Python package and compute the projected correlation function by choosing an integration length of $\pi_{max}=60$ \Mpc. We also tested if a different integration length would change our results, but cannot report any relevant differences when using $\pi_{max}=[40,80,100]$ \Mpc. 

\begin{figure*}
	  \includegraphics[width=9cm]{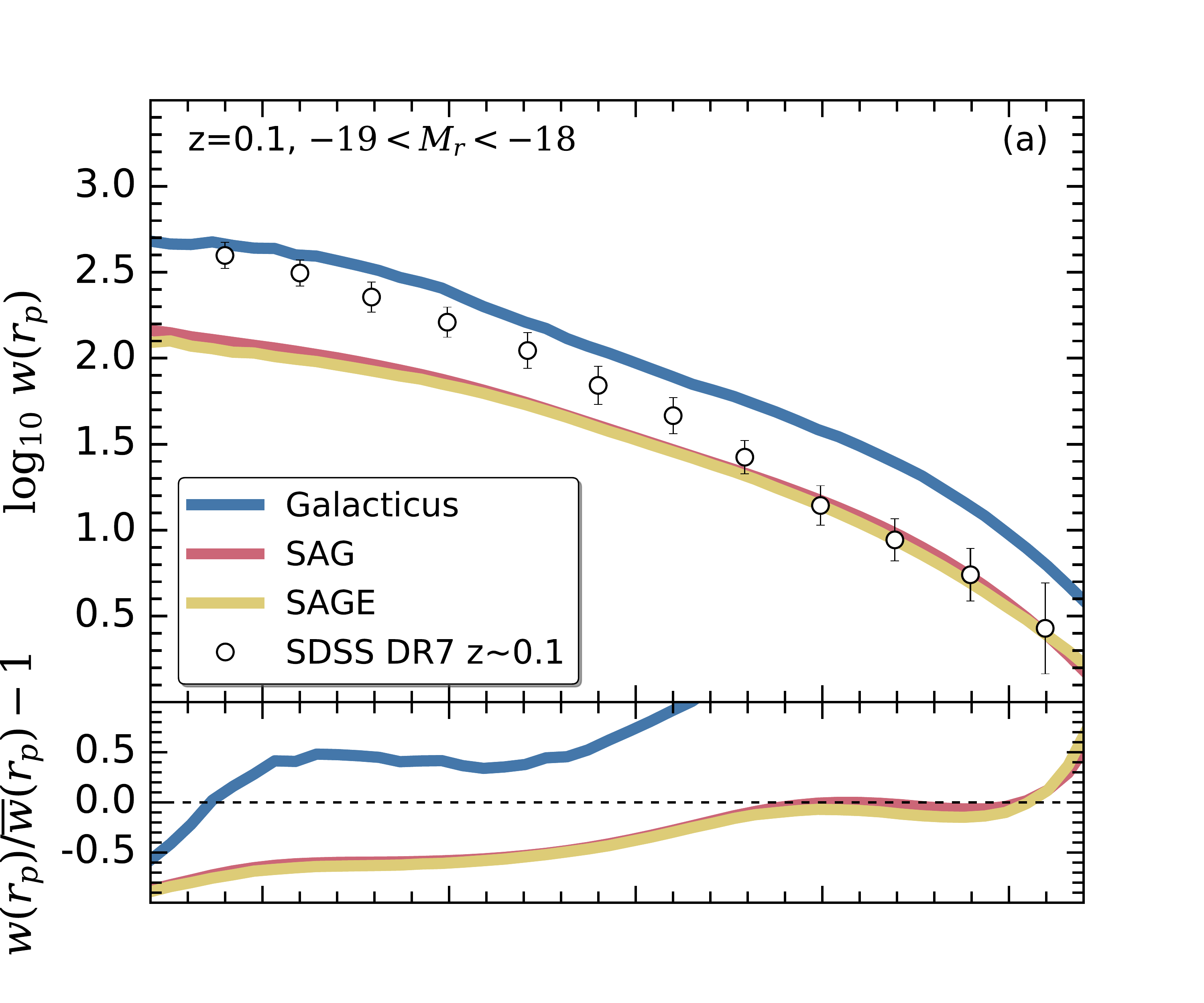}\hspace{-1.9cm}\vspace{-1.4cm}%
	  \includegraphics[width=9cm]{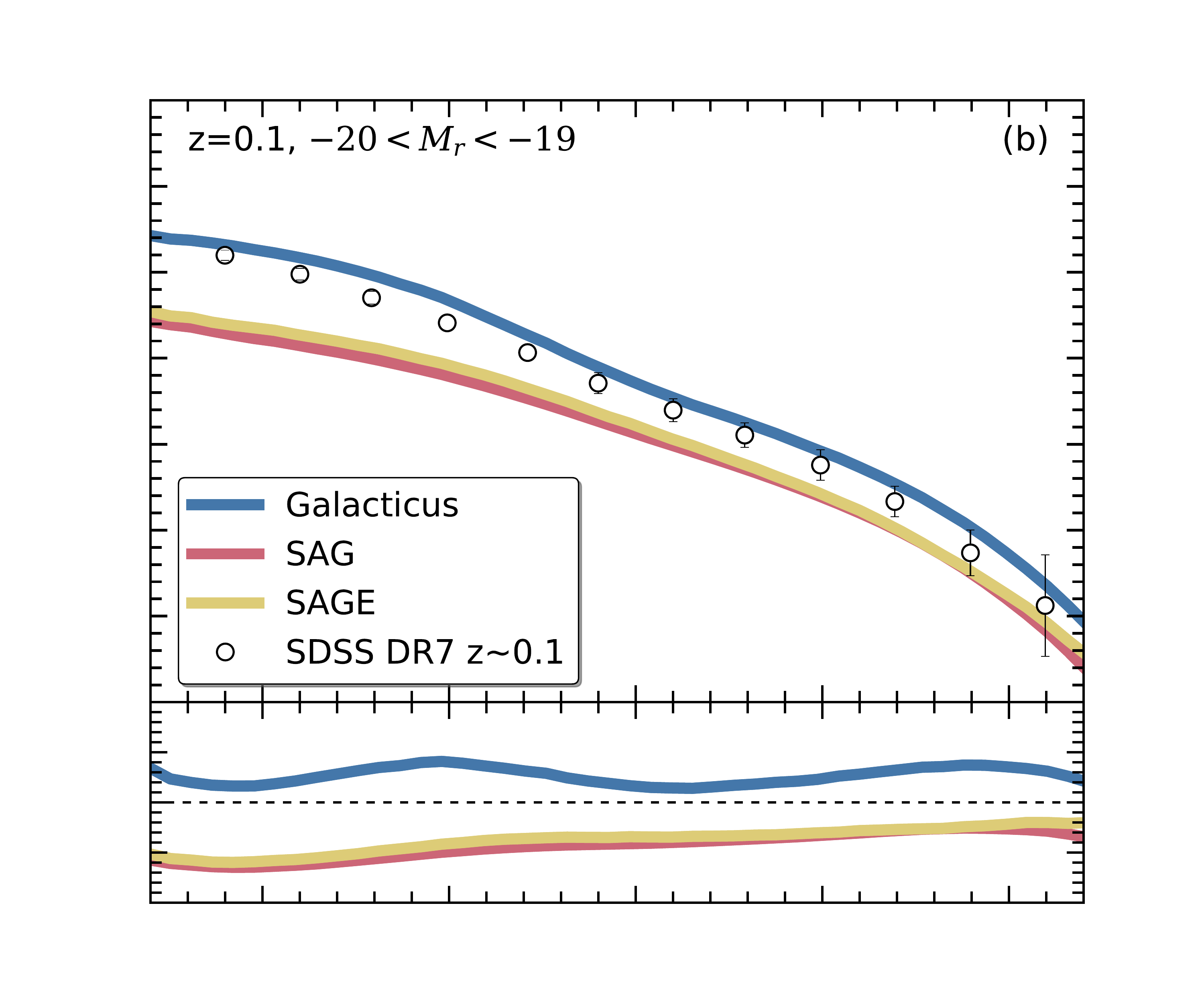}
	  \includegraphics[width=9cm]{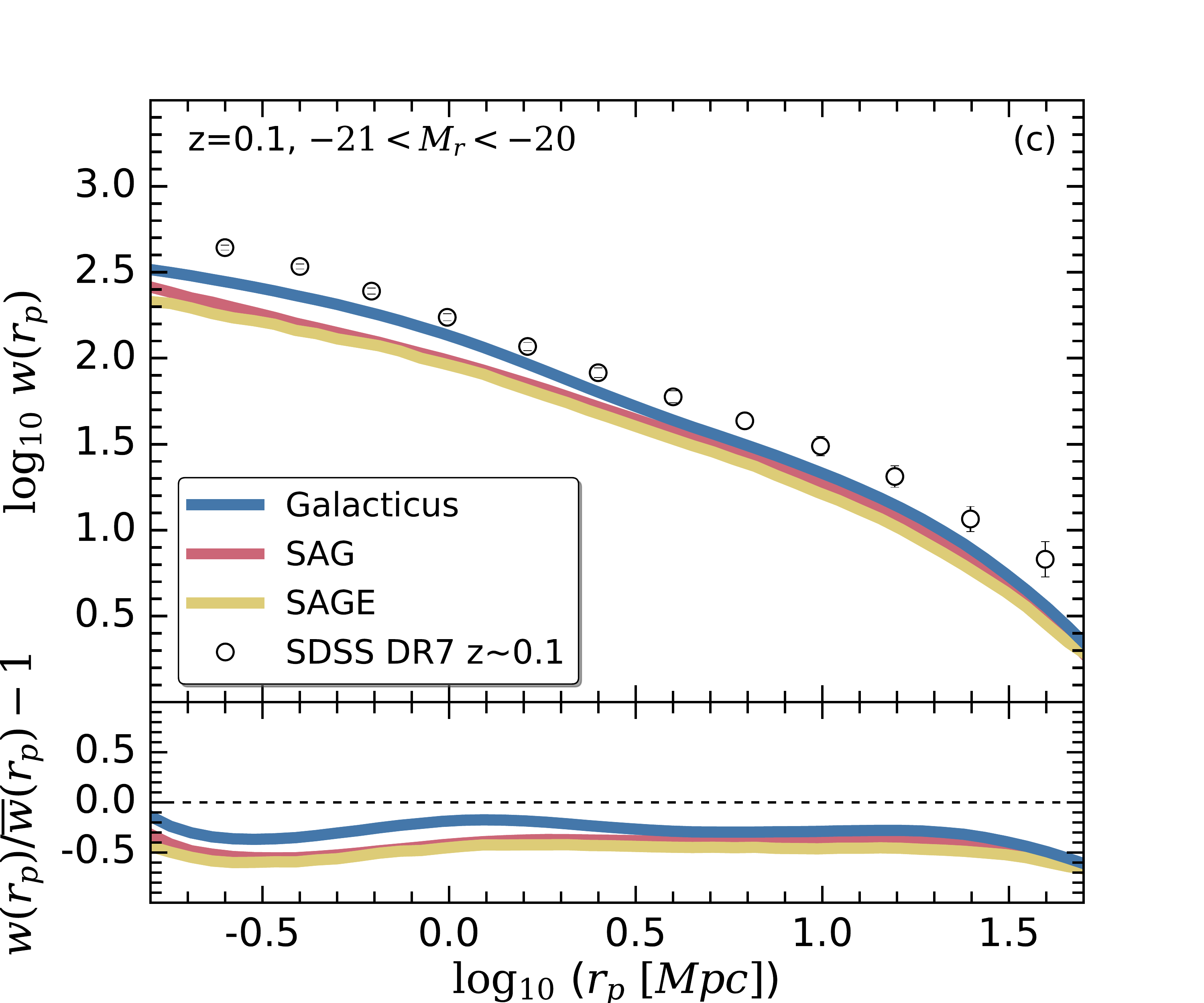}\hspace{-1.9cm}%
	  \includegraphics[width=9cm]{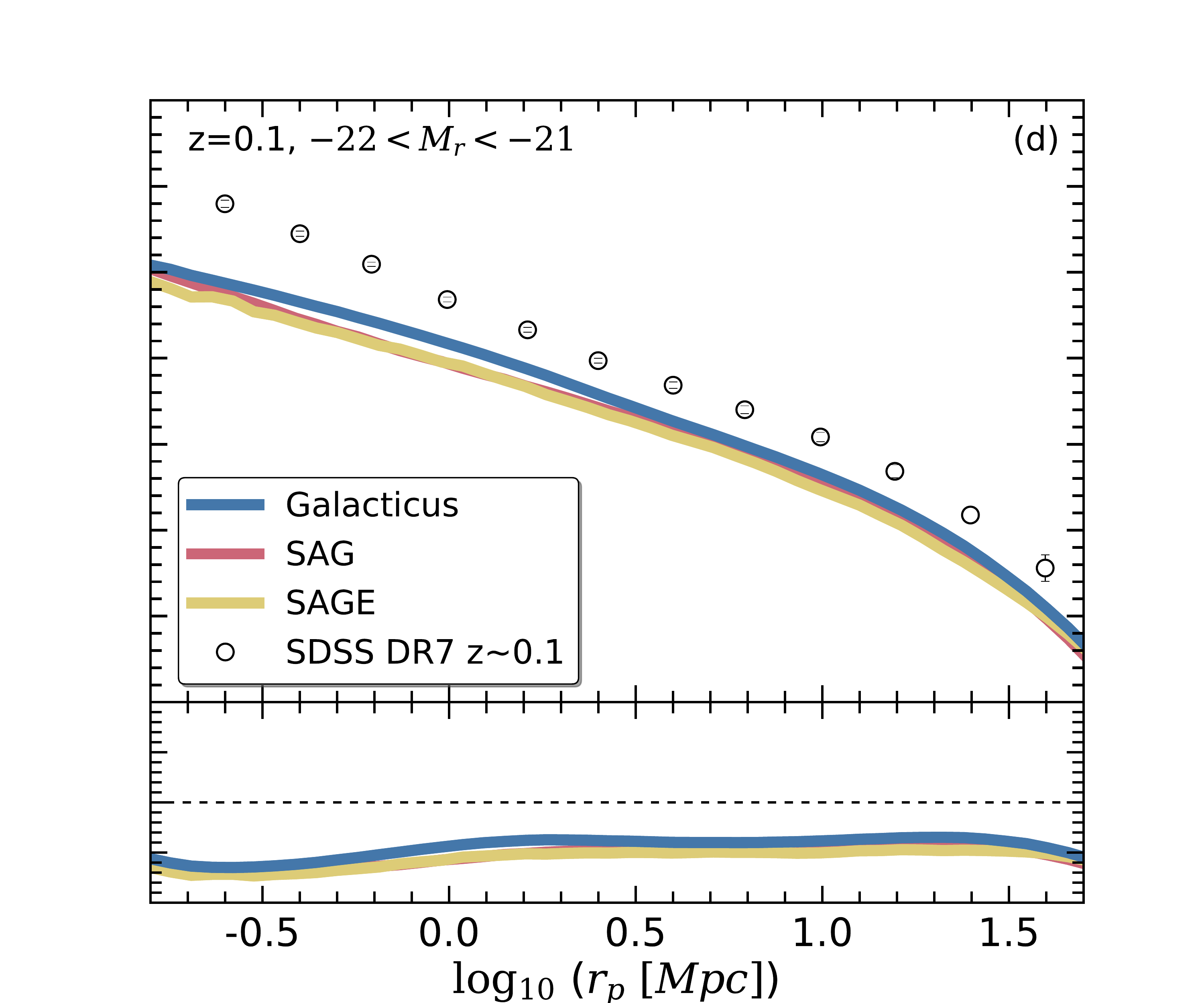}
	\caption{The projected 2PCF for different $r$-band abslute magnitude $M_r$ bins compared to SDSS DR7 observations in the same bins taken from \citet{Zehavi:11}. The bottom panels are again the fractional difference with respect to the mean $\bar{w}(r_p)$ (defined in the same way as for \hyperref[fig:2PCF_CUT1]{\Fig{fig:2PCF_CUT1}}).}
	\label{fig:p2PCF}
	\vspace{-0.3cm}
\end{figure*}

Our results can be viewed in \hyperref[fig:p2PCF]{\Fig{fig:p2PCF}} where show the p2PCF for the aforementioned four magnitude bins. In each of the panels we further compare them to the SDSS results from \citet[Table 7]{Zehavi:11} at $z\sim0.1$, within the same magnitude bins. The upper part of each panel shows the correlation function with the observations as open circles and the lower part represents the residuals with respect to the observations in the respective magnitude bin. In \hyperref[tab:ncut_p2PCF]{\Tab{tab:ncut_p2PCF}} we show the number densities and the fraction of satellites and orphan satellites, respectively, of our samples presented in \hyperref[fig:p2PCF]{\Fig{fig:p2PCF}}. 

All our models reproduce the basic features of the observational p2PCF, and the transition from the one-halo to the two-halo term at around $r_p\sim$1-2~\Mpc\ is well described. Especially in the bin (a) where \sag and \sage reproduce the SDSS clustering signal perfectly for large separations, and in (b) where \galacticus\ describes the observational data best, and within 15-40 per cent. This can be understood if we take a look at the fraction of satellites in \hyperref[tab:ncut_p2PCF]{\Tab{tab:ncut_p2PCF}}. \galacticus\ shows the largest fraction of satellites -- when combining satellites and orphans together -- and the largest fraction of orphans, respectively. As we discussed in previous sections, this again confirms how strong the clustering behaviour correlates with the galaxy type (see Figs.~\ref{fig:2PCF_CUT1}-\ref{fig:2PCF_CUT3}): models (in our case \sag\ and \sage) with smaller satellite fraction lack clustering power on small scales, but nevertheless reproduce the observed p2PCF very well beyond the one-halo term. However, we also need to remind the reader that the positions for the orphans in \galacticus\ coincide with the position of the central galaxy as that model does not integrate the orbits of satellite galaxies once they are stripped off their dark matter halo. And this artificially enhances the clustering signal. But it is remarkable to note that \sage\ -- the model without any orphans -- basically provides identical results to \sag\ -- the model with the most sophisticated treatment of orphan positions. However, we also need to acknowledge that our cuts in magnitude introduce a selection-bias: we have seen in the upper panels of \hyperref[fig:colours]{\Fig{fig:colours}} that while all models feature red and blue galaxies, their exact locii in the colour-magnitude diagram are shifted with respects to each other. Therefore, using fixed bins in magnitude will select different populations.

If we consider the brighter magnitude end, as shown in panels (c) and (d), the clustering signals of the models are almost fully in agreement with each other. However, \galacticus\ always shows the largest clustering strength as seen before in panels (a) and (b). But for all of the SAM models the p2PCF is shifted downwards in amplitude about 50-80 per cent across the whole separation range.

\begin{table}
	\caption{Number density measured in (\hMpc)$^{-3}$, for the selection of galaxy samples for the projected 2PCF calculation and the fractions of satellite and orphan satellites, respectively, in the four distinct magnitude bins used for \hyperref[fig:p2PCF]{\Fig{fig:p2PCF}}.}
	\label{tab:ncut_p2PCF}\vspace{-0.4cm}
	\begin{center}
		\begin{tabular}{p{1.2cm}p{0.6cm}ccc}
		\hline
		\textsc{panel}	& & & & \\
		$M_r$ \textsc{bin}	     &			& \galacticus\ 			& \sag 				& \sage	\\						
		\hline
		\hline
		(a)			     & $n_{\rm gal}$ 	& 12.22{$\times10^{-3}$}	& 17.39{$\times10^{-3}$}	& 18.86{$\times10^{-3}$} 	 \\
		\multirow{3}{*}{$[-19,-18]$} & f$_{\rm sats}$	& 0.11				& 0.15				& 0.12	 \\
		                             & f$_{\rm orphans}$ 	& 0.58				& 0.09 				& -	 \\
		\hline
		(b)			     & $n_{\rm gal}$ 	& 20.34{$\times10^{-3}$}	& 10.65{$\times10^{-3}$}	& 15.39{$\times10^{-3}$} 	\\
		\multirow{3}{*}{$[-20,-19]$} & f$_{\rm sats}$	& 0.18 				& 0.16 				& 0.15	\\
		                             & f$_{\rm orphans}$ 	& 0.28 				& 0.05				& -	\\
		\hline
		(c)			     & $n_{\rm gal}$ 	& 17.36{$\times10^{-3}$}	& 8.07{$\times10^{-3}$}		& 11.10{$\times10^{-3}$}	\\
		\multirow{3}{*}{$[-21,-20]$} & f$_{\rm sats}$	& 0.17				& 0.17				& 0.15	\\
		                             & f$_{\rm orphans}$ 	& 0.10				& 0.03 			& -	\\
		\hline
		(d)			     & $n_{\rm gal}$ 	& 6.48{$\times10^{-3}$}		& 3.49{$\times10^{-3}$}		& 10.01{$\times10^{-3}$}	\\
		\multirow{3}{*}{$[-22,-21]$} & f$_{\rm sats}$	& 0.17				& 0.12				& 0.13	\\
		                             & f$_{\rm orphans}$ 	& 0.03			& 0.02	 			& -	\\
		\hline
		\vspace{-0.8cm}
		\end{tabular}
	\end{center}
\end{table}

%%%%%%%%%%%%%%%%%%%%%%%%%%%%%%%%%%%%%%%%%%%%%%%%%%%%%
\section{Summary and Discussion}\label{sec:discussion}
%%%%%%%%%%%%%%%%%%%%%%%%%%%%%%%%%%%%%%%%%%%%%%%%%%%%%
We present the public data release of three distinct galaxy catalogues from the three semi-analytic models \galacticus, \sag, and \sage\ as applied to the same underlying cosmological dark-matter simulation \MDPL. The two latter models \sag\ and \sage\ have been re-calibrated to the simulation whereas \galacticus\ has been used with its standard choice for the parameters. In the first part of the paper, we compared the model galaxies to observational data. This serves as a gauge for the performance of the models. Even though the general aim of each SAM is to model galaxy formation, it is  important to bear in mind that models might be tuned to serve different purposes. Therefore, our three models perform differently as they put their focus differently: \sage\ fits multiple observables simultaneously, first and foremost the stellar mass function and stellar-to-halo mass relation; \galacticus\ has its strength in the star formation rate function and evolution; and \sag\ is a model with strengths in providing reasonable gas fractions and metallicity relations. Further, the most recent changes implemented into the \sag\ model (cf. \hyperref[sec:sag]{\Sec{sec:sag}}) produce galaxies with properties in excellent agreement with observations such as the galaxy main sequence (sSFR vs. stellar mass, \hyperref[fig:sSFR2Mstar]{\Fig{fig:sSFR2Mstar}}) and the mass-metallicity relation (\hyperref[fig:zgas2mstar]{\Fig{fig:zgas2mstar}}), yet showing an excess of galaxies at the high-mass end of the \SMF\ at redshift $z=0$. These `model priorities' are certainly reflected by the plots presented in \hyperref[sec:comparison]{\Sec{sec:sag}}. We have seen that \sag\ fits the \sSFR-$M_*$ relation of \citet{Elbaz11} much better than both \galacticus\ and \sage, \galacticus fits the cosmic SFR density at low-$z$ better than the other two SAMs, but it does it because its under-efficient star formation (low \sSFR) is compensated by an excessive stellar mass density, and \sage\ fits the SDSS+GALEX data much better than \galacticus\ and \sag. We relate the latter to the distinct treatment of orphans in \sage. This model does not feature any galaxies devoid of a dark matter halo but rather disrupts them adding their stars to an intra-cluster component. While both other models treat such a component differently, it furnishes \sage\ with the possibility to deposit stars that in the other two models find their way into the galaxies and hence leading to a larger stellar mass than for \sage. And while \sag\ also features such a component, the implementation of tidal stripping appears to be too inefficient and hence leading to an under-estimated stellar content in its ICC. While this difference cannot explain all of the deviations seen at the high-$M_{*}$ end in the \SMF\ plot  \hyperref[fig:SMF]{\Fig{fig:SMF}} it certainly plays a significant role. For all these reasons of different model designs, we considered it important to have not only a single but multiple galaxy formation models available exploring different approaches to galaxy formation physics.
 
In the second part of the paper we applied three galaxy number density cuts in stellar mass, cold gas fraction, and star formation rate to define various sub-samples of galaxies for a study of the two-point correlation function. We confirm the results recently reported by \citet{Pujol17}, i.e. even though there might be noteworthy variations of internal properties of galaxies across different SAMs, the positions are stable and there is only very little scatter in the clustering properties of our galaxies -- irrespective of the selection criterion for the chosen sub-sample. The 2PCF shape largely remains the same across all models (at least on scales $\gsim 1$\Mpc) and hence appears to be independent of the implementation of the physical processes. However, its amplitude (and thus any measurement of the galaxy bias) is affected. We further confirm that all our models reproduce the \textit{observed} projected 2PCF albeit again showing model-to-model variations. This might again be attributed to variations in the treatment of orphan galaxies and number densities of galaxies in the respective magnitude bin, but also relates to the fact that the applied magnitude cuts introduce a selection bias. 

We conclude that the models applied here and the galaxy catalogues based upon them will be a valuable asset to the community and can be readily used for science that requires reliable galaxy information in volumes large enough to match on-going and upcoming surveys. And unless SAM models are specifically designed to predict (and/or describe) the same galaxy properties, physical processes are treated identically, and calibration has been performed in an identical manner, model-to-model variations as seen here are expected \citep[][Knebe et al., in prep.]{Lee14,Knebe15}: models perform differently reflecting their individual designs. Therefore, it appears important to not only regard a single model but a selection of models when studying mock galaxies in order to properly capture such scatter. However, one might argue that a better approach would be to fine-tune each model to the actual simulation until the observations used in that calibration procedure are best reproduced. But this becomes intrinsically difficult the larger the simulations are and sub-sets have to be used for the parameter adjustment. Further, even a scrupulous re-calibration will not guarantee that different galaxy formation models will all give the same results (see Knebe et al., in prep.). To achieve perfect agreement a universal protocol would need to be defined that involves using the same observational data sets, the same allowance for scatter during the calibration, the same assumption for initial mass functions, the same yields, the same recycled fractions, etc. But in the end differing implementations of the same physics will eventually leave us with some level of residual variance \citep[e.g.][]{Fontanot09,Lu14}.

We close with the remark that this paper only forms the first in a series where the models and their galaxies will be studied in far more detail. This paper simply introduces the three galaxy catalogues (\galacticus, \sag, and \sage) populating a common dark matter simulation (\MDPL) that is large enough to tackle cosmological questions such as the position and width of the Baryon Acoustic Oscillation peak and how this is affected by baryon physics. Besides of publicly releasing all data, the source code of two of the galaxy models (\galacticus\ and \sage) is open too, allowing the community to explore the impact that the specific modelling of a physical process has on different measurements used in cosmology, open the possibility to also explore the cross correlation of different cosmological tracers

%%%%%%%%%%%%%%%%%%%%%%%%%%%%%%%%%%%%%%%%%%%%%%%%%%%%%
\section*{Acknowledgments}
%%%%%%%%%%%%%%%%%%%%%%%%%%%%%%%%%%%%%%%%%%%%%%%%%%%%%

AK is supported by the {\it Ministerio de Econom\'ia y Competitividad} and the {\it Fondo Europeo de Desarrollo Regional} (MINECO/FEDER, UE) in Spain through grant AYA2015-63810-P.  He also acknowledges support from the {\it Australian Research Council} (ARC) grant DP140100198 and further thanks The Go-Betweens for 16 lovers lane.

DS wants to acknowledge the following software tools and packages the author is using throughout this work: \textsc{matplotlib}\footnote{\url{http://matplotlib.org/}\label{ftn:matplotlib}} 2012-2016, \citet{Matplotlib}; \textsc{Python Software Foundation}\footnote{\url{http://www.python.org}\label{ftn:python}} 1990-2017, version 2.7., \textsc{Pythonbrew}\footnote{\url{https://github.com/utahta/pythonbrew}\label{ftn:pythonbrew}}; \textsc{Cosmolopy}\footnote{\url{http://roban.github.io/CosmoloPy/docAPI/cosmolopy-module.html}\label{ftn:cosmolopy}}; we use whenever possible in this work a colour-blind friendly colour palette\footnote{\url{https://personal.sron.nl/~pault/}\label{ftn:cb-friendly}} for our plots; the author's fellowship is funded by the \textit{Spanish Ministry of Economy and Competitiveness} (MINECO) under the 2014 \textit{Severo Ochoa} Predoctoral Training Programme. This research made use of the \textit{K-corrections calculator} service available at \url{http://kcor.sai.msu.ru/}.

DS and FP acknowledge funding support from the MINECO grant AYA2014-60641-C2-1-P.

SAC acknowledges grants from Consejo Nacional de Investigaciones Cient\'ificas y T\'ecnicas (PIP-112-201301-00387), Agencia Nacional de Promoci\'on Cient\'ifica y Tecnol\'ogica (PICT-2013-0317), and Universidad Nacional de La Plata (UNLP 11-G124), Argentina.

NDP was supported by BASAL PFB-06 CATA, and Fondecyt 1150300.  Part of the calculations presented here were run using the Geryon cluster at the Center for Astro-Engineering at U. Catolica, which received funding from QUIMAL 130008 and Fondequip AIC-57.

CVM was supported by a fellowship from CONICET, Argentina.

PB was supported by program number HST-HF2-51353.001-A, provided by NASA through a Hubble Fellowship grant from the Space Telescope Science Institute, which is operated by the Association of Universities for Research in Astronomy, Incorporated, under NASA contract NAS5-26555.

VGP acknowledges support from the University of Portsmouth through the Dennis Sciama Fellowship award.

GY is supported by the {\it Ministerio de Econom\'ia y Competitividad} and the {\it Fondo Europeo de Desarrollo Regional} (MINECO/FEDER, UE) in Spain through grant AYA2015-63810-P.

We all thank the anonymous referee whose suggestions and remarks greatly helped to improve the paper.

The \textsc{CosmoSim} database\footnote{\url{http://www.cosmosim.org}} used in this paper is a service by the Leibniz-Institute for Astrophysics Potsdam (AIP). The \MD database was developed in cooperation with the Spanish \MD Consolider Project CSD2009-00064.

The \sage\ galaxy model is a publicly available codebase and is available for download at \url{https://github.com/darrencroton/sage}. Magnitudes for \sage\ galaxies were generated using Swinburne University's Theoretical Astrophysical Observatory (TAO). TAO is part of the Australian All-Sky Virtual Observatory (ASVO) and is freely accessible at \url{https://tao.asvo.org.au}.

The authors gratefully acknowledge the Gauss Centre for Supercomputing e.V. (www.gauss-centre.eu) and the Partnership for Advanced Supercomputing in Europe (PRACE, www.prace-ri.eu) for funding the \MD simulation project by providing computing time on the GCS Supercomputer SuperMUC at Leibniz Supercomputing Centre (LRZ, www.lrz.de). The \MDPL\ simulation has been performed under grant pr87yi.

The authors contributed to this paper in the following ways: AK and FP initiated and coordinated the project. DS created all the plots. The three of them wrote the paper (with help from the remaining authors). GY (together with SG and AAK) ran the \MDPL\ simulation. SG, HE, NIL, and MSt provided access to the database to which KR uploaded the simulation, halo catalogues, merger trees, and eventually galaxy catalogues. PB generated the halo catalogues and merger trees. CB and AB ran the \galacticus\ model over the simulation. SAC, NDP, CAVM, and ANR ran the \sag\ model over the simulation. DJC, ARHS, and MSi ran the \sage\ model. All authors proof-read and commented on the paper.

This research has made use of NASA's Astrophysics Data System (ADS) and the arXiv preprint server.

\bibliographystyle{mn2e}
\bibliography{MDarchive}

\appendix

%%%%%%%%%%%%%%%%%%%%%%%%%%%%%%%%%%%%%%%%%%%%%%%%%%%%%%%%%%%%%%%%%%%%%%%%%%%
\section{Database Release} \label{app:database}
%%%%%%%%%%%%%%%%%%%%%%%%%%%%%%%%%%%%%%%%%%%%%%%%%%%%%%%%%%%%%%%%%%%%%%%%%%%
All the data used for this paper are publicly available. While we refer to \hyperref[sec:simulation]{\Sec{sec:simulation}} for a description of the simulation, halo catalogues, and merger trees, we like to present here some of the particulars of the galaxy catalogues. The data can be individually referenced by using a Digital Object Identifier (DOI): we list them in \hyperref[tab:DOI]{\Tab{tab:DOI}} for the three models in the database.

\galacticus\ has been run in its native configuration, whereas \sag\ and \sage\ retuned their parameters to the \MDPL\ simulation. In \hyperref[tab:database]{\Tab{tab:database}} we list those properties that are common to all models and for which we chose identical names in the database. Those properties have also been converted to the same units. For \galacticus\ and \sag\ luminosities/magnitudes have also been uploaded to the database whereas for \sage\ they have to be generated by the user on TAO. We further encourage the reader to visit the database website and the additional documentation provided there as the list of galaxy properties for each model is not limited to what is shown in \hyperref[tab:database]{\Tab{tab:database}}: for each model substantially more information has been added to the database.

\begin{figure}
   \includegraphics[width=\columnwidth]{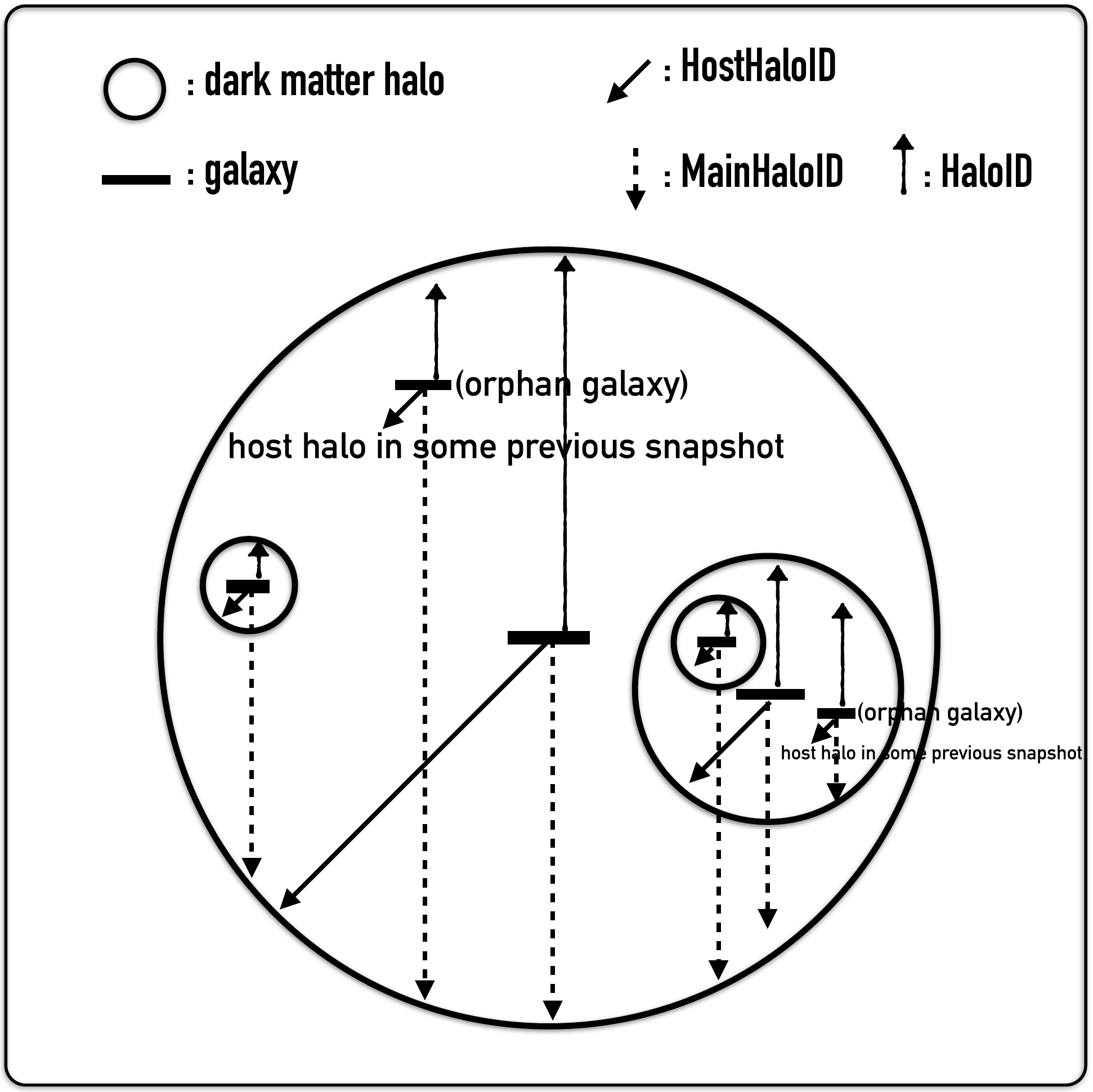}\vspace{-0.2cm}
   \caption{Illustrating the various pointers to haloes in which galaxies are residing.}
 \label{fig:halopointers}
 \vspace{-0.3cm}
 \end{figure}

We further provide in \hyperref[fig:halopointers]{\Fig{fig:halopointers}} the nomenclature for the pointers to the haloes of the galaxies. A \textsc{HostHaloID} will point to the immediate dark matter host halo around the galaxy, which does not exist anymore for orphan galaxies by definition (but points to the last halo to which the galaxy belonged, i.e. a halo from a previous snapshot). The \textsc{MainHaloID} pointer will give access to the top-level dark matter halo in which the galaxy orbits while \textsc{HaloID} points to the lower-level halo around the galaxy. Note that \textsc{HaloID=HostHaloID} for all but orphan galaxies, and that \textsc{HaloID} only exists for \sag\ (which is why it is omitted from the list in \hyperref[tab:database]{\Tab{tab:database}}).

\begin{table}
	\caption{DOI's for the three models.}
	\label{tab:DOI}
	\begin{center} 
		\begin{tabular}{ll}
		\hline
		\textsc{catalogue} & DOI \\
		 \hline \hline
		 \MDPL-\galacticus 		& doi:10.17876/cosmosim/mdpl2/009 \\
		 \hline
		 \MDPL-\sag 			& doi:10.17876/cosmosim/mdpl2/007 \\
		 \hline
		 \MDPL-\sage 			& doi:10.17876/cosmosim/mdpl2/008 \\
		 \hline
		\end{tabular}
	\end{center}
\end{table}

\begin{savenotes}
\begin{table*}
  \caption{Set of galaxy properties common to all semi-analytic galaxy formation models. For a sketch explaining the halo pointers please refer to Fig.1 of \citet{Knebe15}. Note that $x,y,z,v_x,v_y,v_z$ have been integrated for orphans in \sag\ yet are unavailable for \galacticus\ (the \sage\ model does not feature orphans). Please note that many more than the properties listed here have been uploaded to the database; please refer to the database website for more information.}
\label{tab:database}
\begin{center}
\begin{tabular}{lll}
\hline
database name			& unit 			& description\\
\hline
\hline
redshift				& n/a				& redshift $z$\\
HostHaloID			& n/a				& pointer to dark matter halo in which galaxy resides;\\
					&				& not applicable for orphan galaxies\\
MainHaloID			& n/a				& pointer to dark matter halo in which galaxy orbits\\
GalaxyType			& n/a				& 0 = central galaxy\\
					&				& 1 = satellite galaxy\\
					&				& 2 = orphan galaxy (only for \galacticus\ and \sag)\\
X					& comoving \hMpc	& $x$-position of galaxy\\
Y					& comoving \hMpc	& $y$-position of galaxy\\
Z					& comoving \hMpc	& $z$-position of galaxy\\
Vx					& peculiar km/s		& $v_x$-velocity of galaxy \\
Vy					& peculiar km/s		& $v_y$-velocity of galaxy	 \\
Vz					& peculiar km/s		& $v_z$-velocity of galaxy\\
MstarSpheroid			& \hMsun			& stellar mass of bulge component of galaxy\\
MstarDisk				& \hMsun			& stellar mass of disc component of galaxy\\
McoldSpheroid			& \hMsun			& cold gas mass of bulge component of galaxy\\
McoldDisk				& \hMsun			& cold gas mass of disc component of galaxy\\
Mhot					& \hMsun			& total hot gas mass in galaxy\\
Mbh					& \hMsun			& mass of central black hole\\
SFR					& \hMsun/Gy		& total star formation rate\\
SFRspheroid			& \hMsun/Gy		& star formation rate in bulge component of galaxy\\
SFRdisk				& \hMsun/Gy		& star formation rate in disc component of galaxy\\
MeanAgeStars			& Gyrs			& mean age of all stars\\
HaloMass				& \hMsun			& $M_{\rm 200c}$ of galaxy's dark matter halo\\
Vmax				& km/s			& peak circular rotation velocity of galaxy's dark matter halo\\
Vpeak				& km/s			& maximum Vmax across all redshifts\\
NFWconcentration		& n/a				& concentration of galaxy's dark matter halo\\
SpinParameter			& n/a				& spin parameter $\lambda$ of galaxy's dark matter halo\\
MZstarSpheroid		& \hMsun			& mass of metals in stellar component of bulge\\
MZstarDisk			& \hMsun			& mass of metals in stellar component of disc\\
MZgasDisk			& \hMsun			& mass of metals in gas component of disc\\
MZhotHalo			& \hMsun			& mass of metals in hot gas component of halo\\
\hline
\underline{\galacticus\ luminosities\footnote{dust corrected luminosities, band-pass shifted to the emission rest-frame (cf. \hyperref[sec:lf]{\Tab{sec:lf}} for how to convert them to absolute rest-frame magnitudes)} and metalicities:}\\
LstarSDSSu			& $4.4659\times10^{13}$ W/Hz		& total stellar luminosity in SDSS u band\\
LstarSDSSg			& $4.4659\times10^{13}$ W/Hz		& total stellar luminosity in SDSS g band\\
LstarSDSSr			& $4.4659\times10^{13}$ W/Hz		& total stellar luminosity in SDSS r band\\
LstarSDSSi			& $4.4659\times10^{13}$ W/Hz		& total stellar luminosity in SDSS i band\\
LstarSDSSz			& $4.4659\times10^{13}$ W/Hz		& total stellar luminosity in SDSS z band\\
MZgasSpheroid		& \hMsun						& mass of metals in gas component of bulge\\
\hline
\underline{\sag\ magnitudes\footnote{dust corrected absolute rest-frame magnitudes} and metalicities:}\\
MagStarSDSSu		& n/a				& magnitude in SDSS u band\\
MagStarSDSSg		& n/a				& magnitude in SDSS g band\\
MagStarSDSSr			& n/a				& magnitude in SDSS r band\\
MagStarSDSSi			& n/a				& magnitude in SDSS i band\\
MagStarSDSSz			& n/a				& magnitude in SDSS z band\\
MZgasSpheroid		& \hMsun			& mass of metals in gas component of bulge\\
\hline
\underline{\sage\ luminosities and metalicities:}\\
- to be processed via TAO\\
- no additional metalicities\\
\\
\hline
\end{tabular}
\end{center}
\end{table*}
\end{savenotes}

%%%%%%%%%%%%%%%%%%%%%%%%%%%%%%%%%%%%%%%%%%%%%%%%%%%%%%%%%%%%%%%%%%%%%%%%%%%
\section{Summary of plots in Section~3} \label{app:plotsummary}
%%%%%%%%%%%%%%%%%%%%%%%%%%%%%%%%%%%%%%%%%%%%%%%%%%%%%%%%%%%%%%%%%%%%%%%%%%%
To facilitate the reading of \hyperref[sec:comparison]{\Sec{sec:comparison}} and provide more convenient access to the information about the data presented in this paper we summarize in \hyperref[tab:plots]{\Tab{tab:plots}} all the plots to be discussed in that Section. That table lists what galaxy property (or correlation between properties) is presented in which sub-section of the paper. It further indicates whether or not any selection criterion for our model galaxies has been applied. The following columns then provide information about the reference data used for each particular plot, i.e. the actual bibliographic reference, the redshift range of that data, the IMF entering into the derivation of that data.

\begin{savenotes}
\begin{table*}
\begin{center}
    \caption{Here we provide a short description of the plots we present in \hyperref[sec:comparison]{\Sec{sec:comparison}}. The first column `\textsc{property}' corresponds to the physical or statistical property under investigation. The second column points to the `\textsc{sub-section}' where the plot is discussed. The third column indicates whether we applied any cut to the data. The fourth column provides the reference for the observational data or other computations. The fifth and sixth column likewise give the redshift and IMF for the observational/reference data. If that reference data is not based upon a \citet{Chabrier03} IMF, we convert it.}
    \label{tab:plots} \vspace{-0.2cm}
	\setlength{\tabcolsep}{0.5pt}
   \begin{tabular}{l|c|c|c|c|c|}
 \hline
  \textsc{property} 				& \textsc{sub-section} 						& \textsc{selection} 				& \textsc{reference}  				& \textsc{redshift} 	& {\textsc{IMF}} \\
  \hline \hline
	{Stellar mass function} 		& \hyperref[sec:smf]{\ref{sec:smf}} 				&  NO 						& SDSS-Galex  				& 0.1 			& \citet{Chabrier03} \\
	\textsc{(\SMF)} 				& 										& 							& \citep{Moustakas2013_SMF} 		& 				& 				\\
	\hline
	{Star formation rate} 			& \hyperref[sec:SFRF]{\ref{sec:SFRF}} 			& NO 						& GOODS-S+COSMOS/PACS+ 	& $0.0<z<0.3$ 		& \citet{Chabrier03} \\
	\textsc{function (\SFRF)} 		& 										& 							& Herschel \citep{Gruppioni15} 		& 				& \\
	\hline
	{Specific \SFR  to stellar} 		& \hyperref[sec:ssfr2mstar]{\ref{sec:ssfr2mstar}} 	& NO 						& \citet{Elbaz11} 				& 0.0 			& \citet{Salpeter55} \\
    	\textsc{mass function} 		& & & & & \\
	\hline
	{Cosmic star formation} 		& \hyperref[sec:cSFR]{\ref{sec:cSFR}} 			&  \sSFR                   				& \citet{Behroozi13c_cSFRF} 		& $0.0<z<8.0$ 		& \citet{Chabrier03} \\
	{rate density (\cSFRD)} 		&                                                           			& $>10^{-11}yr^{-1}$ 			& 							& 				& \\	
	\hline
	{Black hole to} 				& \hyperref[sec:BHBM]{\ref{sec:BHBM}} 			& NO 						& \citet{Kormendy13} 			& 0.0 			& dynamical zero-point \\
	{bulge mass (\BHBM)} 		& 										& 							& \citet{McConnell13} 			& 0.0 			& - \\
	\hline
	{Cold gas fraction to} 		& \hyperref[sec:mcold2mstar]{\ref{sec:mcold2mstar}} & NO 						& \citet{Boselli14} 				& 0.0 			& \citet{Chabrier03} \\
	{stellar mass (\CGF)} 		& 										 & 							& \citet{Peeples11} 				& 0.0 			& \citet{Chabrier03} \\
	\hline
	{Total gas-phase metallicity} 	& \hyperref[sec:zgas2mstar]{\ref{sec:zgas2mstar}} 	& NO 						& \citet{Tremonti04} 				& 0.1 			& \citet{Kroupa01} \\
	\hline	
	{Stellar to halo mass} 		& \hyperref[sec:SHMF]{\ref{sec:SHMF}} 			& non-orphans 					& \citet{Behroozi10} 				& 0.1 			& \citet{Chabrier03} \\
	{function (\SHMF)} 			& 										&  							& 							&  				& \\
	\hline
	{Luminosity function (\LF)} 	& \hyperref[sec:lf]{\ref{sec:lf}} 					& NO 						& SDSS \citep{Montero09} 		& 0.1 			& - \\
	 \hline
 	{Colour diagrams} 			& \hyperref[sec:lf]{\ref{sec:lf}} 					& \Mstar 						& `red'-`blue' separation  			& 0.1 			& - \\
 							& 										& $>1 \times 10^8$ $[M_{\odot}]$ 	& \citet{Strateva01} 				& 				&\\
	 \hline	
	\vspace{-0.6cm}
	\end{tabular}
\end{center}
\end{table*}
\end{savenotes}

\bsp

\label{lastpage}

\end{document}